\documentclass[twocolumn,amsmath,amssymb,prd,dvipdfmx]{revtex4}

\usepackage{graphicx}
\usepackage{dcolumn}
\usepackage[varg]{txfonts}
\usepackage{bm}

\usepackage{color}

\begin{document}

\title{Impacts of biasing schemes in
  the one-loop integrated perturbation theory}

\author{Takahiko Matsubara} \email{taka@kmi.nagoya-u.ac.jp}
\affiliation{%
  Department of Physics, Nagoya University, Chikusa, Nagoya 464-8602,
  Japan;}%
\affiliation{%
  Kobayashi-Maskawa Institute for the Origin of Particles and the
  Universe (KMI), Nagoya University, Chikusa, Nagoya 464-8602, Japan}%

\author{Vincent Desjacques} \email{Vincent.Desjacques@unige.ch}
\affiliation{D\'epartement de Physique Th\'eorique and Center for Astroparticle Physics (CAP)
Universit\'e de Gen\`eve, 24 quai Ernest Ansermet, CH-1211 Gen\`eve, Switzerland}

\date{\today}

\begin{abstract}

    The impact of biasing schemes on the clustering of tracers of the
    large-scale structure is analytically studied in the weakly
    nonlinear regime. For this purpose, we use the one-loop
    approximation of the integrated perturbation theory together with
    the renormalized bias functions of various, physically motivated
    Lagrangian bias schemes. These include the halo, peaks and
    excursion set peaks model, for which we derive useful formulas for
    the evaluation of their renormalized bias functions. The shapes of
    the power spectra and correlation functions are affected by the
    different bias models at the level of a few percent on weakly
    nonlinear scales. These effects are studied quantitatively both in
    real and redshift space. The amplitude of the scale-dependent bias
    in the presence of primordial non-Gaussianity also depends on the
    details of the bias models. If left unaccounted for, these
    theoretical uncertainties could affect the robustness of the
    cosmological constraints extracted from galaxy clustering data.

\end{abstract}

\maketitle

\section{\label{sec:Introduction}
Introduction
}

The large-scale structure (LSS) of the universe contains rich
information on cosmology. The LSS is mainly probed by the spatial
distributions of astronomical objects, such as galaxies, clusters of
galaxies, or any other tracer that can be observed in the distant
Universe (such as the Lyman-alpha forest etc.) The spatial
distribution of these objects differs from that of the total mass
(which includes the mysterious dark matter), while direct predictions
from cosmological theories are made for the mass distributions. In
fact, except for the lensing shear, essentially all observables of the
LSS are biased tracers of the mass distribution.

Although a relation between the spatial distribution of biased tracers
and that of the matter is not trivial at small scales owing to the
complexity of the physical processes governing star formation etc.,
the large-scale clustering of LSS tracers is much less complicated as
it is ``only'' governed by gravity. On very large scales, the biasing
is simply given by a linear relation \cite{Dav85,Kai84}, and all the
complications which arise from the biasing mechanisms are confined to a
single variable known as the linear bias factor. In particular, the
power spectrum $P_X(k)$ of biased tracers $X$ is linearly related to
that of the mass $P_\mathrm{m}(k)$ through
\begin{equation}
  P_X(k) = {b_X}^2 P_\mathrm{m}(k),
  \label{eq:1-1}
\end{equation}
where $b_X$ is the linear bias factor of $X$. The label $X$ represents
any kind of biased tracers, i.e. a particular type of galaxies or
clusters of galaxies within a certain range of mass for instance. The
correlation function, which is the three-dimensional Fourier transform
of the power spectrum, satisfies a similar relation, $\xi_X(r) =
{b_X}^2\xi_\mathrm{m}(r)$.

In redshift surveys, the radial distances to the objects are measured
by their redshifts. The observed redshifts are contaminated by the
peculiar velocities of the LSS tracers. As a result, clustering
patterns in redshift space are distorted along the lines of sight.
This effect is known as the redshift-space distortions. In the linear
regime, the redshift-space distortions of the power spectrum are
analytically given by Kaiser's formula \cite{Kai87},
\begin{equation}
  P_X(\bm{k}) = {b_X}^2 \left(1 + \beta_X \mu^2\right)^2
  P_\mathrm{m}(k),
  \label{eq:1-2}
\end{equation}
where $\mu = \hat{\bm{z}}\cdot\bm{k}/|\bm{k}|$ is the direction cosine
between the lines of sight $\hat{\bm{z}}$ and the wave vector
$\bm{k}$. The variable $\beta_X = f/b_X$, where $f=\ln D/\ln a$ is the
linear growth rate, is called the redshift-space distortion parameter.
The correlation function in redshift space is given by a Fourier
transform of the Kaiser's formula \cite{Ham92}.

However, the linear theory with linear bias is valid only in the
large-scale limit. It is severely violated at small scales where
nonlinearities induced by gravitational coupling become important, and
exact analytical treatments are extremely difficult. Fortunately,
there is an intermediate range of scales between the linear and the
highly nonlinear regimes where nonlinearities are weak, so that
statistical correlators such as the power spectrum and correlation
function are amenable to a perturbative treatment (for a review of
perturbation theory in LSS, see Ref.~\cite{Ber02}).

The traditional perturbation theory predicts weakly nonlinear
evolutions of unbiased dark matter in real space. The integrated
perturbation theory (iPT) \cite{Mat11, Mat14} is a general framework
to predict the weakly nonlinear power spectra and higher-order
polyspectra of biased tracers both in real space and in redshift
space. This is essential for the analysis of future redshift survey
data. Furthermore, the iPT can also include the effect of a primordial
non-Gaussianity in the curvature perturbation, which the power
spectrum of biased tracers is sensitive to \cite{Dal08}. In principle,
any bias model could be incorporated into the iPT. The dependence of
the polyspectra on the biasing scheme predicted by the theory is
encoded in the so-called {\it renormalized bias functions}. Hereby,
the framework of iPT separates the issue of biasing at small scales
from the weakly nonlinear dynamics at larger scales.

The iPT is based on the Lagrangian perturbation theory
\cite{buc89,mou91,buc92,cat95,bha96,cat98,por98,RB12}, and the renormalized
bias functions are directly calculated from the Lagrangian models of
bias, in which the bias relations are specified in Lagrangian space.
The bias relation is not necessarily a local function of the density
in Lagrangian space. In fact, it will involve e.g.~derivatives of the
linear density if a peak constraint is present \cite{BBKS,LMD15}, as
well as the tidal shear if the collapse is not spherical
\cite{BM96,OKT04,SCS13}. Any kind of bias is represented by a
``nonlocal'' bias in Lagrangian space, because all the structures in
the Universe are formed by a deterministic evolution of the initial
density field.

In this work, we investigate the predictions of one-loop iPT for
observables such as the power spectrum and correlation function with
representative models of Lagrangian bias. The biasing schemes
considered in this paper include the halo bias \cite{MW96,MJW97},
peaks model \cite{BBKS,LMD15}, and excursion set peaks (ESP)
\cite{AP90,PS12}. These Lagrangian biasing schemes are physically
motivated, and the mass scale is the only parameter left (once the
halo mass function or the collapse barrier is known).

The main goal of this paper is to see how differences in the
renormalized bias functions predicted by these models are reflected in
the weakly nonlinear power spectrum and correlation function. It is
not our purpose in this paper to find an accurate model of bias. We
are rather interested in assessing the extent to which observed
quantities are affected by uncertainties in the biasing. We naively
expect that those effects should not be very significant on large
scales, because the characteristic formation scales of astrophysical
objects are small. Furthermore, the large-scale behavior of the power
spectrum and the correlation function is not much affected by
small-scale dynamics, except for the scale-independent, linear bias
factor. However, scale-dependent corrections predicted, e.g., by a
peak constraint can affect the shape of a feature such as the baryon
acoustic oscillation \cite{VD08a,Des10}. These kind of effects cannot
be neglected, should they mimic a signature of fundamental physics
detectable in future LSS data or bias cosmological constraints.

Our paper is organized as follows. In Sec.~\ref{sec:iPT}, the
essential equations of the one-loop iPT used in this paper are
summarized. In Sec.~\ref{sec:seminonlocal}, the renormalized bias
functions in the bias models considered in paper are derived. In
Sec.~\ref{sec:Results}, the resulting predictions of iPT with various
biasing schemes are presented for the power spectra and correlation
functions in real space and redshift space. The impacts on the
scale-dependent bias from primordial non-Gaussianity are indicated.
Conclusions are summarized in Sec.~\ref{sec:Conclusions}.

\section{\label{sec:iPT}
One-loop Integrated perturbation theory in a nutshell}

In this section, we briefly summarize the formulas of one-loop iPT for
the weakly nonlinear power spectra and correlation functions in the
presence of bias in general \cite{Mat14}.

In this section, we adopt the notation
\begin{equation}
  \bm{k}_{1\cdots n} = \bm{k}_1 + \cdots + \bm{k}_n,
\label{eq:2-0-1}
\end{equation}
and
\begin{equation}
  \int_{\bm{k}_{1\cdots n}=\bm{k}}\cdots
  =
  \int \frac{d^3k_1}{(2\pi)^3} \cdots \frac{d^3k_n}{(2\pi)^3}
  (2\pi)^3\delta_{\rm D}^3\left(\bm{k}-\bm{k}_{1\cdots n}\right)
  \cdots.
\label{eq:2-0-2}
\end{equation}
for brevity. The one-loop power spectrum of biased tracers $X$ is
given by the formula
\begin{multline}
  P_X(\bm{k}) = 
  \left[\varGamma^{(1)}_X(\bm{k})\right]^2 P_{\rm L}(k)
\\
  + \frac12 \int_{\bm{k}_{12}=\bm{k}}
  \left[\varGamma^{(2)}_X(\bm{k}_1,\bm{k}_2)\right]^2
  P_{\rm L}(k_1) P_{\rm L}(k_2)
\\
  + \varGamma^{(1)}_X(\bm{k})
  \int_{\bm{k}_{12}=\bm{k}}
  \varGamma^{(2)}_X(\bm{k}_1,\bm{k}_2)
  B_{\rm L}(k,k_1,k_2),
\label{eq:2-1}
\end{multline}
where $P_{\rm L}(k)$ and $B_{\rm L}(k,k_1,k_2)$ are the linear power
spectrum and the linear bispectrum, respectively, and
$\varGamma^{(n)}_X$ is the $n$th-order multipoint propagator of biased
tracers $X$. Although the time dependence is omitted in the notation,
the functions $P_X$, $P_\mathrm{L}$, $B_\mathrm{L}$ and
$\varGamma^{(n)}_X$ depend also on the cosmic time or the redshift of
observed objects. In the notation of this paper, the time variable is
always omitted in the argument of all the functions for shorthand
convenience.

The multipoint propagator of biased tracers can be decomposed into a
vertex resummation factor and a normalized propagator as follows,
\begin{equation}
  \varGamma^{(n)}_X(\bm{k}_1,\ldots,\bm{k}_n)
  = \varPi(\bm{k}_{1\cdots n})
  \hat{\varGamma}^{(n)}_X(\bm{k}_1,\ldots,\bm{k}_n),
\label{eq:2-2}
\end{equation}
where $\varPi(\bm{k}) = \langle e^{-i\bm{k}\cdot\bm{\varPsi}}\rangle$
is the vertex resummation factor and $\bm{\varPsi}$ is a displacement
field in the Lagrangian description of cosmological perturbations. The
propagators are evaluated with Lagrangian perturbation theory in iPT.
The Fourier transform of the displacement field,
$\tilde{\bm{\varPsi}}(\bm{k})$, is expanded by the linear density
contrast $\delta_\mathrm{L}(\bm{k})$ in Fourier space as
\begin{equation}
  \tilde{\bm{\varPsi}}(\bm{k}) =
  \sum_{n=1}^\infty \frac{i}{n!}
  \int_{\bm{k}_{1\cdots n}=\bm{k}}
  \bm{L}^{(n)}(\bm{k}_1,\ldots,\bm{k}_n)
  \delta_{\rm L}(\bm{k}_1)\cdots\delta_{\rm L}(\bm{k}_n),
\label{eq:2-2-1}
\end{equation}
which define the Lagrangian kernel functions $\bm{L}^{(n)}$. The
kernel functions are calculated by the Lagrangian perturbation theory
\cite{buc89,mou91,buc92,cat95,RB12}. They are polynomials of the wave
vectors which make up their arguments. The Lagrangian kernels in
redshift space are obtained by linear transformations of those in real
space. For concrete expressions for the Lagrangian kernels in real
space and in redshift space, see Refs.~\cite{Mat08a,RB12,Mat15}.

Up to the one-loop order in Eq.~(\ref{eq:2-1}), we have
\begin{equation}
  \varPi(\bm{k}) =
  \exp\left\{
    -\frac12 \int\frac{d^3p}{(2\pi)^3}
    \left[\bm{k}\cdot\bm{L}^{(1)}(\bm{p})\right]^2
    P_{\rm L}(p)
  \right\},
\label{eq:2-3}
\end{equation}
\begin{multline}
  \hat{\varGamma}_X^{(1)}(\bm{k})
  = c_X^{(1)}(\bm{k}) + \bm{k}\cdot\bm{L}^{(1)}(\bm{k})
\\
  + \int\frac{d^3p}{(2\pi)^3} P_{\rm L}(p)
  \biggl\{
      c_X^{(2)}(\bm{k},\bm{p})
      \left[\bm{k}\cdot\bm{L}^{(1)}(-\bm{p})\right]
\\
      +\,c_X^{(1)}(\bm{p})
      \left[\bm{k}\cdot\bm{L}^{(1)}(-\bm{p})\right]
      \left[\bm{k}\cdot\bm{L}^{(1)}(\bm{k})\right]
\\
      + \frac12
      \bm{k}\cdot\bm{L}^{(3)}(\bm{k},\bm{p},-\bm{p})
\\
      +\,c_X^{(1)}(\bm{p})
      \left[\bm{k}\cdot\bm{L}^{(2)}(\bm{k},-\bm{p})\right]
\\
      + \left[\bm{k}\cdot\bm{L}^{(1)}(\bm{p})\right]
      \left[\bm{k}\cdot\bm{L}^{(2)}(\bm{k},-\bm{p})\right]
  \biggr\},
\label{eq:2-4}
\end{multline}
and
\begin{multline}
  \hat{\varGamma}^{(2)}_X(\bm{k}_1,\bm{k}_2) = 
  c^{(2)}_X(\bm{k}_1,\bm{k}_2)
  + c^{(1)}_X(\bm{k}_1) \left[\bm{k}\cdot\bm{L}^{(1)}(\bm{k}_2)\right]
\\
  + c^{(1)}_X(\bm{k}_2) \left[\bm{k}\cdot\bm{L}^{(1)}(\bm{k}_1)\right]
  + \left[\bm{k}\cdot\bm{L}^{(1)}(\bm{k}_1)\right]
    \left[\bm{k}\cdot\bm{L}^{(1)}(\bm{k}_2)\right]
\\
  + \bm{k}\cdot\bm{L}^{(2)}(\bm{k}_1,\bm{k}_2), 
\label{eq:2-5}
\end{multline}
where $c^{(1)}_X$ and $c^{(2)}_X$ are the renormalized bias functions.
The third line of Eq.~(\ref{eq:2-4}) is usually zero for
$c^{(1)}_X(\bm{p})$ is only a function of the modulus of $\bm{p}$,
$c^{(1)}_X(p)$. The series of renormalized bias functions is generally
defined by \cite{Mat12}
\begin{equation}
  \left\langle
      \frac{\delta^n \delta^{\rm L}_X(\bm{k})}
      {\delta\delta_{\rm L}(\bm{k}_1)
        \cdots\delta\delta_{\rm L}(\bm{k}_n)}
  \right\rangle = 
  (2\pi)^{3-3n}\delta_{\rm D}^3(\bm{k}-\bm{k}_{1\cdots n})
  c^{(n)}_X(\bm{k}_1,\ldots,\bm{k}_n),
\label{eq:2-6}
\end{equation}
where $\delta^\mathrm{L}_X(\bm{k})$ is the Fourier transform of the
density contrast of biased tracers in Lagrangian space,
$\delta/\delta\delta_\mathrm{L}(\bm{k})$ is the functional derivative
with respect to $\delta_\mathrm{L}$, and $\langle\cdots\rangle$
denotes the statistical average. All the statistical information about
spatial biasing is included in the set of renormalized bias functions.

In Lagrangian biasing schemes in general, the number density
$n^\mathrm{L}_X$ of biased tracers in Lagrangian space is modelled as
a functional of the linear density field,
$n^\mathrm{L}_X={\cal F}[\delta_\mathrm{L}]$. The relation is
generally given by a functional, instead of a function, because the
density of biased tracers at some position is determined by the linear
density field not only at the same position but also at other
positions as well. We thus have functional derivatives as in
Eq.~(\ref{eq:2-6}).

Once the number density of biased tracers $n^\mathrm{L}_X$ is modelled
as a functional of the linear density field, and the statistical
distribution of the linear density field is specified, the
renormalized bias functions are obtained from Eq.~(\ref{eq:2-6}) and
$\delta^\mathrm{L}_X = n^\mathrm{L}_X/\langle n^\mathrm{L}_X \rangle -
1$. In order to evaluate the one-loop power spectrum of
Eq.~(\ref{eq:2-1}), only two functions, $c^{(1)}_X(\bm{k})$ and
$c^{(2)}_X(\bm{k}_1,\bm{k}_2)$, are required. Some of the angular
integrations can be performed analytically, so that Eq.~(\ref{eq:2-4})
reduces to two- and one-dimensional integrals \cite{Mat14}.

In real space, the power spectrum $P_X(k)$ is a function of the
modulus of wave vector $k = |\bm{k}|$ for homogeneous and isotropic 
random fields. In this case, the correlation function is simply given by
\begin{equation}
  \xi_X(r) = \int_0^\infty \frac{k^2dk}{2\pi^2} j_0(kr) P_X(k),
\label{eq:2-7}
\end{equation}
where $j_0(z)$ denotes the spherical Bessel function $j_l(z)$ of order
zero, $l=0$. In redshift space, however, the power spectrum has an
angular dependence as well. Adopting the distant-observer
approximation where all the lines of sight have a common direction,
the power spectrum $P_X(k,\mu)$ is a function of the modulus $k$ and
direction cosine $\mu$ relative to the line of sight. In this case, it
is convenient to expand the angular dependence of the power spectrum
in Legendre polynomials $\mathsf{P}_l(\mu)$ according to
\begin{align}
  P_X(k,\mu) &= \sum_{l=0}^\infty p_X^l(k)\,\mathsf{P}_l(\mu);
\label{eq:2-8a}\\
  p_X^l(k) &= \frac{2l+1}{2} \int_{-1}^1 d\mu\,\mathsf{P}_l(\mu) P_X(k,\mu).
\label{eq:2-8b}
\end{align}
The same expansion of the correlation function is given by
\begin{align}
  \xi_X(r,\mu) &= \sum_{l=0}^\infty \xi_X^l(r)\,\mathsf{P}_l(\mu);
\label{eq:2-205a}\\
  \xi_X^l(r) &=
  \frac{2l+1}{2} \int_{-1}^1 d\mu\,\mathsf{P}_l(\mu) \xi_X(r,\mu).
\label{eq:2-205b}
\end{align}
The relation between the multipole coefficients is
\begin{equation}
  \xi_X^l(r) =
  i^{-l} \int_0^\infty \frac{k^2dk}{2\pi^2} j_l(kr) p_X^l(k).
\label{eq:2-9}
\end{equation}
Thus, once the power spectrum in redshift space $P_X(k,\mu)$ is
calculated by iPT, the multipoles $p_X^l(k)$ and $\xi_X^l(r)$ are
evaluated by Eqs.~(\ref{eq:2-8b}) and (\ref{eq:2-9}). Analytical
integrations of Eq.~(\ref{eq:2-8b}) are also possible \cite{Mat14}.

\section{\label{sec:seminonlocal}
Renormalized bias functions in semilocal models of bias
}

The concept of renormalized bias functions in the formalism of iPT is
applicable to a broad range of generally nonlocal models of bias.
However, most of the bias models that have been proposed in recent years
fall into a category of, what we call in this paper, semilocal models
of bias. In this type of biasing models, the formation sites of LSS 
tracers depend on the local values of the smoothed mass density
field and its spatial derivatives. In this section, we present a general
derivation of the renormalized bias functions for a class of semilocal 
models of Lagrangian bias. To illustrate our method, we compute the 
renormalized bias functions for a few bias models: the halo, peaks and 
ESP models.

\subsection{\label{subsec:semilocalbias}
Semilocal models of Lagrangian bias
}

In the semilocal models, the number density field $n_X(\bm{x})$ of
observable objects $X$ is described by a function of the smoothed
linear density contrast $\delta_s$ and its spatial derivatives
$\partial_i\delta_s$, $\partial_{ij}\delta_s$, etc. In general,
various types of filtering kernels can be simultaneously introduced to
accommodate specific variables. For instance, the linear gravitational
potential can be included in a straightforward manner by adding a
suitable smoothing kernel.

To keep the discussion general, we consider here various smoothing of
the linear density contrast,
\begin{equation}
  \label{eq:3-1}
  \delta_s(\bm{x}) = \int \frac{d^3k}{(2\pi)^3}
  \delta_\mathrm{L}(\bm{k}) W_s(kR_s) e^{i\bm{k}\cdot\bm{x}},
\end{equation}
where the index $s$ refers to the types of smoothing kernel;
$\delta_\mathrm{L}$ is a linear density contrast in Fourier space; and
$W_s$ and $R_s$ are, respectively, a smoothing function and a
smoothing radius for each type $s$ of the smoothing kernel. Popular
kernels include the top-hat ($s=\mathrm{T}$) and Gaussian
($s=\mathrm{G}$) window functions,
\begin{equation}
  \label{eq:3-2}
  W_\mathrm{T}(x) = 3j_1(x)/x, \quad
  W_\mathrm{G}(x) = e^{-x^2/2}.
\end{equation}
The linear gravitational potential $\phi_\mathrm{L}$ can also be
expressed in the form of Eq.~(\ref{eq:3-1}) with a smoothing kernel
$W_\phi(x) = -1/x^2$ and smoothing radius
$R_\phi = a^{-1}(4\pi G \bar{\rho})^{-1/2}$. In this case, we have
$s=\phi$ and $\delta_\phi = \phi_\mathrm{L}$. Another example is the
effective window function
$W_\mathrm{eff}(x) =
W_\mathrm{T}(x)W_\mathrm{G}(f_\mathrm{eff}^{1/2}x/5)$ recently
proposed by Ref.~\cite{CSS15} to model Lagrangian halos. Here,
$f_\mathrm{eff}$ is a free parameter that must be calibrated with
simulations. This effective window function furnishes a good fit to
the small-scale, scale-dependent Lagrangian halo bias measured from
numerical simulations.

While Eq.~(\ref{eq:3-1}) can incorporate many different smoothing
functions such as e.g. $s = \phi$, we specifically consider biasing
models that depend on the spatial derivatives of the smoothed field up
to second order, $\partial_i\delta_s$ and $\partial_{ij}\delta_s$, in
addition to the field values themselves, $\delta_s$. It is convenient
to introduce the spectral moments $\sigma_{s0}= \langle
(\delta_s)^2\rangle^{1/2}$, $\sigma_{s1}= \langle
\bm{\nabla}\delta_s\cdot\bm{\nabla}\delta_s\rangle^{1/2}$, and
$\sigma_{s2}=
\langle(\bm{\nabla}\cdot\bm{\nabla}\delta_s)^2\rangle^{1/2}$ so as to
normalize the linear density fields:
\begin{equation}
  \label{eq:3-3}
  \nu_s(\bm{x}) = \frac{\delta_s(\bm{x})}{\sigma_{s0}},\quad
  \eta_{si}(\bm{x}) = \frac{\partial_i\delta_s(\bm{x})}{\sigma_{s1}},\quad
  \zeta_{sij}(\bm{x})
  = \frac{\partial_i\partial_j\delta_s(\bm{x})}{\sigma_{s2}}.
\end{equation}
The spectral parameters are integrals of the linear power spectrum
$P_\mathrm{L}(k)$,
\begin{equation}
  \label{eq:3-4}
  {\sigma_{sj}}^2 =
  \int \frac{k^2dk}{2\pi^2}  k^{2j} P_\mathrm{L}(k) [W_s(kR_s)]^2.
\end{equation}

The number density field $n_X(\bm{x})$ of biased objects is assumed to
be a multivariate function of $\nu_s$, $\eta_{si}$ and $\zeta_{sij}$,
where the filtering kernels can be
$s=\mathrm{T},\mathrm{G},\mathrm{eff}, \ldots$ and the spatial indices
run over $i=1,2,3$ and $ij = 11,22,33,12,23,13$. These linear field
variables are denoted by $y_\alpha$, where the index $\alpha$
indicates one of the above field variables, such as $\nu_\mathrm{T}$,
$\eta_{\mathrm{G}2}$, $\zeta_{\mathrm{G}13}$, etc.

The Fourier transform of the variables $y_\alpha(\bm{x})$ is of the
form
\begin{equation}
  \label{eq:3-5}
  \tilde{y}_\alpha(\bm{k}) = U_\alpha(\bm{k}) \delta_\mathrm{L}(\bm{k}),
\end{equation}
where the functions $U_\alpha(\bm{k})$ corresponding to the variables
in Eq.~(\ref{eq:3-3}) are given by $W_s(\bm{k})/\sigma_{s0}$,
$ik_iW_s(\bm{k})/\sigma_{s1}$ and $-k_ik_jW_s(\bm{k})/\sigma_{s2}$,
respectively. The renormalized bias functions of iPT are given by
\cite{Mat11}
\begin{multline}
  \label{eq:3-6}
  c_X^{(n)}(\bm{k}_1,\ldots,\bm{k}_n) =
  \frac{1}{\bar{n}_X}
  \sum_{\alpha_1,\ldots,\alpha_n}
  \left\langle
    \frac{\partial^n n_X}
    {\partial y_{\alpha_1} \cdots \partial y_{\alpha_n}}
  \right\rangle
\\
  \times
  U_{\alpha_1}(\bm{k}_1) \cdots U_{\alpha_n}(\bm{k}_n).
\end{multline}
Here, $\bar{n}_X = \langle n_X \rangle$ is the mean number density of
objects $X$. It is convenient to define a differential operator
\begin{align}
  \label{eq:3-7}
  {\cal D}(\bm{k}) &\equiv \sum_\alpha U_\alpha(\bm{k}) 
  \frac{\partial}{\partial y_\alpha}
\nonumber \\
  &= 
  \sum_s W_s(kR_s)
  \left[
    \frac{1}{\sigma_{s0}} \frac{\partial}{\partial\nu_s}
    + \frac{i}{\sigma_{s1}} {\cal D}^s_\eta(\bm{k}) 
    - \frac{1}{\sigma_{s2}} {\cal D}^s_\zeta(\bm{k}) 
  \right],
\end{align}
where
\begin{equation}
  \label{eq:3-8}
  {\cal D}^s_\eta(\bm{k}) = \sum_i k_i\frac{\partial}{\partial\eta_{si}},\quad
  {\cal D}^s_\zeta(\bm{k}) = \sum_{i\leq j} k_i k_j
  \frac{\partial}{\partial\zeta_{sij}}.
\end{equation}
Although the set of variables $\zeta_{sij}$ is a symmetric tensor and
has six independent degrees of freedom, it is useful to introduce a
set of redundant variables
\begin{equation}
  \label{eq:3-9}
  \xi_{sij} \equiv
  \begin{cases}
    \zeta_{sij} & (i \leq j) \\
    \zeta_{sji} & (i > j)
  \end{cases}.
\end{equation}
Any function of $\zeta_{sij}$ ($i \leq j$) can be considered as a
function of $\xi_{sij}$. The differentiation with respect to independent
variables $\zeta_{sij}$ is given by
\begin{equation}
  \label{eq:3-10}
  \frac{\partial}{\partial\zeta_{sij}} = 
  \begin{cases}
    \displaystyle
    \frac{\partial}{\partial\xi_{sii}} & (i = j) \\
    & \\
    \displaystyle
    \frac{\partial}{\partial\xi_{sij}}
    + \frac{\partial}{\partial\xi_{sji}}
    & (i < j)
  \end{cases},
\end{equation}
when it acts on an explicit function of $\xi_{sij}$. With the variables
$\xi_{sij}$, the differential operator ${\cal D}^s_\zeta(\bm{k})$ in
Eq.~(\ref{eq:3-8}) reduces to
\begin{equation}
  \label{eq:3-11}
  {\cal D}^s_\zeta(\bm{k})
  = \sum_{i,j}k_i k_j \frac{\partial}{\partial\xi_{sij}}.
\end{equation}

Using the differential operator ${\cal D}(\bm{k})$, Eq.~(\ref{eq:3-6})
reduces to
\begin{multline}
  \label{eq:3-12}
  c_X^{(n)}(\bm{k}_1,\ldots,\bm{k}_n)
  = \frac{1}{\bar{n}_X}
  \left\langle
    {\cal D}(\bm{k}_1)\cdots{\cal D}(\bm{k}_n) n_X
  \right\rangle
\\
  = \frac{(-1)^n}{\bar{n}_X}
  \int d^Ny\, n_X(\bm{y})
    {\cal D}(\bm{k}_1)\cdots{\cal D}(\bm{k}_n) {\cal P}(\bm{y}),
\end{multline}
where ${\cal P}(\bm{y})$ is the joint probability distribution
function and $N$ is the dimension of $y_\alpha$. Integrations by parts
are applied in the second line. The mean number density is given by
\begin{equation}
  \label{eq:3-13}
  \bar{n}_X =
  \langle n_X \rangle
  = \int d^Ny\, n_X(\bm{y}) {\cal P}(\bm{y}).
\end{equation}
For a given model of bias, the functions $n_X(\bm{y})$ and
$U_\alpha(\bm{k})$ are specified, and the renormalized bias functions
are calculated by Eqs.~(\ref{eq:3-12}) and (\ref{eq:3-13}). The joint
probability distribution function ${\cal P}(\bm{y})$ is determined by
the statistics of the initial density field $\delta_\mathrm{L}$.

Equations (\ref{eq:3-12}) and (\ref{eq:3-13}) are also applicable in
the presence of initial non-Gaussianity. When the initial density
field is random Gaussian, ${\cal P}$ is a multivariate Gaussian
distribution function. In this case, the covariance matrix of the set
of variables $\{y_\alpha\}$,
\begin{equation}
  \label{eq:3-14}
  {\cal M}_{\alpha\beta} =
  \left\langle y_\alpha y_\beta \right\rangle
  = \int\frac{d^3k}{(2\pi)^3}
  U_\alpha^*(\bm{k}) U_\beta(\bm{k}) P_\mathrm{L}(k),
\end{equation}
completely determines the distribution function as
\begin{equation}
  \label{eq:3-15}
  {\cal P}(\bm{y}) =
  \frac{1}{\sqrt{(2\pi)^N \det{\cal M}}}
  \exp\left(-\frac{1}{2}\bm{y}^\mathrm{T} {\cal M}^{-1} \bm{y}\right).
\end{equation}
Generalization of the following analysis in the presence of initial
non-Gaussianity is fairly straightforward by applying the multivariate
Gram-Charlier expansion of the distribution function
\cite{PGP09,GPP12,DGR13}. 

\subsection{\label{subsec:halobias}
Simple halo model
}

The renormalized bias functions in the halo model of bias are derived
in Ref.~\cite{Mat12}. We summarize the results in this subsection. In
the halo model, the smoothing radius $R$ is associated with a mass
scale $M$ by a relation,
\begin{equation}
  \label{eq:3-2-1}
  M = \frac{4\pi\bar{\rho}_0}{3}R^3,
\end{equation} 
where $\bar{\rho}_0$ is the mean matter density at the present time.
The above relation is equivalently represented by
\begin{equation}
  \label{eq:3-2-2}
  R = \left(\frac{M}{1.163\times 10^{12}\,h^{-1} M_\odot
      \varOmega_\mathrm{m0}}\right)^{1/3}h^{-1}\mathrm{Mpc},
\end{equation} 
where $M_\odot=1.989\times 10^{30}\,\mathrm{kg}$ is the solar mass,
$\varOmega_\mathrm{m0}$ is the density parameter of the present
universe, and $h = H_0/(100\,\mathrm{km} \cdot \mathrm{s}^{-1} \cdot
\mathrm{Mpc}^{-1})$ is the dimensionless Hubble parameter.

The mass element at a Lagrangian position $\bm{x}$ is assumed to be
contained in a halo of mass larger than $M$, if the value of linear
density contrast $\delta_M$ smoothed by the mass scale $M$ exceeds a
critical value $\delta_\mathrm{c}$. The critical value is usually
taken to be $\delta_\mathrm{c} = 3(3\pi/2)^{2/3}/5 \simeq 1.686$, which
follows from the spherical collapse calculation. 
The localized differential number density of halos at a Lagrangian 
position $\bm{x}$ is given by
\cite{Mat12}
\begin{equation}
  \label{eq:3-2-3}
  n(\bm{x},M) =
  -\frac{2\bar{\rho}_0}{M}
  \frac{\partial}{\partial M}
  \varTheta\left[\delta_M(\bm{x})-\delta_\mathrm{c}\right],
\end{equation}
where $n(\bm{x},M)$ is the differential mass function of halos, and
$\varTheta(x)$ is the step function. This model is a generalization of
the Press-Schechter (PS) formalism \cite{PS74}. In fact, on taking the
spatial average of the above equation, the number density of halos
$n(M)$ in the original PS formalism is recovered.

When the initial condition is Gaussian, and the smoothed mass density
contrast $\delta_M(\bm{x})$ is a Gaussian field, the spatial average
of the step function
$\langle\varTheta(\delta_M(\bm{x})-\delta_\mathrm{c})\rangle$ is given
by the complementary error function. In this case, the global
(spatially averaged) mass function has the form,
\begin{equation}
  \label{eq:3-2-4}
  n(M)dM = \frac{\bar{\rho}_0}{M} f(\nu) \frac{d\nu}{\nu},
\end{equation}
where $\nu = \delta_\mathrm{c}/\sigma_M$, $\sigma_M = \langle
(\delta_M)^2\rangle^{1/2}$ are functions of mass $M$, and $f(\nu) =
(2/\pi)^{1/2} \nu e^{-\nu^2/2}$. The function $f(\nu)$ is called the
``multiplicity function'' [Note that another convention defines
$f(\nu)$ as $n(M)dM = (\bar{\rho}_0/M) f(\nu)d\nu$].

While the mass function of dark matter halos identified in $N$-body
simulations broadly agrees with the PS prediction, the agreement is
far from perfect. Recent studies have shown that using multiplicity
functions different from the PS mass function provides better models
for halo statistics. One of the simplest models is given by a
Sheth-Tormen mass function \cite{ST99}, for which the multiplicity
function reads
\begin{equation}
  \label{eq:3-2-5}
  f(\nu) = A(p) \sqrt{\frac{2}{\pi}}
  \left[
    1 + \frac{1}{(q\nu^2)^p}
  \right] \sqrt{q}\,\nu\,e^{-q\nu^2/2},
\end{equation}
where $p=0.3$, $q=0.707$, and $A(p) =
[1+\pi^{-1/2}2^{-p}\varGamma(1/2-p)]^{-1}$ is a normalization factor.

When the mass function is changed from the PS one,
Eq.~(\ref{eq:3-2-3}) should be simultaneously changed in order to be
compatible with Eq.~(\ref{eq:3-2-4}). This can be achieved by
substituting the step function $\varTheta(\delta_M -
\delta_\mathrm{c})$ with an auxiliary function $\varXi(\delta_M -
\delta_\mathrm{c},\sigma_M)$. This function should explicitly depend
on the mass $M$ through $\sigma_M$. Otherwise, if the mass dependence
is only implicit through the smoothing kernel of $\delta_M$, the
resulting mass function is only compatible with the PS mass function.
More details on the relation between the multiplicity function and the
auxiliary function is discussed in Appendix~\ref{app:XiFunc}.

The relation between the multiplicity function $f(\nu)$
and the new function $\varXi$ is given by
\begin{equation}
  \label{eq:3-2-6}
  \left\langle
    \varXi\left(\delta_M - \delta_\mathrm{c}, \sigma_M\right)
  \right\rangle
  = \frac{1}{2} \int_\nu^\infty \frac{f(\nu)}{\nu}d\nu,
\end{equation}
and the local mass function is given by
\begin{align}
  n(\bm{x},M) &=
  -\frac{2\bar{\rho}_0}{M}
  \frac{\partial}{\partial M}
  \varXi\left[\delta_M(\bm{x})-\delta_\mathrm{c},\sigma_M\right]
  \nonumber
\\
  &=
  \frac{2\bar{\rho}_0}{M}
  \left\{
    \frac{\partial\delta_M(\bm{x})}{\partial M}
    \frac{\partial}{\partial\delta_\mathrm{c}}
    \varXi\left[\delta_M(\bm{x})-\delta_\mathrm{c},\sigma_M\right]
  \right.
  \nonumber
\\
  & \hspace{3.2pc}
  \left.
    -\, \frac{d\sigma_M}{dM}
    \frac{\partial}{\partial\sigma_M}
    \varXi\left[\delta_M(\bm{x})-\delta_\mathrm{c},\sigma_M\right]
  \right\}.
  \label{eq:3-2-7}
\end{align}

The model of Eq.~(\ref{eq:3-2-7}) for the number density field depends
on the linear density field through two variables, $\delta_M(\bm{x})$
and $\partial\delta_M(\bm{x})/\partial M$, which corresponds to the
variables $y_\alpha$ in Sec.~\ref{subsec:semilocalbias}. The window
functions for these variables $U_\alpha(\bm{k})$ are given by $W(kR)$
and $\partial W(kR)/\partial M$, respectively, where $R$ and $M$ are
related by Eq.~(\ref{eq:3-2-1}) or (\ref{eq:3-2-2}) . The renormalized
bias functions in this model are derived by Eqs.~(\ref{eq:3-6}) and
(\ref{eq:3-2-7}). The unknown function $\varXi$ can be removed from
the resulting expressions thanks to the relation of
Eq.~(\ref{eq:3-2-6}). Closed forms for all the renormalized bias
functions are derived in Ref.~\cite{Mat12}. (In the notation of
Ref.~\cite{Mat12}, the dependence of $\sigma_M$ in the function
$\varXi$ is implicit, but it is actually assumed.) The results are
given by
\begin{multline}
  \label{eq:3-2-8}
  c_X^{(n)}(\bm{k}_1,\ldots,\bm{k}_n)
  = b_n^\mathrm{L} W(k_1R)\cdots W(k_nR)
\\
  + \frac{A_{n-1}(M)}{{\delta_\mathrm{c}}^n}
  \frac{d}{d\ln\sigma_M}
  \left[
     W(k_1R)\cdots W(k_nR)
  \right],
\end{multline}
where 
\begin{align}
  \label{eq:3-2-9}
  b_n^\mathrm{L}(M) &\equiv
  \left(-\frac{1}{\sigma_M}\right)^n \frac{f^{(n)}(\nu)}{f(\nu)},
\\
  \label{eq:3-2-10}
  A_n(M) &\equiv
  \sum_{m=0}^n \frac{n!}{m!}{\delta_\mathrm{c}}^m b_m^\mathrm{L}(M).
\end{align}

In this paper, we need only the first two functions, $c_X^{(1)}$ and
$c_X^{(2)}$, which are explicitly given by
\begin{align}
  \label{eq:3-2-11a}
  c_X^{(1)}(k) &= b^\mathrm{L}_1 W(kR) + \frac{1}{\delta_\mathrm{c}}
  \frac{dW(kR)}{d\ln\sigma_M},
  \\
  \label{eq:3-2-11b}
  c_X^{(2)}(\bm{k}_1,\bm{k}_2) &=
  b^\mathrm{L}_2 W(k_1R) W(k_2R)
  \nonumber\\
&  \qquad
  + \frac{1+\delta_\mathrm{c} b^\mathrm{L}_1}{{\delta_\mathrm{c}}^2}
  \frac{d\left[W(k_1R)W(k_2R)\right]}{d\ln\sigma_M}.
\end{align}

\subsection{\label{subsec:peaksbias}
Peaks model
}

In the peaks model, the formation sites of dark matter halos are
identified with density peaks in Lagrangian space. The peaks are
described by field values with up to second derivatives of a smoothed
density field, $\nu_s$, $\eta_{si}$ and $\zeta_{sij}$. While the
choice of smoothing kernel $s$ is arbitrary (so long as the
convergence of the spectral moments is ensured), the Gaussian kernel
($s=\mathrm{G}$) is frequently adopted. In the peaks model, only a
single kind of smoothing kernel is involved. Therefore, we omit the
subscript $s$ in this subsection below and use notations like $\nu$,
$\eta_i$, $\zeta_{ij}$, $\sigma_0$, $\sigma_1$, $\sigma_2$, etc.

\subsubsection{\label{subsubsec:peaksRBF}
Derivation of renormalized bias functions in the peaks model
}

The differential number density of discrete peaks with a peak height
$\nu_\mathrm{c}$ is given by \cite{BBKS}
\begin{equation}
  \label{eq:3-3-1}
  n_\mathrm{pk} = 
  \frac{3^{3/2}}{{R_*}^3}
  \delta_\mathrm{D}(\nu - \nu_\mathrm{c})
  \delta_\mathrm{D}^3(\bm{\eta})
  \varTheta(\lambda_3)
  \left|\det\bm{\zeta}\right|,
\end{equation}
where $R_* = \sqrt{3}\sigma_1/\sigma_2$ is a characteristic radius and
$\lambda_3$ is the smallest eigenvalue of the $3\times 3$ matrix
$(-\zeta_{ij})$. The number density of peaks with peak height between
$\nu_\mathrm{c}$ and $\nu_\mathrm{c} + d\nu_\mathrm{c}$ is given by
$n_\mathrm{pk}d\nu_\mathrm{c}$.

The variables $(y_\alpha)$ consists of ten variables, $(\nu, \eta_i,
\zeta_{ij})$ with $1 \leq i\leq j \leq 3$, and the corresponding
kernels $(U_\alpha)$ are $[W(kR)/\sigma_0, ik_iW(kR)/\sigma_1,
-k_ik_jW(kR)/\sigma_2]$.

When the linear density field $\delta_\mathrm{L}$ is statistically
isotropic, the joint probability distribution function ${\cal
  P}(\bm{y})$ only depends on rotationally invariant quantities
\cite{Dor70,PGP09,GPP12}. Using the redundant variables $\xi_{ij}$ defined in
Eq.~(\ref{eq:3-9}), these are
\begin{equation}
  \label{eq:3-3-2}
  \eta^2 \equiv \bm{\eta}\cdot\bm{\eta}, \quad
  J_1 \equiv - \xi_{ii}, \quad 
  J_2 \equiv \frac{3}{2} \tilde{\xi}_{ij} \tilde{\xi}_{ji}, \quad
  J_3 = \frac{9}{2} \tilde{\xi}_{ij} \tilde{\xi}_{jk} \tilde{\xi}_{ki},
\end{equation}
where repeated indices are summed over, and
\begin{equation}
  \label{eq:3-3-3}
  \tilde{\xi}_{ij} \equiv \xi_{ij} + \frac{1}{3}\delta_{ij} J_1,
\end{equation}
is the traceless part of $\xi_{ij}$.
Covariances among the field
variables are given by \cite{BBKS}
\begin{align}
  \label{eq:3-3-4a}
  \langle \nu^2 \rangle &= 1,\;
  \langle \nu \eta_i\rangle = 0,\;
  \langle \nu \xi_{ij} \rangle = - \frac{\gamma}{3}\delta_{ij},\;
  \langle \eta_i \eta_j\rangle = \frac{1}{3}\delta_{ij},
\\
  \label{eq:3-3-4b}
  \langle\eta_i\xi_{jk}\rangle &= 0,\;\;
  \langle\xi_{ij}\xi_{kl}\rangle = 
  \frac{1}{15}
  \left(
    \delta_{ij}\delta_{kl} + \delta_{ik}\delta_{jl} +
    \delta_{il}\delta_{jk} 
  \right),
\end{align}
where
\begin{equation}
  \label{eq:3-3-5}
  \gamma \equiv \frac{{\sigma_1}^2}{\sigma_0\sigma_2}
\end{equation}
characterizes the broadband shape of the smoothed linear power spectrum.
Adopting the above covariances, the multivariate distribution function of
Eq.~(\ref{eq:3-15}) reduces to \cite{Dor70,BBKS,PGP09,GPP12,Des13,LMD15}
\begin{equation}
  {\cal P}(\bm{y}) \propto
  \exp\left[
    - \frac{\nu^2 + {J_1}^2 - 2\gamma \nu J_1}{2(1-\gamma^2)}
    - \frac{3}{2} \eta^2 - \frac{5}{2} J_2
  \right]
  \label{eq:3-3-6}
\end{equation}
up to a normalization constant, which is irrelevant for our applications
in the following. The distribution function above is still for linear
variables $\bm{y}$, and not for rotationally invariant variables.

Since the distribution function ${\cal P}(\bm{y})$ depends only on
four rotationally invariant variables $\nu$, $J_1$, $\eta^2$, and
$J_2$, the first-order derivatives are given by
\begin{equation}
  \label{eq:3-3-7}
  \frac{\partial}{\partial\eta_i} {\cal P}
  = 2 \eta_i \frac{\partial}{\partial(\eta^2)}{\cal P}, \quad
  \frac{\partial}{\partial\xi_{ij}} {\cal P}
  = \left[
    - \delta_{ij} \frac{\partial}{\partial J_1} 
    + 3 \tilde{\xi}_{ij} \frac{\partial}{\partial J_2}
    \right] {\cal P},
\end{equation}
for which the relations
\begin{equation}
  \label{eq:3-3-8}
  \frac{\partial(\eta^2)}{\partial\eta_i}= 2\eta_i, \quad
  \frac{\partial J_1}{\partial\xi_{ij}} = -\delta_{ij}, \quad
  \frac{\partial J_2}{\partial\xi_{ij}} = 3\tilde{\xi}_{ij}
\end{equation}
are used. Further differentiating the above equations, we have
\begin{align}
  \label{eq:3-3-9a}
  \frac{\partial^2}{\partial\eta_i\partial\eta_j} {\cal P} &=
  \left[
    2 \delta_{ij} \frac{\partial}{\partial(\eta^2)}
  + 4 \eta_i \eta_j \frac{\partial^2}{\partial(\eta^2)^2}
  \right] {\cal P},
\\
  \frac{\partial^2}{\partial\xi_{ij}\partial\xi_{kl}} {\cal P}
  &=
  \left[
  \delta_{ij} \delta_{kl} \frac{\partial^2}{\partial {J_1}^2}
  - 3 \left(
    \delta_{ij} \tilde{\xi}_{kl} + \delta_{kl}
    \tilde{\xi}_{ij} \right)
  \frac{\partial^2}{\partial J_1 \partial J_2}
  \right.
\nonumber\\
  \label{eq:3-3-9b}
  &
  \quad\left. +\:
  9 \tilde{\xi}_{ij} \tilde{\xi}_{kl} 
  \frac{\partial^2}{\partial {J_2}^2}
  + \left(3 \delta_{ik} \delta_{jl} - \delta_{ij} \delta_{kl}\right)
  \frac{\partial}{\partial J_2}
  \right] {\cal P},
\end{align}
where a relation
$\partial\tilde{\xi}_{kl}/\partial\xi_{ij}=\delta_{ik}\delta_{jl} -
\delta_{ij}\delta_{kl}/3$ is used.

The number density of peaks $n_\mathrm{pk}(\bm{y})$ and the
distribution function ${\cal P}(\bm{y})$ both depend only on
rotationally invariant variables. Thus, the differential operators
${\cal D}(\bm{k}_1)\cdots{\cal D}(\bm{k}_n)$ in Eq.~(\ref{eq:3-12})
can be replaced by those averaged over the rotation of coordinates,
$\langle\cdots\rangle_\Omega$. For that purpose, we have
\begin{align}
  \label{eq:3-3-10a}
  &
  \left\langle \eta_i \right\rangle_\Omega = 0, \quad
  \left\langle \eta_i \eta_j \right\rangle_\Omega = \frac{1}{3}
  \delta_{ij} \eta^2, \quad
  \left\langle \tilde{\xi}_{ij} \right\rangle_\Omega = 0,
\\
  \label{eq:3-3-10b}
  &
  \left\langle \tilde{\xi}_{ij} \tilde{\xi}_{kl}
  \right\rangle_\Omega =
  \frac{1}{15}
  \left(
      \delta_{ik} \delta_{jl} +  \delta_{il} \delta_{jk}
      - \frac{2}{3} \delta_{ij} \delta_{kl}
    \right) J_2,
\end{align}
and so forth.

Combining Eqs.~(\ref{eq:3-7}), (\ref{eq:3-8}), (\ref{eq:3-11}), and
(\ref{eq:3-3-7})--(\ref{eq:3-3-10b}), we have
\begin{align}
  \label{eq:3-3-11a}
&
  \left\langle {\cal D}(\bm{k}) \right\rangle_\Omega {\cal P}
  = W(kR)\left(
    \frac{1}{\sigma_0} \frac{\partial}{\partial\nu} 
    + \frac{k^2}{\sigma_2} \frac{\partial}{\partial J_1}
    \right) {\cal P},
\\
  \label{eq:3-3-11b}
&
  \left\langle
    {\cal D}(\bm{k}_1) {\cal D}(\bm{k}_2)
  \right\rangle_\Omega
  {\cal P}
  = W(k_1R)  W(k_2R)
\nonumber\\
& \qquad \times
  \left\{
    \left(
      \frac{1}{\sigma_0} \frac{\partial}{\partial\nu} 
      + \frac{{k_1}^2}{\sigma_2} \frac{\partial}{\partial J_1}
    \right)
    \left(
      \frac{1}{\sigma_0} \frac{\partial}{\partial\nu} 
      + \frac{{k_2}^2}{\sigma_2} \frac{\partial}{\partial J_1}
    \right)
  \right.
\nonumber\\
  & \hspace{4pc}
  \left. -\:
    \frac{2(\bm{k}_1\cdot\bm{k}_2)}{{\sigma_1}^2}
    \left[
      1 + \frac{2}{3} \eta^2 \frac{\partial}{\partial (\eta^2)}
    \right] \frac{\partial}{\partial(\eta^2)}
  \right.
\nonumber\\
  & \hspace{4pc}
  \left. +\:
    \frac{3(\bm{k}_1\cdot\bm{k}_2)^2 - {k_1}^2 {k_2}^2}{{\sigma_2}^2}
      \left[
        1 + \frac{2}{5} \zeta^2 \frac{\partial}{\partial J_2}
      \right] \frac{\partial}{\partial J_2 }
  \right\} {\cal P}.
\end{align}
Derivatives with respect to variables $\nu$ and $J_1$ in
Eqs.~(\ref{eq:3-3-11a}) and (\ref{eq:3-3-11b}) can be represented by
bivariate Hermite polynomials \cite{Des13},
\begin{equation}
  \label{eq:3-3-12}
  H_{ij}(\nu,J_1) \equiv
  \frac{(-1)^{i+j}}{{\cal N}(\nu,J_1)}
  \left(\frac{\partial}{\partial \nu}\right)^i
  \left(\frac{\partial}{\partial J_1}\right)^j
  {\cal N}(\nu,J_1),
\end{equation}
where
\begin{equation}
  \label{eq:3-3-13}
  {\cal N}(\nu,J_1) \equiv
  \frac{1}{2\pi\sqrt{1-\gamma^2}}
  \exp\left[
    - \frac{\nu^2 + {J_1}^2 - 2\gamma\nu J_1}{2(1-\gamma^2)}
  \right],
\end{equation}
is the bivariate normal distribution function. Derivatives with
respect to variables $\eta^2$ and $J_2$ are straightforwardly
obtained as
\begin{align}
  \label{eq:3-3-14a}
    \left[
      1 + \frac{2}{3} \eta^2 \frac{\partial}{\partial (\eta^2)}
    \right] \frac{\partial}{\partial(\eta^2)} e^{-3\eta^2/2}
    &= \frac{3}{2} \left(\eta^2 - 1 \right) e^{-3\eta^2/2}
\nonumber\\
    &= - L^{(1/2)}_1\left(\frac{3}{2}\eta^2\right) e^{-3\eta^2/2},
\\
  \label{eq:3-3-14b}
    \left[
      1 + \frac{2}{5} J_2 \frac{\partial}{\partial J_2}
    \right] \frac{\partial}{\partial J_2} e^{-5J_2/2}
    &= \frac{5}{2} \left(J_2 - 1 \right) e^{-5J_2/2}
\nonumber\\
    &= - L^{(3/2)}_1\left(\frac{5}{2} J_2\right) e^{-5J_2/2},
\end{align}
where
\begin{equation}
  \label{eq:3-3-15}
  L^{(\alpha)}_n(x) =
  \frac{x^{-\alpha} e^x}{n!}
  \frac{d^n}{dx^n}\left( x^{n+\alpha} e^{-x}\right)
\end{equation}
are the generalized Laguerre polynomials.

Substituting Eqs.~(\ref{eq:3-3-11a}) and (\ref{eq:3-3-11b}) into
the integrand of Eq.~(\ref{eq:3-12}), we obtain
\begin{align}
  \label{eq:3-3-16a}
  c_X^{(1)}(k) &= 
  \left( b_{10} + b_{01} k^2 \right) W(kR),
\\
  \label{eq:3-3-16b}
  c_X^{(2)}(\bm{k}_1, \bm{k}_2) &=
  \biggl\{
    b_{20} + b_{11}({k_1}^2 + {k_2}^2) + b_{02} {k_1}^2 {k_2}^2
   - 2 \chi_1 (\bm{k}_1\cdot\bm{k}_2)
\nonumber\\
& \qquad
    + \omega_{10}\left[3(\bm{k}_1\cdot\bm{k}_2)^2 - {k_1}^2{k_2}^2\right]
  \biggr\}
  W(k_1R)W(k_2R),
\end{align}
where
\begin{align}
  \label{eq:3-3-17a}
  b_{ij} &\equiv \frac{1}{{\sigma_0}^i{\sigma_2}^j\bar{n}_\mathrm{pk}}
  \int d^{10}y\, n_\mathrm{pk} H_{ij}(\nu,J_1)\,{\cal P},
\\
  \label{eq:3-3-17b}
  \chi_k &\equiv
  \frac{(-1)^k}{{\sigma_1}^{2k}\bar{n}_\mathrm{pk}}
  \int d^{10}y\, n_\mathrm{pk}
  L^{(1/2)}_k\left(\frac{3}{2}\eta^2\right)
  \,{\cal P},
\\
  \label{eq:3-3-17c}
  \omega_{l0} &\equiv
  \frac{(-1)^l}{{\sigma_2}^{2l}\bar{n}_\mathrm{pk}}
  \int d^{10}y\, n_\mathrm{pk}
  L^{(3/2)}_l\left(\frac{5}{2}J_2\right)
  \,{\cal P}.
\end{align}
The higher-order renormalized bias functions $c_X^{(n)}$ can be
similarly obtained by further differentiating Eqs.~(\ref{eq:3-3-9a})
and (\ref{eq:3-3-9b}) and following similar procedures as above.

The above results have exactly the same form as the peak bias
functions, which have been derived in Refs.~\cite{Des13,LMD15}. These
authors generalized the peak-background split and argued that the peak
bias factors indeed are the ensemble average of orthogonal
polynomials. However, they did not explicitly demonstrate that their
generalized polynomial expansion holds beyond second order. In
Appendix~\ref{app:PeaksiPT}, we briefly sketch how this could be done
and emphasize the connection between the peak approach and the iPT.

Note that, as the peak constraint $n_\mathrm{pk}$ has a factor
$\delta_\mathrm{D}^3(\bm{\eta})$, only the constant term of the
generalized Laguerre polynomials $L^{(\alpha)}_n(0) =
\varGamma(n+\alpha+1)/[\varGamma(n+1)\varGamma(\alpha+1)]$ appears.
Therefore, Eq.~(\ref{eq:3-3-17b}) reduces to
\begin{equation}
  \label{eq:3-3-18}
  \chi_k = \frac{(2k+1)!!}{2^k k!}
  \frac{(-1)^k}{{\sigma_1}^{2k}}.
\end{equation}
The integrals Eqs.~(\ref{eq:3-3-17a}), (\ref{eq:3-3-17b}) and
(\ref{eq:3-3-17c}) appear up to second order, i.e., in the functions
$c_X^{(1)}$ and $c_X^{(2)}$. Note, however, that the bias coefficients
will generically take the form \cite{GPP12}
\begin{equation}
  \label{eq:3-3-19}
  \int d^{10}y\, n_\mathrm{pk} H_{ij}(\nu,J_1)
  L^{(1/2)}_k\left(\frac{3}{2}\eta^2\right)
  F_{lm}\left(5J_2, J_3\right)
  \,{\cal P},
\end{equation}
in the renormalized bias functions $c_X^{(n)}$ with $n \geq 3$
\cite{LMD15}, where
\begin{multline}
  \label{eq:3-3-20}
  F_{lm}\left(5J_2, J_3\right)
  \equiv (-1)^l\sqrt{\frac{\varGamma(5/2)}{2^{3m}\varGamma(3m+5/2)}}
\\
  \times L_l^{(3m+3/2)}\left(\frac{5}{2}J_2\right)
  P_m\left(\frac{J_3}{{J_2}^{3/2}}\right),
\end{multline}
are polynomials of $J_2$ and $J_3$, orthogonalized with the
Gram-Schmidt procedure, and $P_m(x)$ are Legendre polynomials. The
appearance of $P_m(x)$ reflects the fact that $J_3$ is an ``angular''
variable. This is the reason why we adopt the notation $\chi_k$ and
$\omega_{l 0}$ of Ref.~\cite{LMD15}. We refer the reader to this work
for more details.

\subsubsection{\label{subsubsec:peakscoef}
Bias coefficients of peaks model
}

Even though the bias coefficients $b_{ij}$, $\chi_k$ and $\omega_{l 0}$ 
are explicitly defined as ten-dimensional integrals, they can be reduced
to one-dimensional integrals at most. Explicit formulas of
the coefficients are derived below.

To begin with, we define a set of integrals:
\begin{align}
  \label{eq:3-3-50a}
  A^\mathrm{pk}_n(\nu_\mathrm{c})
  &\equiv \frac{1}{\bar{n}_\mathrm{pk}}
  \int d^{10}y\,n_\mathrm{pk}\,{J_1}^n\,{\cal P},
\\
  \label{eq:3-3-50b}
  B^\mathrm{pk}_n(\nu_\mathrm{c})
  &\equiv \frac{1}{\bar{n}_\mathrm{pk}}
  \int d^{10}y\,n_\mathrm{pk}\,{J_2}^n\,{\cal P}.
\end{align}
All the bias coefficients defined in Eqs.~(\ref{eq:3-3-17a}) and
(\ref{eq:3-3-17b}) can be represented by the above functions
$A^\mathrm{pk}_n$ and $B^\mathrm{pk}_n$ of Eqs.~(\ref{eq:3-3-50a}) and
(\ref{eq:3-3-50b}), because $H_{ij}$ and $L^{(\alpha)}_n$ are just
polynomials of their arguments, and peak constraints in
$n_\mathrm{pk}$ contain delta functions as
$\delta_\mathrm{D}(\nu-\nu_\mathrm{c})\,\delta_\mathrm{D}^3(\bm{\eta})$.
Defining invariant variables
\begin{equation}
  \label{eq:3-3-51}
  x = \lambda_1 + \lambda_2 + \lambda_3,\;\;
  y = \frac{1}{2}\left(\lambda_1 - \lambda_2\right),\;\;
  z = \frac{1}{2}\left(\lambda_1 - 2\lambda_2 + \lambda_3\right),
\end{equation}
where $\lambda_1$, $\lambda_2$, $\lambda_3$ are eigenvalues of
$-\zeta_{ij}$ with a descending order
($\lambda_1 \geq \lambda_2 \geq \lambda_3$), the peak number density
of Eq.~(\ref{eq:3-3-1}) reduces to \cite{BBKS}
\begin{multline}
  \label{eq:3-3-52}
  n_\mathrm{pk} =
  \frac{2}{\sqrt{3}{R_*}^3}
    \delta_\mathrm{D}(\nu-\nu_\mathrm{c})\, 
    \delta_\mathrm{D}^3(\bm{\eta})\,
    (x-2z) \left[(x+z)^2-(3y)^2\right]
    \\
    \times
    \varTheta(y-z) \varTheta(y+z)\varTheta(x-3y+z).
\end{multline}
Other variables in Eqs.~(\ref{eq:3-3-50a}) and (\ref{eq:3-3-50b})
correspond to $J_1 = x$, $J_2 = 3y^2 + z^2$. Following similar
calculations in Ref.~\cite{BBKS}, and defining a function
\begin{equation}
  \label{eq:3-3-53}
  F(x,y,z) \equiv
  (x-2z)\left[(x+z)^2 - (3y)^2\right]y(y^2-z^2),
\end{equation}
Eqs.~(\ref{eq:3-3-50a}) and (\ref{eq:3-3-50b}) reduce to
\begin{align}
  \label{eq:3-3-54a}
  A^\mathrm{pk}_n(\nu_\mathrm{c}) &=
  \frac{\displaystyle
    \int_0^\infty dx\, x^n f_0(x) {\cal N}(\nu_\mathrm{c},x)
  }
  {\displaystyle
    \int_0^\infty dx\, f_0(x) {\cal N}(\nu_\mathrm{c},x)
  },
  \\  
  \label{eq:3-3-54b}
  B^\mathrm{pk}_n(\nu_\mathrm{c}) &=
  \frac{\displaystyle
    \int_0^\infty dx\, f_n(x) {\cal N}(\nu_\mathrm{c},x)
  }
  {\displaystyle
    \int_0^\infty dx\, f_0(x) {\cal N}(\nu_\mathrm{c},x)
  },
\end{align}
where the function ${\cal N}$ is given by Eq.~(\ref{eq:3-3-13}), and
\begin{multline}
  \label{eq:3-3-55}
  f_n(x) \equiv
  \frac{3^2 5^{5/2}}{\sqrt{2\pi}}
  \left(
    \int_0^{x/4} dy \int_{-y}^y dz 
    + \int_{x/4}^{x/2} dy \int_{3y-x}^y dz 
  \right)
\\
  (3y^2+z^2)^n\,F(x,y,z)\,e^{-5(3y^2+z^2)/2}.
\end{multline}
The function $f_0(x)$ is identical to the function $f(x)$ defined by
Eq.~(A.15) of Ref.~\cite{BBKS}:
\begin{multline}
  \label{eq:3-3-56}
  f_0(x) =
  \frac{x}{2}\left(x^2-3\right)
  \left[
    \mathrm{erf}\left(\frac{1}{2}\sqrt{\frac{5}{2}}\,x\right) 
    + \mathrm{erf}\left(\sqrt{\frac{5}{2}}\,x\right) 
  \right]
\\
  + \sqrt{\frac{2}{5\pi}}
  \left[
    \left(\frac{x^2}{2} - \frac{8}{5}\right) e^{-5x^2/2}
    + \left(\frac{31}{4}x^2 + \frac{8}{5}\right) e^{-5x^2/8}
  \right].
\end{multline}
With the same consideration in Ref.~\cite{Des13}, the analytically
closed form of Eq.~(\ref{eq:3-3-55}) is derived from $f_0(x)$ as
\begin{equation}
  \label{eq:3-3-58}
  f_n(x) = \left.\left(-\frac{2}{5}\frac{\partial}{\partial\alpha}\right)^n
  \left[\frac{f_0(\alpha^{1/2}x)}{\alpha^4}\right]\right|_{\alpha = 1}.
\end{equation}
For example, the explicit form of $n=1$ is given by
\begin{multline}
  \label{eq:3-3-59}
  f_1(x) = 
  \frac{x}{2}\left(x^2-\frac{21}{5}\right)
  \left[
    \mathrm{erf}\left(\frac{1}{2}\sqrt{\frac{5}{2}}\,x\right) 
    + \mathrm{erf}\left(\sqrt{\frac{5}{2}}\,x\right) 
  \right]
\\
  + \sqrt{\frac{2}{5\pi}}
  \left[
    \left(\frac{x^2}{2} - \frac{64}{25}\right) e^{-5x^2/2}
    + \left(
      \frac{27}{16}x^4 + \frac{209}{20}x^2 + \frac{64}{25}
    \right) e^{-5x^2/8}
  \right].
\end{multline}
Thus, the originally ten-dimensional integrals of
Eqs.~(\ref{eq:3-3-50a}) and (\ref{eq:3-3-50b}) reduce to just
one-dimensional ones of Eqs.~(\ref{eq:3-3-54a}) and
(\ref{eq:3-3-54b}), for which numerically evaluations are straightforward.

Equations (\ref{eq:3-3-17a}) and (\ref{eq:3-3-18}) can be
straightforwardly represented by $A^\mathrm{pk}_n$ and
$B^\mathrm{pk}_n$, using explicit expressions for the polynomials
$H_{ij}$ and $L^{(\alpha)}_n$. The results are given by
\begin{align}
  \label{eq:3-3-60a}
  b_{10} &= \frac{1}{\sigma_0}\,
  \frac{\nu_\mathrm{c} - \gamma A^\mathrm{pk}_1(\nu_\mathrm{c})}{1-\gamma^2},
\\
  \label{eq:3-3-60b}
  b_{01} &= \frac{1}{\sigma_2}\,
  \frac{-\gamma \nu_\mathrm{c} + A^\mathrm{pk}_1(\nu_\mathrm{c})}{1-\gamma^2},
\\
  \label{eq:3-3-60c}
  b_{20} &= \frac{1}{{\sigma_0}^2}\,
  \frac{1}{1-\gamma^2}\,
  \left[
    \frac{\nu_\mathrm{c}^2 - 2\gamma\nu_\mathrm{c} A^\mathrm{pk}_1(\nu_\mathrm{c})
      + \gamma^2 A^\mathrm{pk}_2(\nu_\mathrm{c})}{1-\gamma^2} - 1
  \right],
\\
  \label{eq:3-3-60d}
  b_{11} &= \frac{1}{\sigma_0\sigma_2}\,
  \frac{1}{1-\gamma^2}
\nonumber\\
  &\quad \times
  \left[
    \frac{-\gamma\nu_\mathrm{c}^2
      + (1+\gamma^2)\nu_\mathrm{c} A^\mathrm{pk}_1(\nu_\mathrm{c})
      - \gamma A^\mathrm{pk}_2(\nu_\mathrm{c})}{1-\gamma^2} + \gamma
  \right],
\\
  \label{eq:3-3-60e}
  b_{02} &= \frac{1}{{\sigma_2}^2}\,
  \frac{1}{1-\gamma^2}\,
  \left[
    \frac{\gamma^2\nu_\mathrm{c}^2 -2 \gamma\nu_\mathrm{c} A^\mathrm{pk}_1(\nu_\mathrm{c})
     + A^\mathrm{pk}_2(\nu_\mathrm{c})}{1-\gamma^2} - 1
  \right],
\end{align}
and
\begin{align}
  \label{eq:3-3-61a}
  \chi_1 &= - \frac{3}{2{\sigma_1}^2},
\\
  \label{eq:3-3-61b}
  \omega_{10} &= - \frac{5}{2{\sigma_2}^2}
  \left[ 1 - B^\mathrm{pk}_1(\nu_\mathrm{c}) \right].
\end{align}
The quantities $A^\mathrm{pk}_1(\nu_\mathrm{c})$,
$A^\mathrm{pk}_2(\nu_\mathrm{c})$ and
$B^\mathrm{pk}_1(\nu_\mathrm{c})$ are given by one-dimensional
integrals of Eqs.~(\ref{eq:3-3-54a}) and (\ref{eq:3-3-54b}) with
Eqs.~(\ref{eq:3-3-13}), (\ref{eq:3-3-56}) and (\ref{eq:3-3-59}).

The above results for $b_{ij}$ can be conveniently represented by
matrix notation as follows. We note that Eq.~(\ref{eq:3-3-13}) is a
multivariate Gaussian function with a covariance matrix,
\begin{equation}
  \label{eq:3-3-62}
  \bm{M} =
  \left(
  \begin{matrix}
    1 & \gamma \\
    \gamma & 1
  \end{matrix}
  \right).
\end{equation}
Defining
\begin{equation}
  \label{eq:3-3-63a}
  \bm{b}^{(1)} \equiv
  \left(
    \begin{matrix}
      \sigma_0 b_{10} \\
      \sigma_2 b_{01}
    \end{matrix}
  \right), \quad
  \bm{b}^{(2)} \equiv
  \left(
    \begin{matrix}
      {\sigma_0}^2 b_{20} &
      \sigma_0\sigma_2 b_{11} \\
      \sigma_0\sigma_2 b_{11} &
      {\sigma_2}^2 b_{02}
    \end{matrix}
  \right),
\end{equation}
and
\begin{equation}
  \label{eq:3-3-64a}
  \bm{A}^{(1)} \equiv
  \left(
    \begin{matrix}
      {\nu_\mathrm{c}} \\
      A^\mathrm{pk}_1({\nu_\mathrm{c}})
    \end{matrix}
  \right), \quad
  \bm{A}^{(2)} \equiv
  \left(
    \begin{matrix}
      {\nu_\mathrm{c}}^2 &
      {\nu_\mathrm{c}} A^\mathrm{pk}_1({\nu_\mathrm{c}}) \\
      {\nu_\mathrm{c}} A^\mathrm{pk}_1({\nu_\mathrm{c}}) &
      A^\mathrm{pk}_2({\nu_\mathrm{c}})
    \end{matrix}
  \right),
\end{equation}
Eqs.~(\ref{eq:3-3-60a})--(\ref{eq:3-3-60e}) are equivalently
represented by
\begin{equation}
  \bm{b}^{(1)} = \bm{M}^{-1} \bm{A}^{(1)}, \quad
  \bm{b}^{(2)} = \bm{M}^{-1} \bm{A}^{(2)} \bm{M}^{-1} - \bm{M}^{-1}.
  \label{eq:3-3-65}
\end{equation}

\subsection{\label{subsec:ESPbias}
Excursion set peaks
}

The ESP model extends the peaks model with another constraint that the
smoothed linear density field should increase when the mass scale
decreases, $\partial\delta_s/\partial R_s<0$, in order to avoid the
cloud-in-cloud problem. We define the normalized slope of the smoothed
linear density field with respect to the smoothing radius,
\begin{equation}
  \label{eq:3-4-1}
  \mu_s = - \frac{1}{\varDelta_{s0}}\frac{\partial\delta_s}{\partial R_s},
\end{equation}
where
\begin{equation}
  \label{eq:3-4-2}
  \varDelta_{s0} =
  \left\langle
    \left(\frac{\partial\delta_s}{\partial R_s}\right)^2
  \right\rangle^{1/2}.
\end{equation}
The constraint of the ESP model is to require an inequality
$\mu_s > 0$. The differential number density of the ESP model is given
by \cite{AP90,PS12,DGR13}
\begin{equation}
  \label{eq:3-4-3}
  n_\mathrm{ESP} = -\left(\frac{d\sigma_{s0}}{dR_s}\right)^{-1}
  \varDelta_{s0}
  \frac{\mu_s}{\nu_s}
  \varTheta(\mu_s)\, n_\mathrm{pk},
\end{equation}
where $n_\mathrm{pk}$ is the differential number density of discrete
peaks given by Eq.~(\ref{eq:3-3-1}). This implies that the
multiplicity function of the excursion reads
\begin{equation}
f_\mathrm{ESP}(\nu_\mathrm{c}) \equiv V \nu_\mathrm{c}
\int d^{11}y\, n_\mathrm{ESP}\,{\cal P},
\end{equation}
where $V=M/\bar{\rho}_0$ is the Lagrangian volume of a halo of mass $M$
and the vector $(y_\alpha)$ now consists of the 11 variables 
$(\nu, \mu, \eta_i, \zeta_{ij})$.

Although it would be desirable to use the same window function (such
as the window shape of Ref.~\cite{CSS15} measured directly from
simulations), for all the relevant fields, our approach remains
perfectly consistent when different filters are applied. For instance,
top-hat smoothing is not appropriate to define density peaks because
the window function does not vanish sufficiently fast at high $k$. As
a result, spectral moments like $\sigma_2$ do not converge for a cold
dark matter (CDM) power spectrum. However, since top-hat smoothing is
the natural choice to relate the peak height to the spherical collapse
expectation, Refs.~\cite{PSD13,BCDP14} suggested applying the top-hat
window $W_\mathrm{T}$ the variables $\nu_s$ and $\mu_s$ and a Gaussian
filter $W_\mathrm{G}$ to the variables $\eta_{si}$ and $\zeta_{sij}$.
In the following, we denote the window function for $\nu_s$ and
$\mu_s$ by $W(kR)$ and that for $\eta_{si}$ and $\zeta_{sij}$ by
$\bar{W}(k\bar{R})$. When a single window function is applied, one can
simply set $\bar{R} = R$ and $\bar{W}(k\bar{R}) = W(kR)$.
In the following, we omit the subscript $s$ in this subsection below
and use notations such as $\nu$, $\mu$, $\eta_i$, $\zeta_{ij}$. The
quantity $\sigma_0$ is associated with the window function of $W(kR)$
and $\bar{\sigma}_1$, $\bar{\sigma}_2$ are associated with
$\bar{W}(k\bar{R})$. The rms of Eq.~(\ref{eq:3-4-2}) is represented by
$\varDelta_0$ with a window function of $W$ and explicitly given by
\begin{equation}
  \label{eq:3-4-5}
  {\varDelta_0}^2 = \int\frac{k^2dk}{2\pi^2}
  k^2 \left[W'(kR)\right]^2 P_\mathrm{L}(k),
\end{equation}
where $W'(x) = dW(x)/dx$ is the first derivative of the window function.

\subsubsection{\label{subsubsec:ESPRBF} Derivation of renormalized
  bias functions in the ESP model }

We define rotationally invariant quantities $\eta^2$, $J_1$, $J_2$ and
$J_3$ as in Eq.~(\ref{eq:3-3-2}). For a Gaussian initial condition,
the joint probability distribution function is given by
\begin{equation}
  \label{eq:3-4-10}
  {\cal P}(\bm{y}) \propto
  {\cal N}(\nu,J_1,\mu)
  \exp\left(
    - \frac{3}{2} \eta^2 - \frac{5}{2} J_2
  \right),
\end{equation}
where ${\cal N}(\nu,J_1,\mu)$ is the trivariate distribution function,
which is given by
\begin{equation}
  \label{eq:3-4-11}
  {\cal N}(\nu,J_1,\mu) = 
  \frac{1}{\sqrt{(2\pi)^3 |\bm{M}|}}
  \exp\left(-\frac{1}{2}\bm{a}^\mathrm{T} \bm{M}^{-1} \bm{a}\right),
\end{equation}
where
\begin{equation}
  \label{eq:3-4-12}
  \bm{a} = 
  \left(
  \begin{matrix} \nu \\ J_1 \\ \mu \end{matrix}
  \right), \quad
  \bm{M} =
  \left(
  \begin{matrix}
    1 & \gamma_{12} & \gamma_{13} \\
    \gamma_{12} & 1 &  \gamma_{23} \\
    \gamma_{13} & \gamma_{23} & 1
  \end{matrix}
\right).
\end{equation}
The matrix $\bm{M}$ is the covariance matrix of $\bm{a}$:
$M_{ij} = \langle a_i a_j \rangle$. The variables are normalized so as
to have the diagonal elements of this matrix unity. The off-diagonal
elements are given by
\begin{align}
  \label{eq:3-4-13a}
  \gamma_{12} &= \langle \nu J_1 \rangle =
  \frac{1}{\sigma_0\bar{\sigma}_2}
  \int\frac{k^2dk}{2\pi^2} k^2W(kR)\bar{W}(k\bar{R}) P_\mathrm{L}(k),
\\
  \label{eq:3-4-13b}
  \gamma_{13} &= \langle \nu\mu \rangle =
  - \frac{1}{\sigma_0\varDelta_0}
  \int\frac{k^2dk}{2\pi^2} kW(kR)W'(kR) P_\mathrm{L}(k),
\\
  \label{eq:3-4-13c}
  \gamma_{23} &= \langle J_1\mu \rangle =
  - \frac{1}{\bar{\sigma}_2\varDelta_0}
  \int\frac{k^2dk}{2\pi^2} k^3 \bar{W}(k\bar{R}) W'(kR) P_\mathrm{L}(k).
\end{align}
The determinant $|\bm{M}|$ and the inverse matrix
$\bm{M}^{-1}$ are given by
\begin{align}
  \label{eq:3-4-14a}
  |\bm{M}| &= 1 - {\gamma_{12}}^2 - {\gamma_{23}}^2
  - {\gamma_{13}}^2 + 2\gamma_{12}\gamma_{23}\gamma_{13},
  \\
  \label{eq:3-4-14b}
  \bm{M}^{-1} &=
  \frac{1}{|\bm{M}|}
  \left(
  \begin{matrix}
    1 - {\gamma_{23}}^2
    & \gamma_{23}\gamma_{13} - \gamma_{12}
    & \gamma_{12}\gamma_{23} - \gamma_{13} \\
    \gamma_{23}\gamma_{13} - \gamma_{12}
    &1 - {\gamma_{13}}^2
    & \gamma_{13}\gamma_{12} - \gamma_{23} \\
    \gamma_{12}\gamma_{23} - \gamma_{13}
    & \gamma_{13}\gamma_{12} -  \gamma_{23}
    & 1 - {\gamma_{12}}^2
  \end{matrix}
  \right).
\end{align}

Choosing a Gaussian filter for both windows, i.e.
$W(kR) = \bar{W}(k\bar{R}) = W_\mathrm{G}(kR)$, leads to
$-kW_\mathrm{G}'(kR) = R k^2W_\mathrm{G}(kR)$ and
$\mu = (R\bar{\sigma}_2/\varDelta_0) J_1$, which signifies that $\mu$
and $J_1$ are redundant variables. In this special case, the third
variable in $\bm{a}$ is not necessary and we only need a
two-dimensional covariance matrix. We will not consider this simpler
case in what follows.

Using the fact that ${\cal P}$ is a function of only $\nu$, $\mu$,
$\eta^2$, $J_1$ and $J_2$, and following the same steps of
Eqs.~(\ref{eq:3-3-9a})--(\ref{eq:3-3-11b}), we have
\begin{widetext}
\begin{align}
  \label{eq:3-4-15a}
  \left\langle {\cal D}(\bm{k}) \right\rangle_\Omega {\cal P}
  &= \left[
    \frac{W(kR)}{\sigma_0}
      \frac{\partial}{\partial\nu} 
    + \frac{k^2\bar{W}(k\bar{R})}{\bar{\sigma}_2}
    \frac{\partial}{\partial J_1}
    -\: \frac{kW'(kR)}{\varDelta_0} \frac{\partial}{\partial\mu} 
    \right] {\cal P},
\\
  \label{eq:3-4-15b}
  \left\langle
    {\cal D}(\bm{k}_1) {\cal D}(\bm{k}_2)
  \right\rangle_\Omega
  {\cal P}
  &=
  \left\{
    \left[
      \frac{W(k_1R)}{\sigma_0}
      \frac{\partial}{\partial\nu}
      + \frac{{k_1}^2\bar{W}(k_1\bar{R})}{\bar{\sigma}_2}
      \frac{\partial}{\partial J_1}
      - \frac{k_1W'(k_1R)}{\varDelta_0} \frac{\partial}{\partial\mu} 
    \right]
    \left[
      \frac{W(k_2R)}{\sigma_0}
      \frac{\partial}{\partial\nu}
      + \frac{{k_2}^2\bar{W}(k_2\bar{R})}{\bar{\sigma}_2}
      \frac{\partial}{\partial J_1}
      - \frac{k_2W'(k_2R)}{\varDelta_0} \frac{\partial}{\partial\mu} 
    \right]
  \right.
\nonumber\\
  & \qquad
  \left.
    -\:
    \frac{2(\bm{k}_1\cdot\bm{k}_2)
      \bar{W}(k_1\bar{R})\bar{W}(k_2\bar{R})}
    {{\bar{\sigma}_1{}}^2}
    \left[
      1 + \frac{2}{3} \eta^2 \frac{\partial}{\partial (\eta^2)}
    \right] \frac{\partial}{\partial(\eta^2)}
  \right.
\nonumber\\
  & \qquad
  \left.
    +\: \frac{\left[3(\bm{k}_1\cdot\bm{k}_2)^2 - {k_1}^2 {k_2}^2\right]
      \bar{W}(k_1\bar{R})\bar{W}(k_2\bar{R})}{{\bar{\sigma}_2{}}^2}
      \left[
        1 + \frac{2}{5} J_2 \frac{\partial}{\partial(J_2)}
      \right] \frac{\partial}{\partial(J_2)}
  \right\} {\cal P},
\end{align}
Substituting Eqs.~(\ref{eq:3-4-15a}) and (\ref{eq:3-4-15b}) into
the integrand of Eq.~(\ref{eq:3-12}), we have
\begin{align}
  \label{eq:3-4-16a}
  c_X^{(1)}(k) &= 
  b_{100} W(kR)
  + b_{010} k^2 \bar{W}(k\bar{R})
  - b_{001} k W'(kR),
  \\
  \nonumber
  c_X^{(2)}(\bm{k}_1, \bm{k}_2) &=
  b_{200} W(k_1R) W(k_2R)
  + b_{110}
  \left[{k_2}^2
    W(k_1R) \bar{W}(k_2\bar{R}) + (1\leftrightarrow 2)
  \right]
  \\
  \nonumber
  & \quad +
  \left\{
    b_{020} {k_1}^2 {k_2}^2
    + \omega_{10}
    \left[3(\bm{k}_1\cdot\bm{k}_2)^2 - {k_1}^2 {k_2}^2\right]
    - 2 \chi_1 (\bm{k}_1\cdot\bm{k}_2)
  \right\} \bar{W}(k_1\bar{R}) \bar{W}(k_2\bar{R})
  \\
  \label{eq:3-4-16b}
  & \quad
  - b_{101}
  \left[
    k_1 W'(k_1R) W(k_2R) + (1\leftrightarrow 2)
  \right]
  - b_{011}
  \left[
    k_1 {k_2}^2 W'(k_1R) \bar{W}(k_2\bar{R}) + (1\leftrightarrow 2)
  \right]
  + b_{002} k_1 k_2 W'(k_1R) W'(k_2R),
\end{align}
\end{widetext}
where
\begin{align}
  \label{eq:3-4-17a}
  b_{ijk} &=
  \frac{1}{{\sigma_0}^i{{\bar{\sigma}_2{}}^j{\varDelta_0}^k\bar{n}_\mathrm{ESP}}}
  \int d^{11}y\, n_\mathrm{ESP} H_{ijk}(\nu,J_1,\mu)\,{\cal P},
  \\
  \label{eq:3-4-17b}
  \chi_k &= \frac{(2k+1)!!}{2^k k!}\frac{(-1)^k}{{\bar{\sigma}_1{}}^{2k}},
  \\
  \label{eq:3-4-17c}
  \omega_{l0} &= \frac{(-1)^l}{\bar{\sigma}_2{}^{2l}\,\bar{n}_\mathrm{ESP}}
  \int d^{11}y\, n_\mathrm{ESP}
  L^{(3/2)}_l\left(\frac{5}{2}J_2\right)  \,{\cal P}.
\end{align}
Here, $H_{ijk}$ are trivariate Hermite polynomials
\begin{equation}
  \label{eq:3-4-18}
  H_{ijk}(\nu,J_1,\mu) \equiv
  \frac{(-1)^{i+j+k}}{{\cal N}(\nu,J_1,\mu)}
  \left(\frac{\partial}{\partial \nu}\right)^i
  \left(\frac{\partial}{\partial J_1}\right)^j
  \left(\frac{\partial}{\partial \mu}\right)^k
  {\cal N}(\nu,J_1,\mu),
\end{equation}
and we have exploited the fact that $n_\mathrm{ESP}$ contains a delta
function $\delta^3_\mathrm{D}(\bm{\eta})$ to simplify $\chi_i$.

Again, Eqs.~(\ref{eq:3-4-16a}) and (\ref{eq:3-4-16b}) exactly agree
with the results derived independently in Refs.~\cite{LMD15,Dizgah+15}
in a fairly different manner.

\subsubsection{\label{subsubsec:ESPcoef} Bias coefficients of the ESP
  model }

The coefficients $b_{ijk}$ and $\omega_{l0}$ also reduce to
expressions with up to one-dimensional integrals, extending the method
of Sec.~\ref{subsubsec:peakscoef}. For this purpose, we define
integrals,
\begin{align} 
  \label{eq:3-4-100a}
  A^\mathrm{ESP}_{nm}(\nu_\mathrm{c}) &\equiv
  \frac{1}{\bar{n}_\mathrm{ESP}}
  \int d^{11}y\,n_\mathrm{ESP}
  \,{J_1}^n\,\mu^m\, {\cal P},
\\
  \label{eq:3-4-100b}
  B^\mathrm{ESP}_n(\nu_\mathrm{c}) &\equiv
  \frac{1}{\bar{n}_\mathrm{ESP}}
  \int d^{11}y\,n_\mathrm{ESP}
  \,{J_2}^n\,{\cal P}.
\end{align}
Just in a similar manner of deriving Eqs.~(\ref{eq:3-3-54a}) and
(\ref{eq:3-3-54b}), Eqs.~(\ref{eq:3-4-100a}) and (\ref{eq:3-4-100b})
reduce to
\begin{align}
  \label{eq:3-4-101a}
  A^\mathrm{ESP}_{nm}(\nu_\mathrm{c}) &=
  \frac{\displaystyle
    \int_0^\infty dx\,x^n\,f_0(x)\,g_m(\nu_\mathrm{c},x)
  }
  {\displaystyle
    \int_0^\infty dx\,f_0(x)\,g_0(\nu_\mathrm{c},x)
  },\quad
\\
  \label{eq:3-4-101b}
  B^\mathrm{ESP}_n(\nu_\mathrm{c}) &=
  \frac{\displaystyle
    \int_0^\infty dx\,f_n(x)\,g_0(\nu_\mathrm{c},x)
  }
  {\displaystyle
    \int_0^\infty dx\,f_0(x)\,g_0(\nu_\mathrm{c},x)
  },
\end{align}
where
\begin{equation}
  \label{eq:3-4-102}
  g_m(\nu_\mathrm{c},x) =
  \int_0^\infty d\mu\,\mu^{m+1}\,
  {\cal N}(\nu_\mathrm{c},x,\mu).
\end{equation}

The function $g_m(\nu_\mathrm{c},x)$ is analytically represented by
the parabolic cylinder function $D_\lambda(z)$ which has an integral
representation,
\begin{equation}
  \label{eq:3-4-103}
  D_\lambda(z) =
  \frac{e^{-z^2/4}}{\varGamma(-\lambda)}
  \int_0^\infty  e^{-zt-t^2/2} t^{-\lambda-1} dt.
\end{equation}
For our convenience, we define a function
\begin{equation}
  \label{eq:3-4-104}
  H_\lambda(z) \equiv e^{z^2/4} D_\lambda(z).
\end{equation}
When $\lambda=n$ is a non-negative integer, this function reduces to
Hermite polynomials $H_n(z)$. When
$\lambda=-n$ is a negative integer, integral representation of
$H_{-n}(z)$ is given by
\begin{align}
  \label{eq:3-4-105}
  H_{-n}(z) &=
  \frac{1}{(n-1)!}
  \int_0^\infty  e^{-zt-t^2/2} t^{n-1} dt
\nonumber\\
  &= \frac{\sqrt{\pi/2}}{(n-1)!}
  \left(-\frac{d}{dz}\right)^{n-1}
  \left[
    e^{z^2/2} \mathrm{erfc}\left(\frac{z}{\sqrt{2}}\right)
  \right].
\end{align}
First several functions are explicitly given by
\begin{align}
  \label{eq:3-4-106a}
  H_{-1}(z) &=
  \sqrt{\frac{\pi}{2}}\, e^{z^2/2}\,
  \mathrm{erfc}\left(\frac{z}{\sqrt{2}}\right),
\\
  \label{eq:3-4-106b}
  H_{-2}(z) &=
  1 - 
  \sqrt{\frac{\pi}{2}}\,z\,e^{z^2/2}\,
  \mathrm{erfc}\left(\frac{z}{\sqrt{2}}\right),
\\
  \label{eq:3-4-106c}
  H_{-3}(z) &=
  \frac{1}{2}
  \left[
    -z
    + \sqrt{\frac{\pi}{2}}\,(z^2+1)\,e^{z^2/2}\,
    \mathrm{erfc}\left(\frac{z}{\sqrt{2}}\right)
  \right],
\\
  \label{eq:3-4-106d}
  H_{-4}(z) &=
  \frac{1}{6}
  \left[
    z^2 + 2
    - \sqrt{\frac{\pi}{2}}\,(3z + z^3)\,e^{z^2/2}\,
    \mathrm{erfc}\left(\frac{z}{\sqrt{2}}\right)
  \right].
\end{align}

Using the function $H_{-n}(z)$ defined above, an integration by $\mu$
in Eq.~(\ref{eq:3-4-102}) can be analytically performed, resulting in,
\begin{multline}
  \label{eq:3-4-107}
  g_m(\nu_\mathrm{c},x) =
  \frac{(m+1)!}{\sqrt{(2\pi)^3 |\bm{M}|}\,(M^{-1}_{33})^{m/2+1}}
  \\
  \times
  \exp\left[
    - \frac{1}{2}
    \left(M^{-1}_{11}{\nu_\mathrm{c}}^2
      + 2M^{-1}_{12}\nu_\mathrm{c} x
      +  M^{-1}_{22} x^2\right)
  \right]
  \\
  \times
  H_{-(m+2)}
  \left(
    \frac{M^{-1}_{13}\nu_\mathrm{c}
      + M^{-1}_{23} x}{\sqrt{M^{-1}_{33}}}
  \right),
\end{multline}
where $M^{-1}_{ij} = [\bm{M}^{-1}]_{ij}$ are matrix elements of the
inverse matrix $\bm{M}^{-1}$ given by Eq.~(\ref{eq:3-4-14b}).
Substituting Eqs.~(\ref{eq:3-4-105}) and (\ref{eq:3-4-107}) into
Eqs.~(\ref{eq:3-4-101a}) and (\ref{eq:3-4-101b}), only one-dimensional
numerical integrations of smooth functions are required.

Equations (\ref{eq:3-4-17a}), (\ref{eq:3-4-17b}) and
(\ref{eq:3-4-17c}) can be straightforwardly represented by functions
$A^\mathrm{ESP}_{nm}(\nu_\mathrm{c})$ and
$B^\mathrm{ESP}_n(\nu_\mathrm{c})$, using explicit forms of
polynomials $H_{ijk}$, $L^{(\alpha)}_i$. As in
Eqs.~(\ref{eq:3-3-62})--(\ref{eq:3-3-65}) of the peaks model, the
results for $b_{ijk}$ are conveniently represented by matrix notation.
Defining
\begin{align}
  \label{eq:3-4-109a}
  \bm{b}^{(1)} &\equiv
  \left(
    \begin{matrix}
      \sigma_0 b_{100} \\
      \bar{\sigma}_2 b_{010} \\
      \varDelta_0 b_{001}
    \end{matrix}
  \right),
  \\
  \label{eq:3-4-109b}
  \bm{b}^{(2)} &\equiv
  \left(
    \begin{matrix}
      {\sigma_0}^2 b_{200} &
      \sigma_0\bar{\sigma}_2 b_{110} &
      \sigma_0\varDelta_0 b_{101} \\
      \sigma_0\bar{\sigma}_2 b_{110} &
      {\bar{\sigma}_2{}}^2 b_{020} &
      \bar{\sigma}_2\varDelta_0  b_{011} \\
      \sigma_0\varDelta_0  b_{101} &
      \bar{\sigma}_2\varDelta_0  b_{011} &
      {\varDelta_0}^2 b_{002}
    \end{matrix}
  \right),
\end{align}
and
\begin{align}
  \label{eq:3-4-110a}
  \bm{A}^{(1)} &\equiv
  \left(
    \begin{matrix}
      {\nu_\mathrm{c}} \\
      A^\mathrm{ESP}_{10}({\nu_\mathrm{c}}) \\
      A^\mathrm{ESP}_{01}({\nu_\mathrm{c}})
    \end{matrix}
  \right),
  \\
  \label{eq:3-4-110b}
  \bm{A}^{(2)} &\equiv
  \left(
    \begin{matrix}
      {\nu_\mathrm{c}}^2 &
      {\nu_\mathrm{c}} A^\mathrm{ESP}_{10}({\nu_\mathrm{c}}) &
      {\nu_\mathrm{c}} A^\mathrm{ESP}_{01}({\nu_\mathrm{c}}) \\
      {\nu_\mathrm{c}} A^\mathrm{ESP}_{10}({\nu_\mathrm{c}}) &
      A^\mathrm{ESP}_{20}({\nu_\mathrm{c}}) &
      A^\mathrm{ESP}_{11}({\nu_\mathrm{c}}) \\
      {\nu_\mathrm{c}} A^\mathrm{ESP}_{01}({\nu_\mathrm{c}}) &
      A^\mathrm{ESP}_{11}({\nu_\mathrm{c}}) &
      A^\mathrm{ESP}_{02}({\nu_\mathrm{c}})
    \end{matrix}
  \right),
\end{align}
we have
\begin{equation}
  \bm{b}^{(1)} = \bm{M}^{-1} \bm{A}^{(1)}, \quad
  \bm{b}^{(2)} = \bm{M}^{-1} \bm{A}^{(2)} \bm{M}^{-1} - \bm{M}^{-1},
  \label{eq:3-4-111}
\end{equation}
where $\bm{M}^{-1}$ is given by Eq.~(\ref{eq:3-4-14b}). All the
coefficients to evaluate the renormalized bias functions up to second
order in Eqs.~(\ref{eq:3-4-16a}) and (\ref{eq:3-4-16b}) for the ESP
model are thus obtained. The results for $\chi_1$ and $\omega_{10}$
are
\begin{align}
  \label{eq:3-4-112a}
  \chi_1 &= - \frac{3}{2{\bar{\sigma}_1{}}^2},
\\
  \label{eq:3-4-112b}
  \omega_{10} &= - \frac{5}{2{\bar{\sigma}_2{}}^2}
  \left[ 1 - B^\mathrm{ESP}_1(\nu_\mathrm{c}) \right].
\end{align}

\section{\label{sec:Results}
Results
}

In this section, all the formulas in previous sections are put
together, and the results of power spectra and correlation functions
with various biasing schemes are presented. In the following, the flat
$\Lambda$CDM model with cosmological parameters
$\varOmega_\mathrm{m0} = 0.3089$, $\varOmega_\mathrm{b0} = 0.0486$,
$h = 0.6774$, $n_\mathrm{s} = 0.9667$, $\sigma_8 = 0.8159$ (Planck2015
\cite{Planck2015}) is assumed. We will hereafter present for the
representative redshifts $z=1,2,3$, which are of particular interest
because currently planned, forthcoming redshift surveys will harvest
this redshift range. We have checked that results at lower redshift,
such as $z=0.5$, are qualitatively similar to those at $z=1$. Another
reason for focusing at $z\geq 1$ is the fact that the applicability
range of the perturbation theory decreases noticeably for $z\ll 1$.

\subsection{\label{subsec:ResultsModels}
Bias models
}

Four different models of bias are considered in this section. The
``halo model'' refers to a model described in
Sec.~\ref{subsec:halobias}, and the renormalized bias functions are
given by Eqs.~(\ref{eq:3-2-11a}) and (\ref{eq:3-2-11b}) with
coefficients of Eq.~(\ref{eq:3-2-9}). The top-hat window function
$W_\mathrm{T}(kR)$, and the Sheth-Tormen mass function,
Eq.~(\ref{eq:3-2-5}), are adopted in this model. The only parameter in
this model is a smoothing radius $R$, or equivalently a mass scale $M$
of Eq.~(\ref{eq:3-2-1}).

The ``local halo'' refers to a model with scale-independent values of
renormalized bias functions, $c_X^{(1)} = b^\mathrm{L}_1$,
$c_X^{(2)} = b^\mathrm{L}_2$, where $b^\mathrm{L}_n$ are given by the
halo model above. This model is a simplified version of the halo
model, in which the renormalized bias functions are replaced by their
low-$k$ limits. Hence, this is equivalent to completely neglecting the
effects of the window function in Eqs.~(\ref{eq:3-2-11a}) and
(\ref{eq:3-2-11b}). Scale-independent bias functions correspond to a
bias model in which the number density of biased tracers
$n^\mathrm{L}_X(\bm{x})$ solely is a function of linear density field
$\delta_\mathrm{L}(\bm{x})$ at the same Lagrangian position $\bm{x}$.
We consider this model for the purpose of assessing the importance of
the window functions in the halo model.

The ``peaks model'' refers to the model described in
Sec.~\ref{subsec:peaksbias}. Its renormalized bias functions are given
by Eqs.~(\ref{eq:3-3-16a}) and (\ref{eq:3-3-16b}) with coefficients
calculated by Eqs.~(\ref{eq:3-3-60a})--(\ref{eq:3-3-61b}). A Gaussian
window function $W_\mathrm{G}(kR_\mathrm{G})$ is adopted throughout.
While the threshold $\nu_\mathrm{c}$ is originally a free parameter of
the peaks model, we fix its value with a relation
$\nu_\mathrm{c} = \delta_\mathrm{c}/\sigma_\mathrm{G0}(R_\mathrm{G})$,
where $\sigma_\mathrm{G0}(R_\mathrm{G}) = \sigma_0(R_\mathrm{G})$ is
the rms of variance. Therefore, the Gaussian smoothing radius
$R_\mathrm{G}$ is the only parameter in this model.

The ``ESP model'' refers to a model described in
Sec.~\ref{subsec:ESPbias}, and the renormalized bias functions are
given by Eqs.~(\ref{eq:3-4-16a}) and (\ref{eq:3-4-16b}) with
coefficients calculated by
Eqs.~(\ref{eq:3-4-109a})--(\ref{eq:3-4-112b}). There are two kinds of
window functions in this model: a top-hat and Gaussian, which we
denote as $W(kR) = W_\mathrm{T}(kR_\mathrm{T})$ and
$\bar{W}(k\bar{R}) = W_\mathrm{G}(kR_\mathrm{G})$, respectively. These
smoothing radii are related by $R_\mathrm{G} = 0.46 R_\mathrm{T}$
\cite{PSD13}. Furthermore, the threshold value is fixed by
$\nu_\mathrm{c} = \delta_\mathrm{c}/\sigma_\mathrm{T0}(R)$, where
$\sigma_\mathrm{T0}(R) = \sigma_0(R)$ is the rms of variance
with the top-hat window function. Hence, the top-hat smoothing radius
$R$ is the only free parameter of this model.

Each bias model has a unique parameter in our settings described
above. To make comparisons among various biasing schemes, the
parameter of each model is adjusted so as to give the same value for
the first-order renormalized bias function in the low-$k$ limit,
$\lim_{k\rightarrow 0}c_X^{(1)}(k)$. This limiting value is the bias
parameter $b^\mathrm{L}_1$, $b_{10}$ or $b_{100}$, depending on the
model details. For the purpose of presentation, we define the value by
the parameter $b^\mathrm{L}_1(M)$ with the top-hat window function and
a mass scale $M=1\times 10^{13} \,h^{-1}M_\odot$ in
Eq.~(\ref{eq:3-2-2}). The resulting values are
$b^\mathrm{L}_1 = 1.053$ ($z=1$), $2.694$ ($z=2$), $5.039$ ($z=3$).
The smoothing radii of peaks and ESP models are adjusted to reproduce
the same values in $b_{10}$ and $b_{100}$. The corresponding mass
scale varies in the range
$M=0.7$--$1.8 \times 10^{13} \,h^{-1}M_\odot$ for the peaks and ESP
models, with a slight dependence on redshift.

\subsection{\label{subsec:ResCns}
Renormalized bias functions
}

\begin{figure*}
\begin{center}
\includegraphics[height=25.2pc]{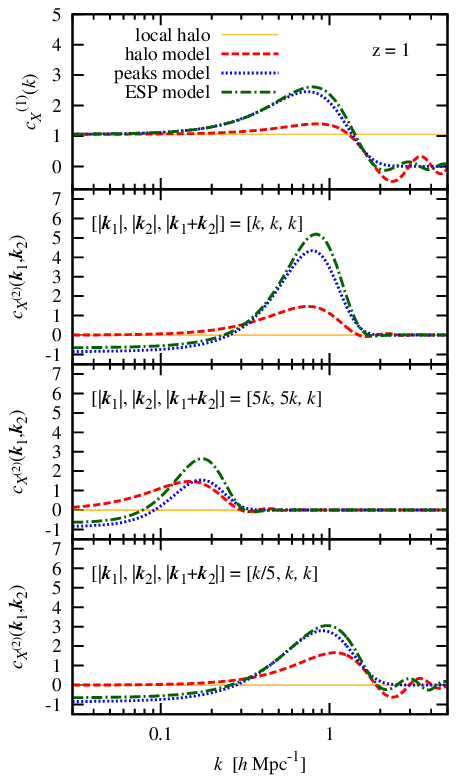}
\hspace{.2pc}
\includegraphics[height=25.2pc]{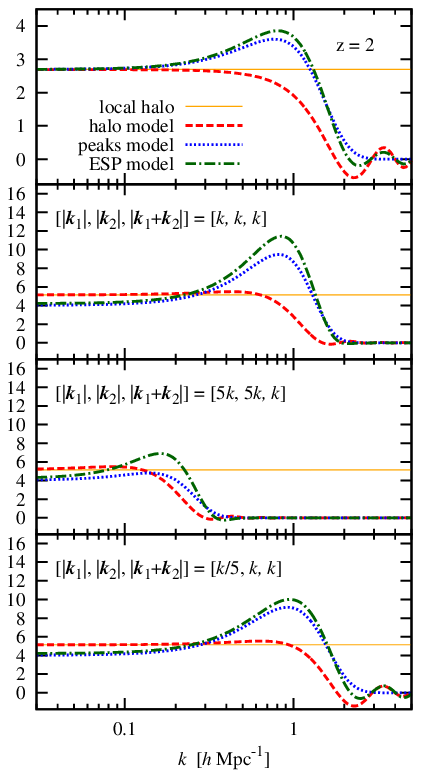}
\hspace{.2pc}
\includegraphics[height=25.2pc]{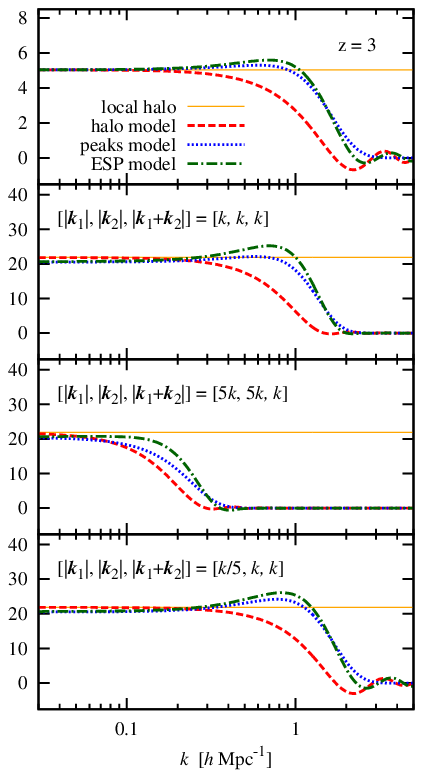}
\caption{\label{fig:cnsplot} Renormalized bias functions, $c_X^{(1)}$
  and $c_X^{(2)}$. The results for three redshifts $z=1,2,3$ are
  shown as indicated in each figure. Four models of bias are plotted
  in different lines: local halo (solid, orange), halo model (dashed,
  red), peaks model (dotted, blue), and ESP model (dot-dashed,
  green).}
\end{center}
\end{figure*}

The renormalized bias functions, $c_X^{(1)}(k)$ and
$c_X^{(2)}(\bm{k}_1,\bm{k}_2)$ are shown in Fig.~\ref{fig:cnsplot}.
For the second-order functions, the horizontal axis corresponds to the
amplitude of $|\bm{k}_1+\bm{k}_2| \equiv k$, which is relevant to the
scale of power spectrum $P(k)$. Three different shapes corresponding
to the triangles
$[|\bm{k}_1|,|\bm{k}_2|,|\bm{k}_1+\bm{k}_2|] = [k,k,k], [5k,5k,k],
[k/5,k,k]$ are plotted to illustrate the characteristic behaviors.
These configurations correspond to equilateral, folded, and squeezed
shapes of a triangle, respectively.

The local halo model has a constant value in each panel by definition.
Other models have asymptotes $c_X^{(n)} \rightarrow 0$ in large $k$,
because the window functions vanish in this limit. This reflects the
fact that the halo centers cannot have clustering power on scales
smaller than the halo mass. The value of second-order parameter
$b^\mathrm{L}_2$ turns out to be very close to zero at redshift $z=1$
for our cosmology and fiducial mass function. Consequently, the
low-$k$ limit of the renormalized bias function in the halo model also
is very close to zero.

A striking feature in the scale dependence of the renormalized bias 
functions is the appearance of peaks before the $c_X^{(n)}$ decay
to zero in the large-$k$ limit. The height of these peaks is generally
larger at lower redshift. 
However, the amplitudes depend strongly on bias models. The
peak height of the halo model is lower than those of peaks and ESP
models. There are oscillations around the asymptote in the large-$k$
tails for halo and ESP models. These oscillations reflect the property
of top-hat window function. Such oscillations are not seen in peaks
model in which only Gaussian window functions are used.

The first-order renormalized bias function $c_X^{(1)}$ has recently
been measured from the analysis of halos in $N$-body simulations
\cite{BDS15,CSS15}. The appearance of peaks at around $kR \sim 2.5$
and oscillating features in high-$k$ tails are clearly observed. For
instance, the behavior of the numerical results in the $z=0.95$ sample
of Ref.~\cite{CSS15} (see its Fig.~5) lies somewhere between the
predictions of the halo model and ESP model in the $z=1$ plot of our
Fig.~\ref{fig:cnsplot}: the peak height in the numerical simulations
is larger than the halo model and smaller than the ESP model, and the
amplitude of oscillations in the high-$k$ tail is smaller than the
halo model and larger than the ESP model. The authors of
Ref.~\cite{CSS15} use an effective window function $W_\mathrm{eff}$
and a model which is similar to our Eq.~(\ref{eq:3-2-11a}) but
consider the coefficients $b^\mathrm{L}_1$ and $1/\delta_\mathrm{c}$
as free parameters. Fitting the three parameters $R$, $b^\mathrm{L}_1$
and $1/\delta_\mathrm{c}$ to their numerical results, the
scale dependence of Lagrangian bias factor is nicely accounted for.

One should, however, bear in mind that the precise shapes of
renormalized bias functions depend on the details of the halo
identification procedure. While the numerical simulations mentioned
above use the ``Friends-of-Friends'' algorithm \cite{DEFW85}, one
should naturally expect that other methods, such as the ``Spherical
Overdensity'' algorithm \cite{LC94}, yield different results. Since
the purpose of this paper is to investigate the impacts of different
biasing schemes rather than fit our models to numerical results based
on a specific halo-finding algorithm, we will keep on investigating
how the four different models affect the predictions of iPT for the
power spectra and correlation functions.

\subsection{\label{subsec:ResultPSCF}
Power spectra and correlation functions in real space
}

\begin{figure*}
\begin{center}
\includegraphics[height=16.8pc]{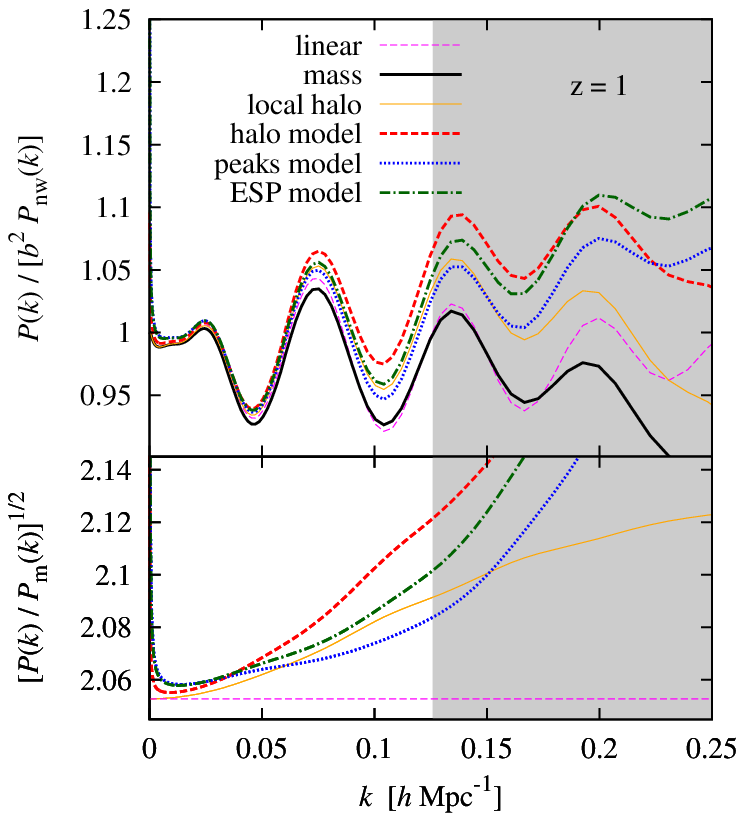}
\includegraphics[height=16.8pc]{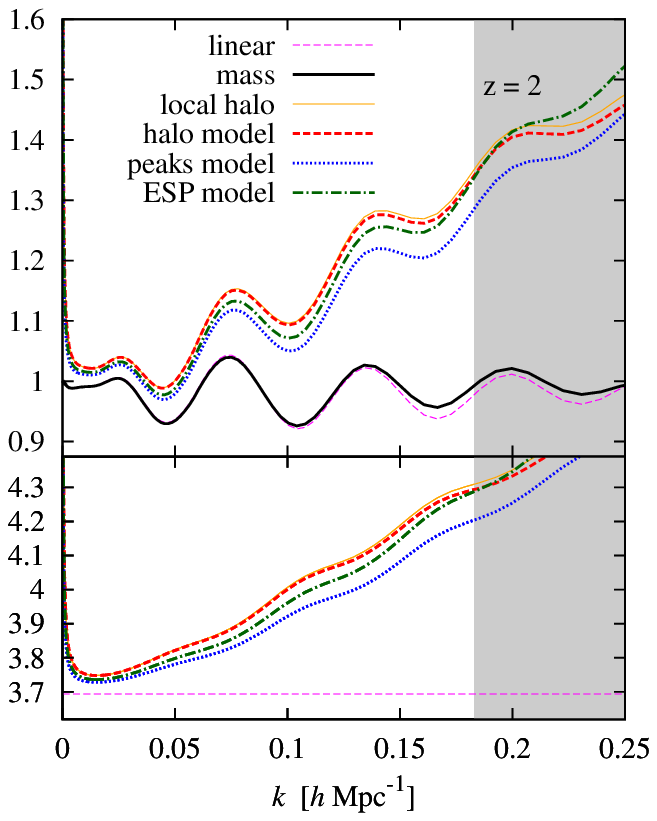}
\includegraphics[height=16.8pc]{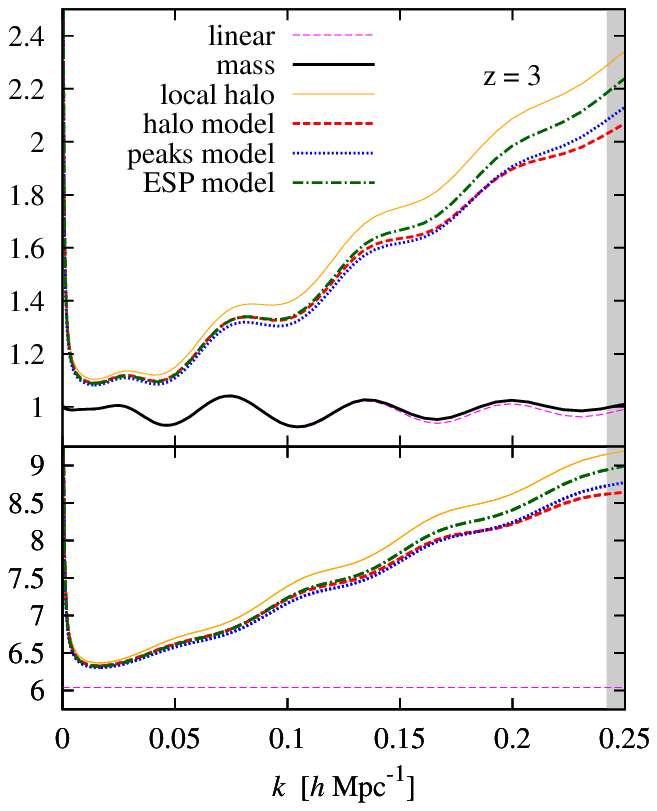}
\caption{\label{fig:psplot2} The one-loop power spectra in real space
  with different biasing schemes. Upper panels show the power spectra
  divided by the linear, no-wiggle power spectrum with the linear
  bias, $P_X(k)/[b^2P_\mathrm{nw}(k)]$. The lower panels show the
  scale-dependent bias, $[P_X(k)/P_\mathrm{m}(k)]^{1/2}$. The meanings
  of different lines are indicated in the panels. Shaded regions
  represent rough estimates where the one-loop perturbation theory is
  expected to be inaccurate.}
\end{center}
\end{figure*}

Our predictions for the one-loop power spectra in real space are shown
in Fig.~\ref{fig:psplot2}. The upper panels show the power spectra
divided by a no-wiggle linear power spectrum $P_\mathrm{nw}(k)$
\cite{EH99} and by the square of Eulerian linear bias parameter,
$b^2 = (1 + b^\mathrm{L}_1)^2$. The lower panels show the
scale-dependent bias, which is defined as the square root of the ratio
between the power spectrum of biased tracers and that of the mass
distribution, $[P_X(k)/P_\mathrm{m}(k)]^{1/2}$. Horizontally straight
lines in bottom panels indicate the linear bias factor $b$. Here and
henceforth, the shaded region in each figure corresponds to a rough
estimate of the $k$ range in which the one-loop iPT is inaccurate at
the level of a few percent. In this figure, they are given by
$k \gtrsim 0.45/\sigma_\mathrm{d}$, where
$\sigma_\mathrm{d} = \langle |\bm{\varPsi}_\mathrm{Zel}|^2
\rangle^{1/2}$ is the rms of the displacement field evaluated with the
Zel'dovich approximation. Our estimate is fairly reasonable when
comparison between the iPT and numerical simulations is available
\cite{SM11,SM13,CRW13}.

There are deviations from the predictions of linear theory even in the
large-scale limit, $k < 0.01\,h\,\mathrm{Mpc}^{-1}$, owing mainly to a
white-noise-like contribution generated by second-order Lagrangian
bias \cite{Mat08b}; the contribution of the first term on the rhs of
Eq.~(\ref{eq:2-5}) to the biased power spectrum of Eq.~(\ref{eq:2-1})
is given by
\begin{equation}
  P_X(\bm{k}) \supset
  \frac{1}{2} \int_{\bm{k}_{12}=\bm{k}}
  \left[c^{(2)}_X(\bm{k}_1,\bm{k}_2)\right]^2
  P_{\rm L}(k_1) P_{\rm L}(k_2).
  \label{eq:4-1}
\end{equation}
The second-order bias function $c^{(2)}_X(\bm{k}_1,\bm{k}_2)$ does not
generally approach zero in the large-scale limit of
$\bm{k}=\bm{k}_1+\bm{k}_2\rightarrow \bm{0}$, and therefore the above
term approaches a positive constant in the same limit. As a result,
the nonlinear power spectra of biased tracers in the large-scale limit
are always larger than the predictions of linear theory. At redshift
$z=1$, the second-order function is coincidentally close to zero in
the large-scale limit, so that this white-noise-like term is small.

In each of our bias models, the power spectra are systematically
larger than the predictions of linear theory toward small scales.
Consequently, the nonlinear scale-dependent bias
$[P_X(k)/P_\mathrm{m}(k)]^{1/2}$ increases at small scales. This
property is not solely due to the scale dependencies of the
first-order bias function, $c^{(1)}_X(k)$ since the local halo model,
in which $c_X^{(1)}$ does not have any scale dependence, exhibits the
same behavior. Therefore, the second-order effects are important to
account for the scale-dependent enhancements of the power spectrum in
the presence of bias.

The qualitative behavior of the power spectrum does not vary
significantly among the different biasing schemes. Except for the
simplistic local halo, the differences between the models are at the
level of 2\%--4\% at $k\lesssim 0.2\,h\,\mathrm{Mpc}^{-1}$. Although the
renormalized bias functions behave fairly differently among different
biasing schemes, these deviations do not have a pronounced impact on
the shape of the power spectrum. The reason is that the biasing
schemes start deviating significantly from each other on scales
smaller than the halo mass $M=1\times 10^{13} \,h^{-1}M_\odot$, which
corresponds to $R\simeq 3\,h^{-1}\mathrm{Mpc}$ or
$k \sim 1\,h\,\mathrm{Mpc}^{-1}$, on which perturbation theory cannot
be applied. It is the asymptotic value of the renormalized bias
functions $c^{(n)}_X$ in the large-scale limit $k\rightarrow 0$ which
determines the overall shape of the nonlinear power spectrum. Clearly,
however, these subtle differences will be important to determine the
shape of $P_X(k)$ at the percent level.

\begin{figure*}
\begin{center}
\includegraphics[height=16.7pc]{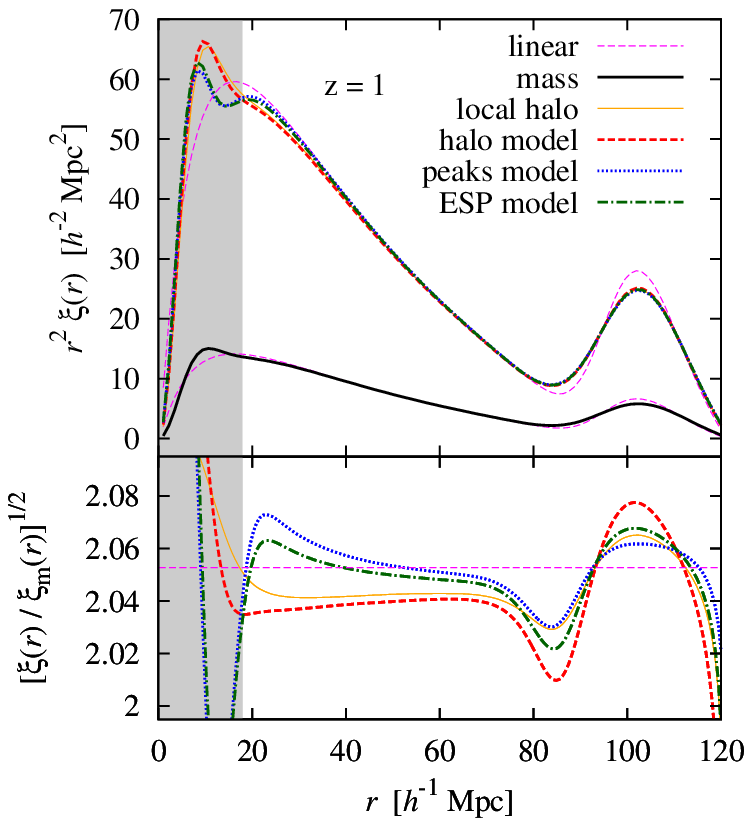}
\includegraphics[height=16.7pc]{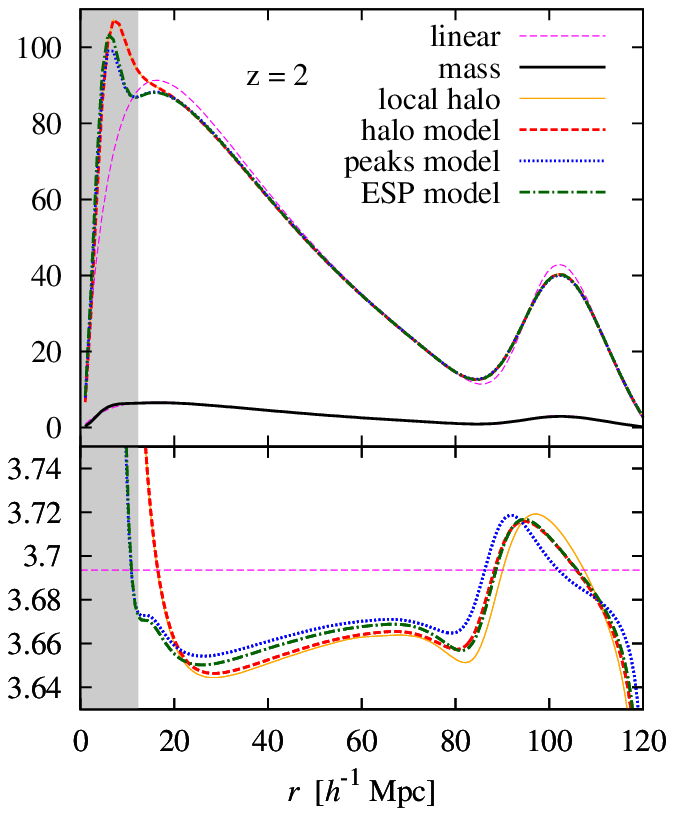}
\includegraphics[height=16.7pc]{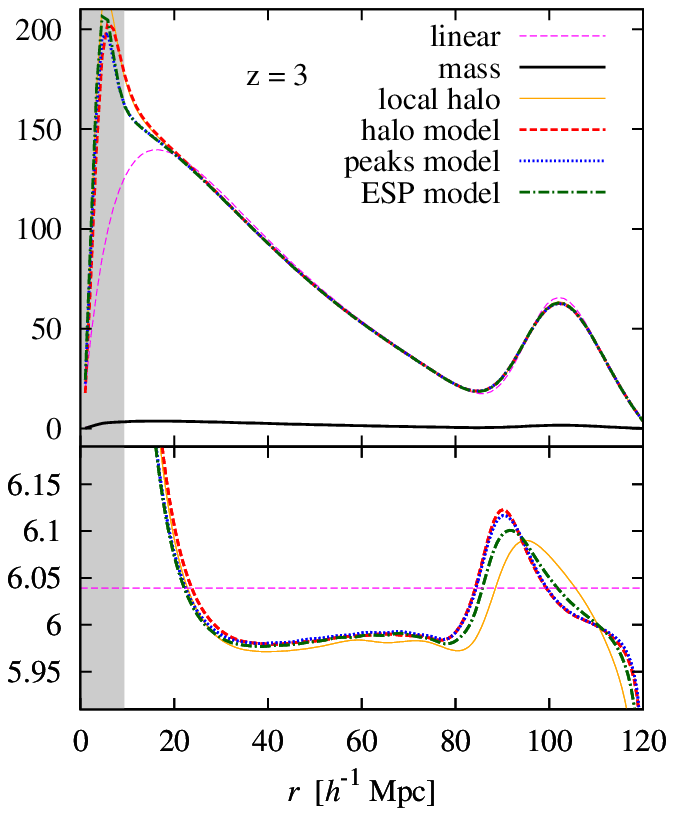}
\caption{\label{fig:xiplot2} The one-loop correlation functions in
  real space with different biasing schemes. Upper panels show the
  correlation functions multiplied by the square of distance,
  $r^2\xi_X(r)$. The lower panels show the scale dependent bias,
  $[\xi_X(r)/\xi_\mathrm{m}(r)]^{1/2}$. Meanings of different lines
  are indicated in the panels. Shaded regions represent rough
  estimates where the one-loop perturbation theory is expected to be
  inaccurate. }
\end{center}
\end{figure*}

The one-loop correlation function in real space, $\xi(r)$, is plotted
in Fig.~\ref{fig:xiplot2}. In the upper panels, the correlation has
been multiplied by the square of the separation $r^2$ to highlight the
shape of the baryon acoustic oscillation (BAO), as is common practice
in the literature. In the lower panels, the nonlinear scale-dependent
bias $[\xi_X(r)/\xi_\mathrm{m}(r)]^{1/2}$ is shown as a function of
distance. Shaded regions correspond to the region $r \lesssim
5\sigma_\mathrm{d}$ where the one-loop iPT is expected to fail at a
level of a few percent at least.

The upper panels indicate that the shape of the BAO peak is not
significantly affected by the choice of biasing scheme. The
differences on scales $r\gtrsim 20\,h^{-1}\mathrm{Mpc}$ are as small
as 1\% at $z=1$ and the subpercent level at $z=2$ and $3$, except for
the simplistic local halo. As seen in the lower panels with $z=1$, the
BAO peaks of biased tracers are slightly sharper than that of mass by
a few percent. However, shapes of the peak for $z=2$ and $3$ are
slightly distorted by a few percent in nontrivial ways. At redshift
$z=2$ and $3$, the scale-dependent bias on scales
$30$--$80\,h^{-1}\mathrm{Mpc}$ is slightly lower than the predictions
of linear theory by about 1\%.

\subsection{\label{subsec:ResultRSD} Power spectra and correlation
  functions in redshift space }

\begin{figure*}
\begin{center}
\includegraphics[height=16.8pc]{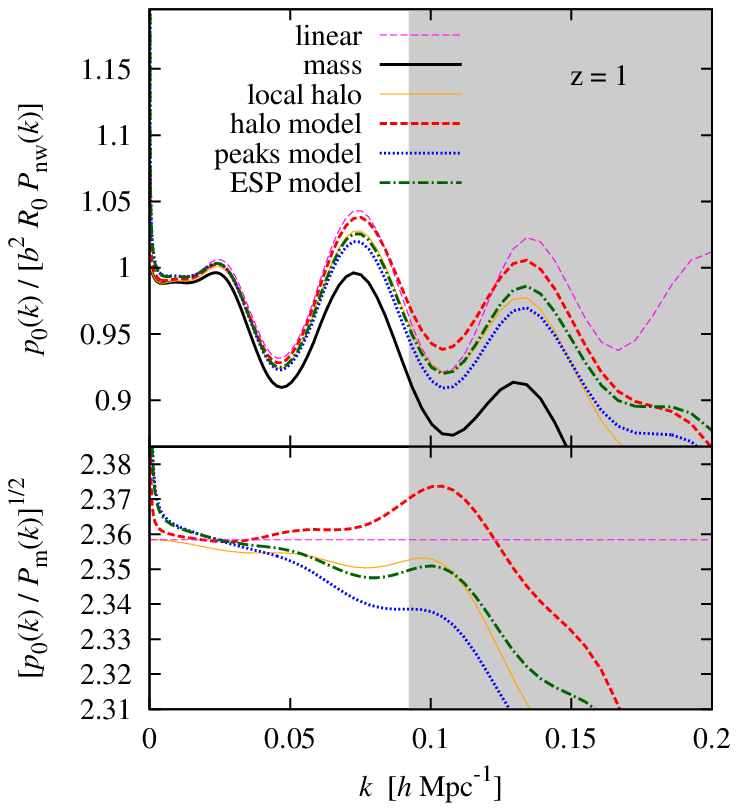}
\includegraphics[height=16.8pc]{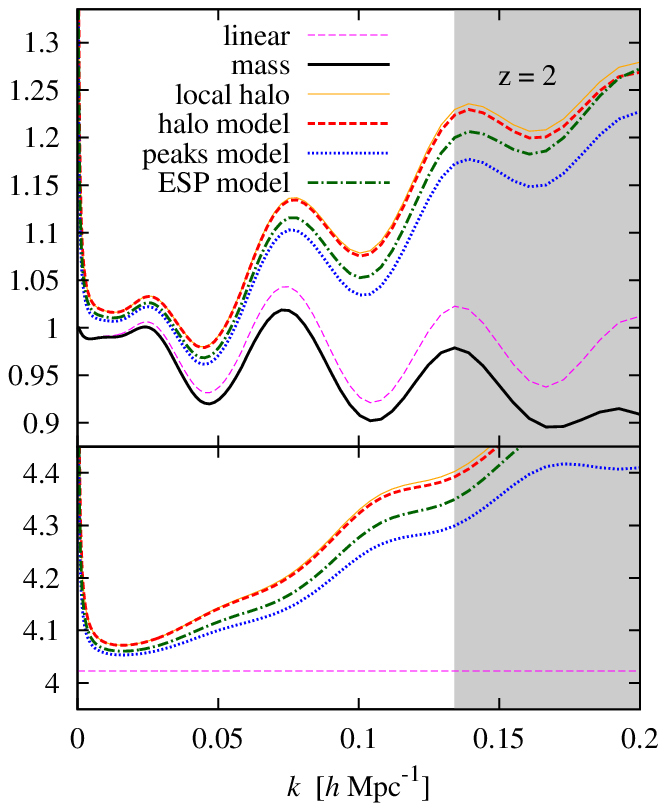}
\includegraphics[height=16.8pc]{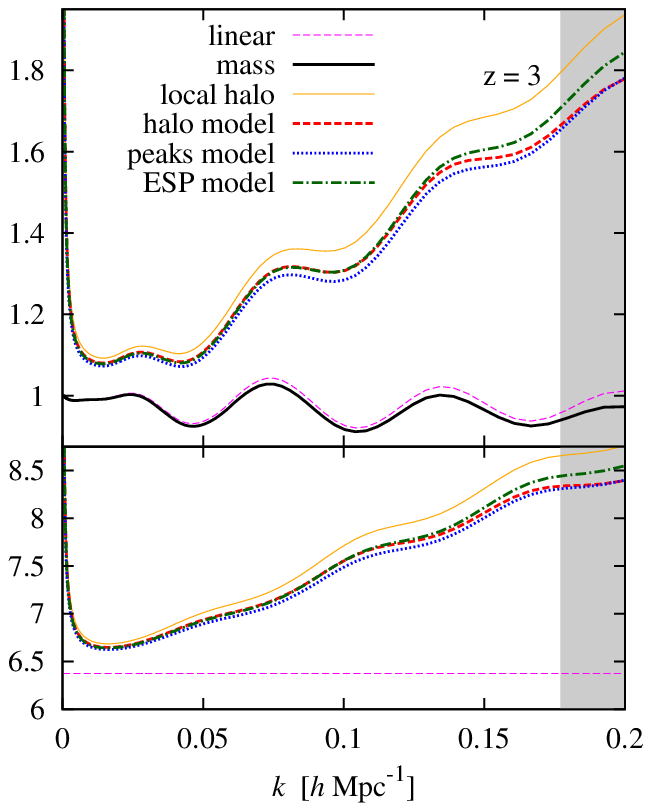}
\caption{\label{fig:psplot2.red0} The monopole components of one-loop
  power spectra in redshift space with different biasing schemes.
  Upper panels show the power spectra divided by the no-wiggle power
  spectrum, the linear bias parameter and the linear redshift-space
  distortion factor, $p_0(k)/[b^2 R_0P_\mathrm{nw}(k)]$. The lower
  panels show the scale-dependent bias in redshift space,
  $[p_0(k)/P_\mathrm{m}(k)]^{1/2}$. Meanings of different lines are
  indicated in the panels. Shaded regions represent rough estimates
  where the one-loop perturbation theory is expected to be inaccurate.
}
\end{center}
\end{figure*}

The monopole components of the one-loop power spectra in redshift space
are plotted in Fig.~\ref{fig:psplot2.red0}. In the upper panels, the results are
normalized by the no-wiggle power spectrum with a linear enhancement 
factor $b^2 R_0$, where $R_0 = 1 + 2\beta/3 + \beta^2/5$ is the
redshift-space enhancement factor of the monopole component in linear
theory \cite{Kai87}. Again, the shaded regions correspond to $k \gtrsim
0.33/\sigma_\mathrm{d}$, for which the one-loop iPT is not expected
to apply at the level of a few percent.

Comparing the behaviors of monopole components in redshift space with
those of Fig.~\ref{fig:psplot2} in real space shows that the nonlinear
enhancements at smaller scales are less pronounced in redshift space.
Overall, however, the impact of nonlinearities is similar to that in
real space. The differences among different biasing schemes are about
2\%--4\% at $k\lesssim 0.2\,h\,\mathrm{Mpc}^{-1}$ except for the
simplistic local halo, i.e. at the same level as in real space.

\begin{figure*}
\begin{center}
\includegraphics[height=16.6pc]{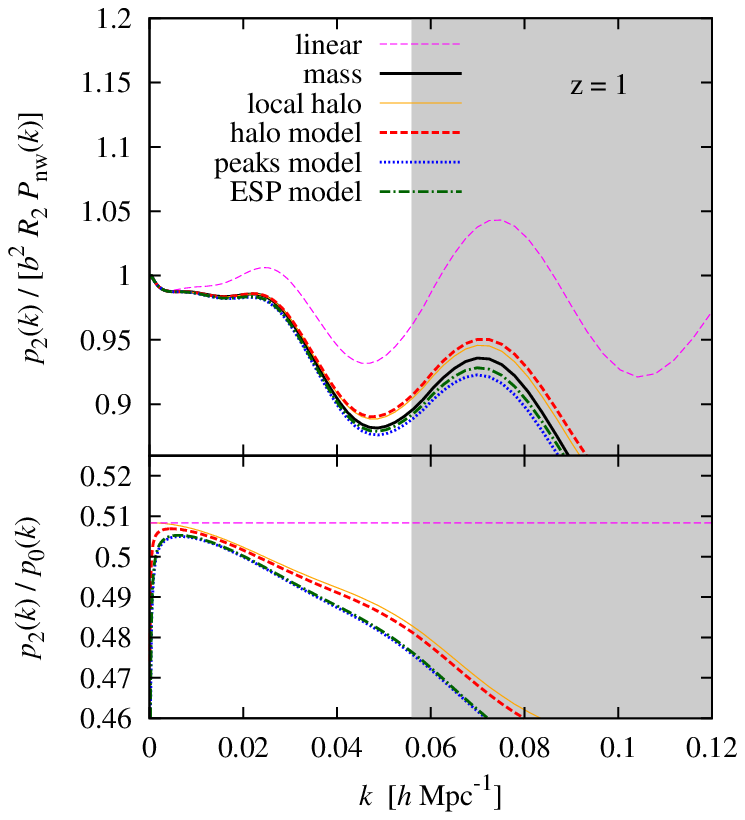}
\includegraphics[height=16.6pc]{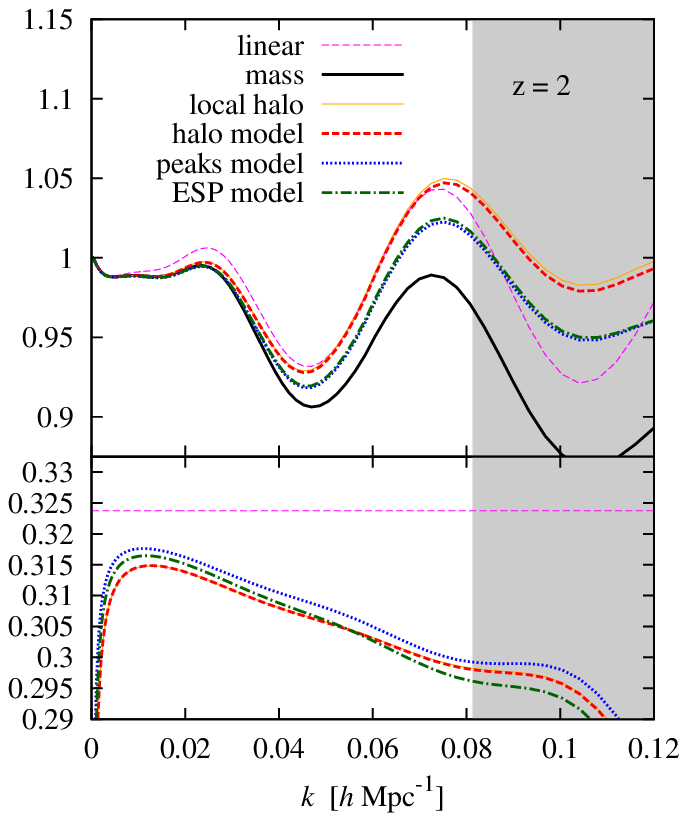}
\includegraphics[height=16.6pc]{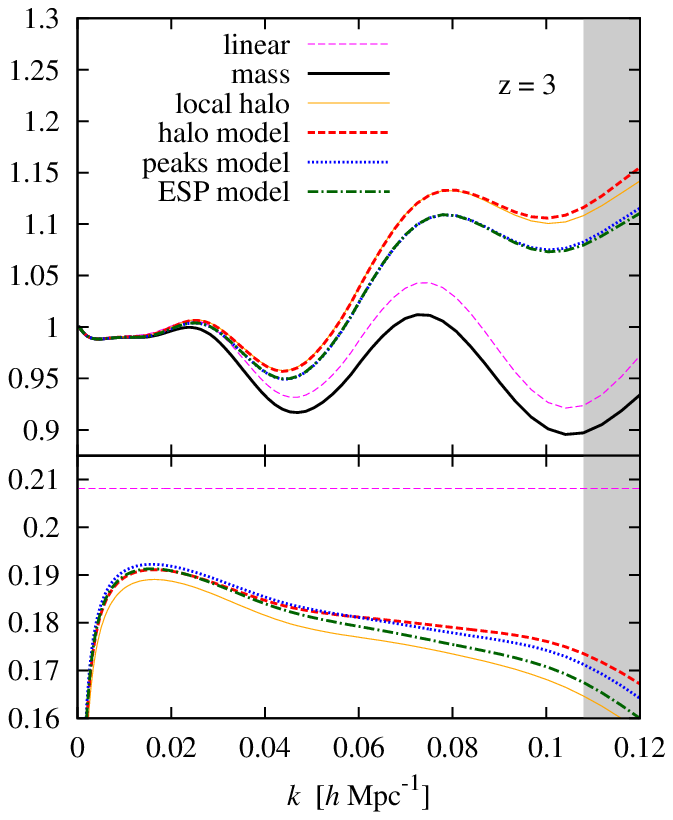}
\caption{\label{fig:psplot2.red2} The quadrupole components of
  one-loop power spectra in redshift space with different biasing
  schemes. Upper panels show the power spectra divided by the
  no-wiggle power spectrum, the linear bias parameter and the linear
  redshift-space distortion factor, $p_0(k)/[b^2
  R_2P_\mathrm{nw}(k)]$. The lower panels show the ratio between the
  quadrupole components and the monopole components. }
\end{center}
\end{figure*}

\begin{figure*}
\begin{center}
\includegraphics[height=16.1pc]{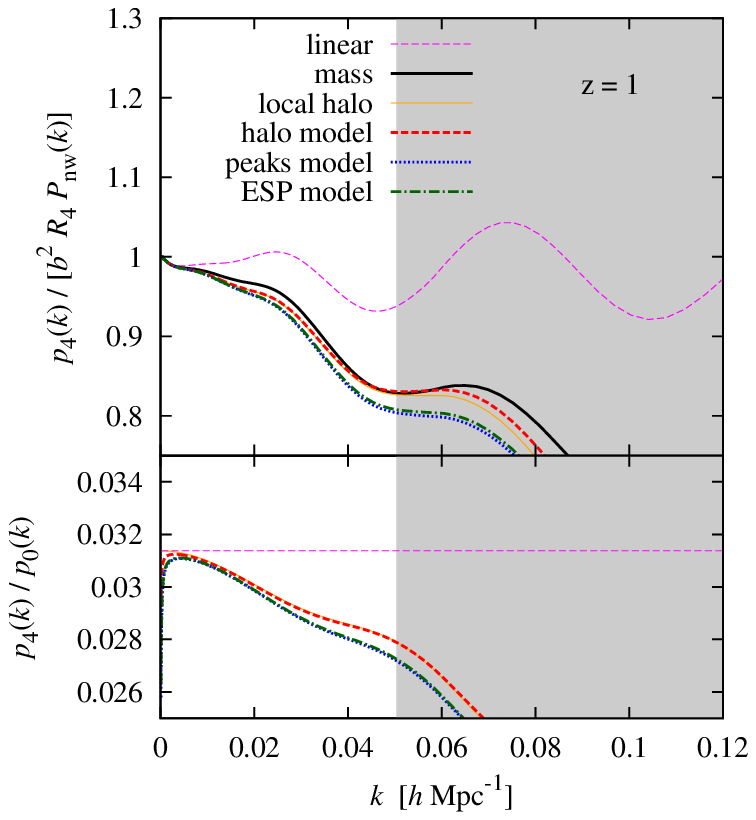}
\includegraphics[height=16.1pc]{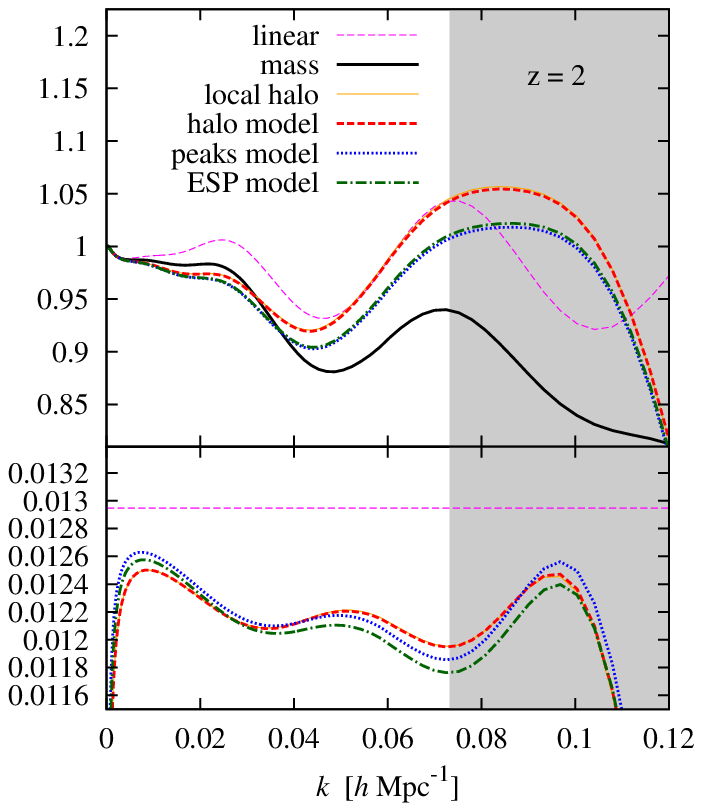}
\includegraphics[height=16.1pc]{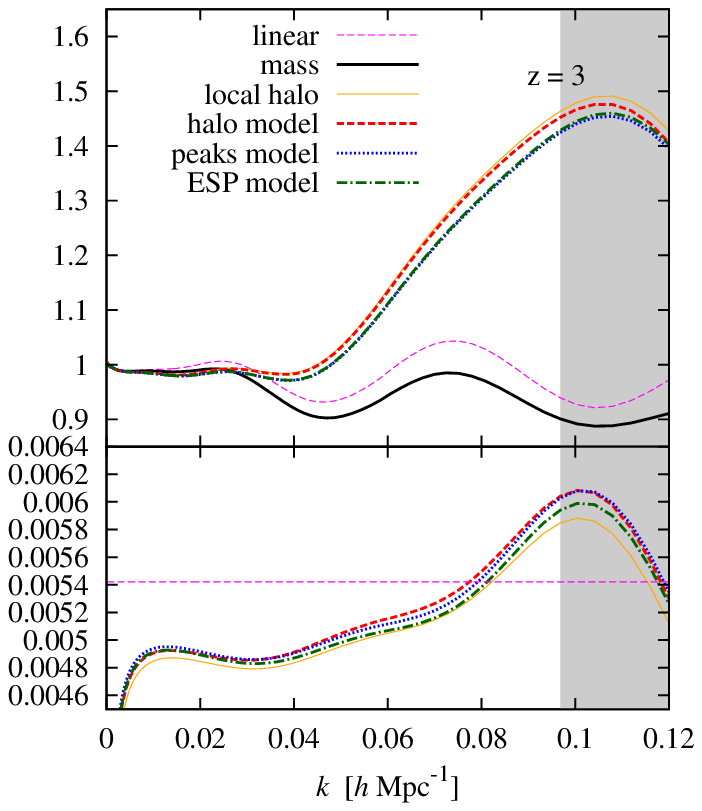}
\caption{\label{fig:psplot2.red4} Same as Fig~\ref{fig:psplot2.red2},
  but for the hexadecapole components. }
\end{center}
\end{figure*}

The quadrupole and hexadecapole components of the one-loop power
spectra in redshift space are shown in Figs.~\ref{fig:psplot2.red2}
and \ref{fig:psplot2.red4}. In the upper panels, the additional
normalization factor induced by the linear redshift-space distortions
are $R_2 = 4\beta/3 + 4\beta^2/7$ and $R_4 = 8\beta^2/35$. In the
lower panels, ratios of the quadrupole and hexadecapole to the
monopole component are shown. These ratios are commonly used for
constraining the nature of gravity through a measurement of the
redshift-space distortion parameter $\beta$ (e.g.,
Refs.~\cite{Guz08,Oku15}). Estimates of the applicability of iPT for
the quadrupole and hexadecapole components are relatively uncertain,
because a detailed comparison between the iPT and numerical
simulations is not available in the literature. Therefore, we have
tentatively defined the confidence region as
$k < 0.2/\sigma_\mathrm{d}$ for the quadrupole, and
$k < 0.18/\sigma_\mathrm{d}$ for the hexadecapole. Although the
multipole components appear to behave strangely at smaller scales, we
warn the reader that our criteria may be inaccurate.

The variances among different biasing schemes are at most at the level
of a few percent, as is the case of the monopole component. The
multipole-to-monopole ratios show relatively large deviations from the
predictions of linear theory, $R_l/R_0$. The nonlinear ratios are
smaller than the linear predictions by 5\%--15\% even on a scale as
large as $k\simeq 0.06\,h\,\mathrm{Mpc}^{-1}$ usually considered to
belong to the linear regime. When the bias factor is large, which is
the case at redshift $z=2$ and $3$, the ratios never attain the linear
values at any scale. Since the ratios of linear theory, $R_2/R_0$ and
$R_4/R_0$ are increasing functions of $\beta$, a blind application of
linear theory to the power spectrum in redshift space would result in
an underestimation of the $\beta$ parameter if the bias factor were
fixed (in actual analyses, however, the bias parameter is
simultaneously fitted to the data). Notwithstanding, the deviations
from the linear ratios are much larger than the variances among
biasing schemes. The iPT provides a way to quantify the systematic
effects produced by the weakly nonlinear evolution fairly
independently of the biasing schemes.

\begin{figure*}
\begin{center}
\includegraphics[height=16.6pc]{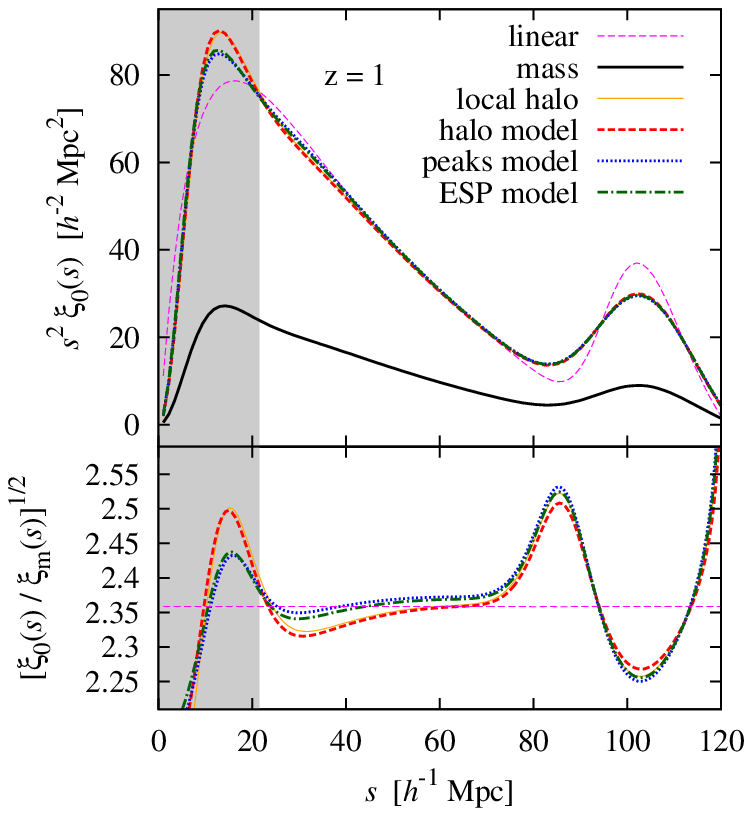}
\includegraphics[height=16.6pc]{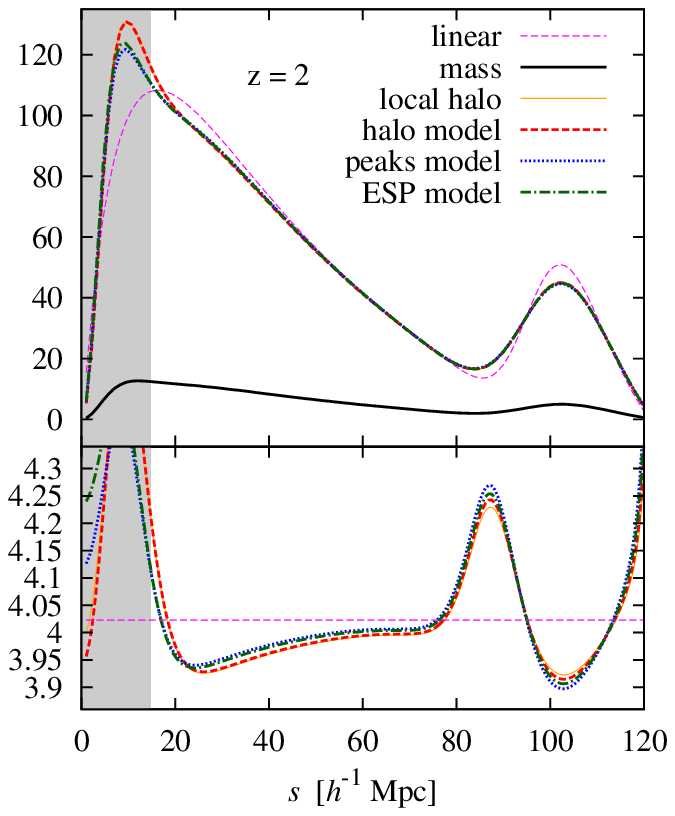}
\includegraphics[height=16.6pc]{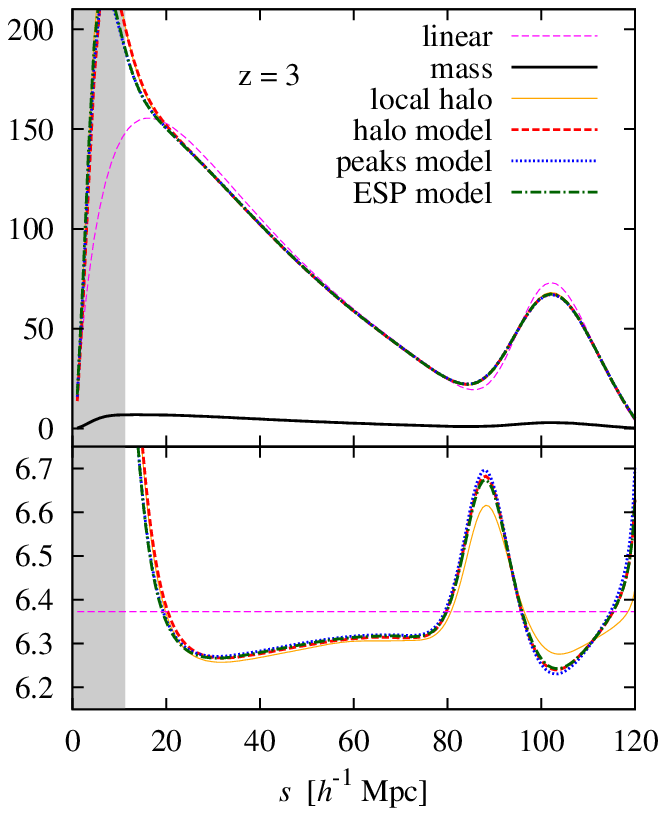}
\caption{\label{fig:xiplot2.red0} The monopole components of one-loop
  correlation function in redshift space with different biasing
  schemes. Upper panels show the monopole functions multiplied by the
  square of distance, $s^2 \xi_0(s)$. The lower panels show the
  scale-dependent bias in redshift space,
  $[\xi_0(s)/\xi_\mathrm{m}(s)]^{1/2}$. }
\end{center}
\end{figure*}

\begin{figure*}
\begin{center}
\includegraphics[height=17.2pc]{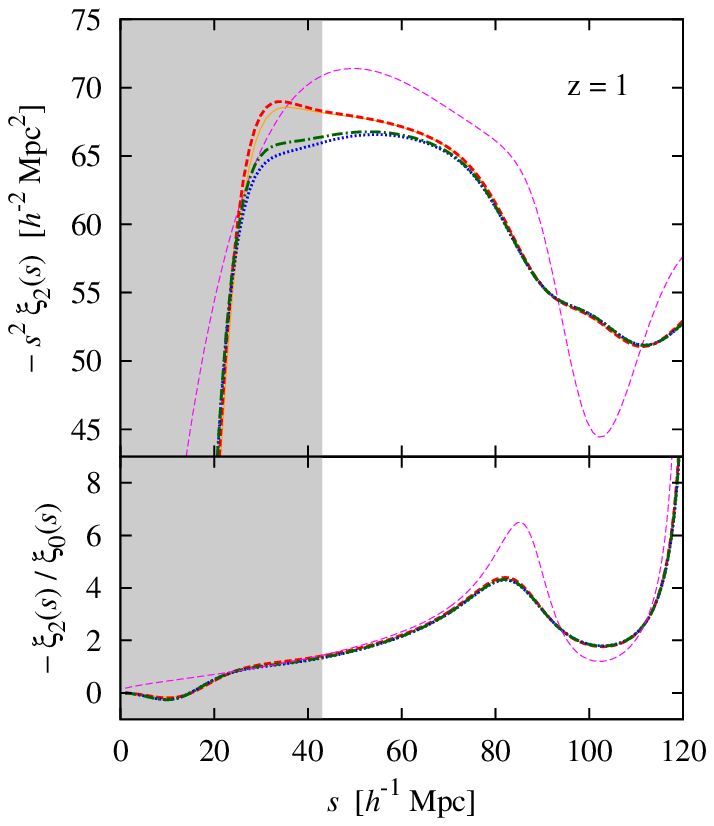}
\includegraphics[height=17.2pc]{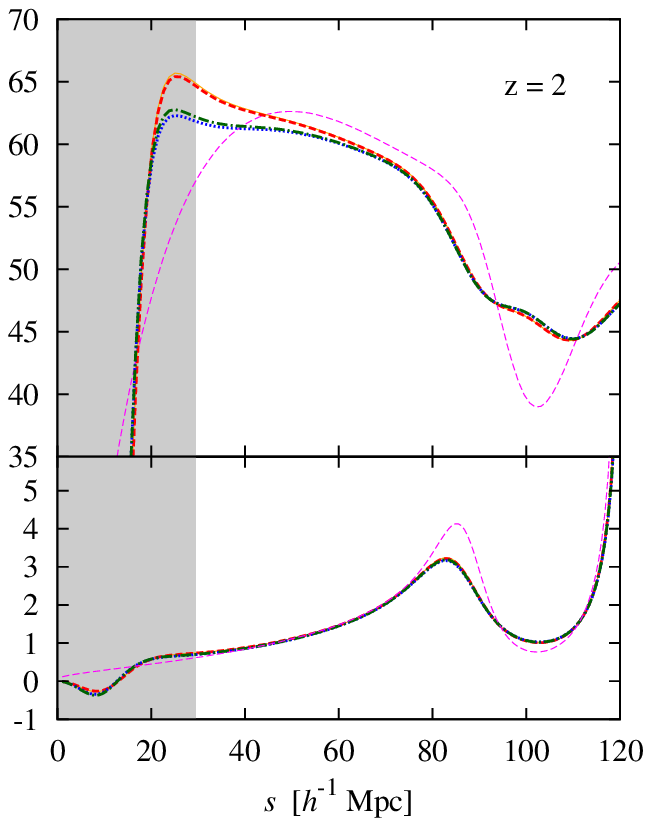}
\includegraphics[height=17.2pc]{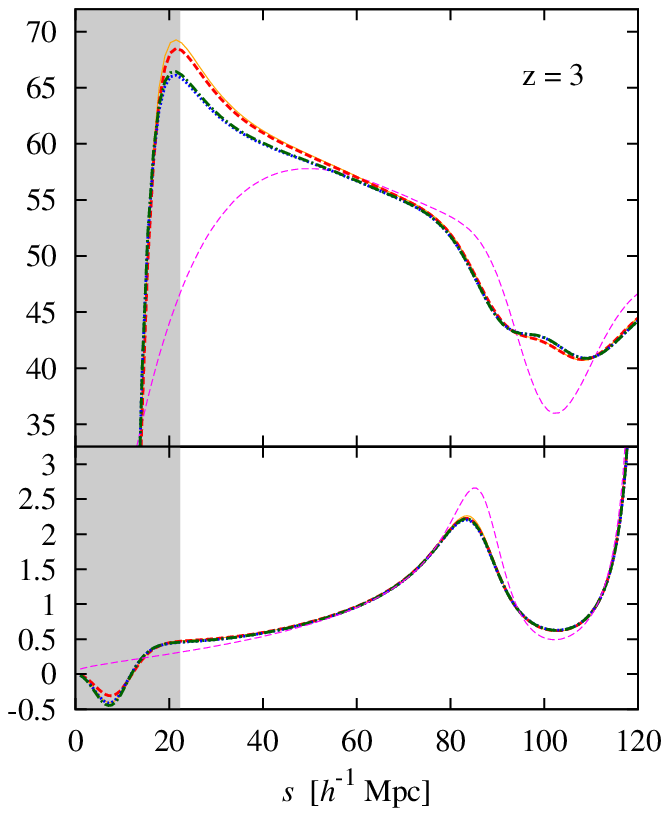}
\caption{\label{fig:xiplot2.red2} The quadrupole components of
  one-loop correlation functions in redshift space with different
  biasing schemes. Upper panels show the quadrupole functions
  multiplied by the minus square of distance , $-s^2\xi_2(r)$. The
  lower panels show the ratio between the quadrupole components and
  the monopole components.}
\end{center}
\end{figure*}

\begin{figure*}
\begin{center}
\includegraphics[height=16.6pc]{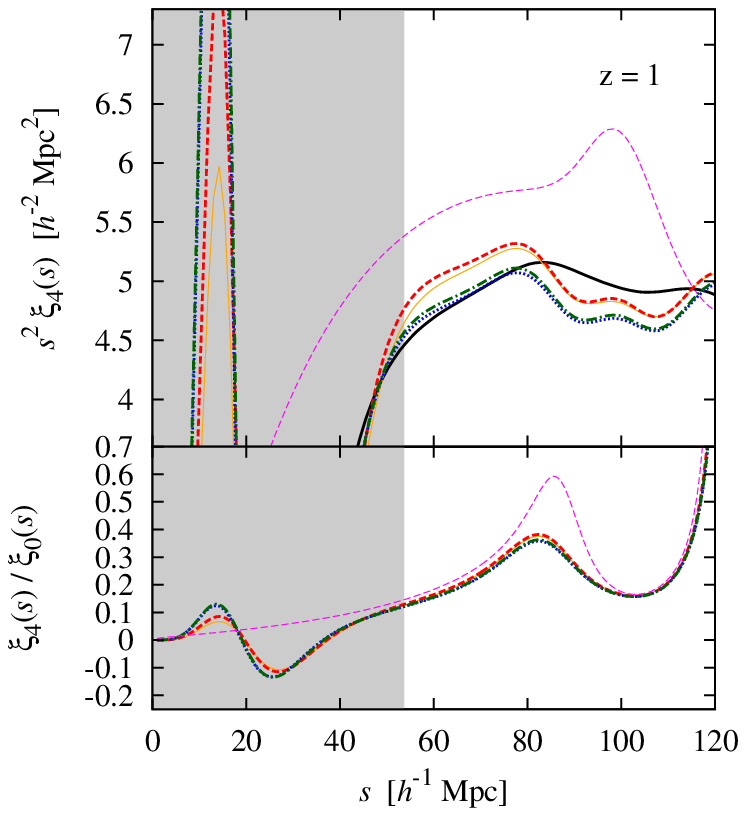}
\includegraphics[height=16.6pc]{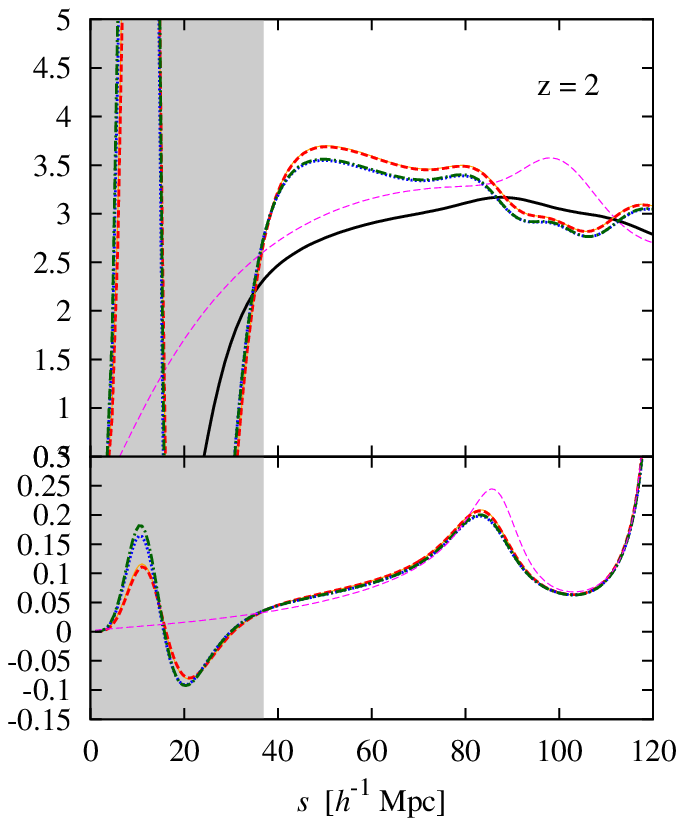}
\includegraphics[height=16.6pc]{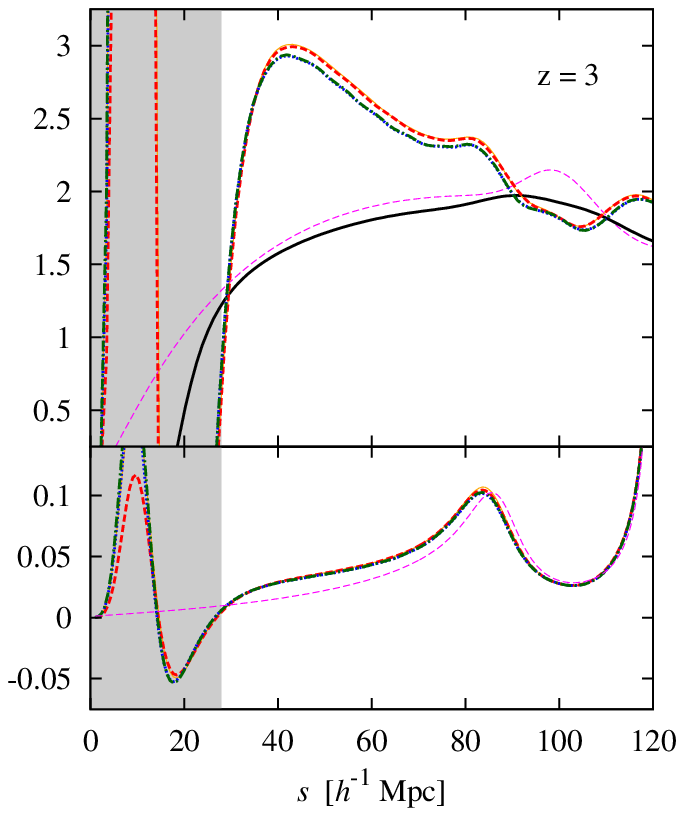}
\caption{\label{fig:xiplot2.red4} Same as Fig~\ref{fig:xiplot2.red2},
  but for the hexadecapole components.}
\end{center}
\end{figure*}

In Figs.~\ref{fig:xiplot2.red0}--\ref{fig:xiplot2.red4}, the monopole,
quadrupole and hexadecapole of the halo correlation functions in
redshift space are plotted. Our estimates for the applicability of our
one-loop iPT prediction are $r < 6\sigma_\mathrm{d}$,
$r < 12 \sigma_\mathrm{d}$, and $r < 15\sigma_\mathrm{d}$ for the
monopole, quadrupole, and hexadecapole components, respectively. While
these bounds are estimated by extrapolating the comparisons of
Ref.~\cite{CRW13}, they could be inaccurate, especially in the case of
hexadecapole.

The variances of different biasing schemes are within a few percent as
in the case of the previous figures. The BAO peaks of the monopole
components in redshift space are smoother than those in real space.
Accordingly, the scale-dependent bias
$[\xi_0(r)/\xi_\mathrm{m}(r)]^{1/2}$ varies more than those in real
space. This effect of BAO smoothing does not significantly depend on
the biasing schemes (except for the simplistic local halo, as usual).

Differences between the quadrupole and hexadecapole predicted by the
halo and peaks/ESP models can be seen in the upper panels of
Figs.~\ref{fig:xiplot2.red2} and \ref{fig:xiplot2.red4}. However, they
have a similar degree of deviations as that seen in the monopole
components in Fig.~\ref{fig:xiplot2.red0}, where it is less apparent
because the scales of vertical axes are much larger. The lower panels
show that deviations in the quadrupole-to-monopole and
hexadecapole-to-monopole ratios among the different biasing schemes
are extremely small in the correlation functions in redshift space.

\subsection{\label{subsec:ResultNG}
Scale-dependent bias in the presence of non-Gaussianity
}

If some amount of inflationary non-Gaussianity is imprinted in the
initial cosmological perturbations, then the bispectrum of the linear
density field, receives a nontrivial primordial contribution,
$B_\mathrm{L}(k_1,k_2,k_3)$. When this primordial bispectrum is
strongly scale dependent, as in, for instance, the case for local-type
non-Gaussianity, Fourier modes of the density fluctuations with long
and short wavelengths, i.e., with wave numbers $k_l\ll k_s$, are
coupled to each other. As a result, the power spectrum of biased
tracers is affected on very large scales as it depends on the biasing
processes which are small-scale phenomena \cite{Dal08,MV08,Slo08}. In
the iPT formalism, the contributions are given by the last term in
Eq.~(\ref{eq:2-1}), general implications of which are discussed in
Ref.~\cite{Mat12}.

The primordial non-Gaussianity also changes the precise shapes of the
renormalized bias functions through the multivariate distribution
function ${\cal P}(\bm{y})$; see Eqs.~(\ref{eq:3-12}) and
(\ref{eq:3-13}). However, this effect is small enough because the
shapes of the renormalized bias functions are dominantly determined by
Gaussian components \cite{Mat12}. For instance, the non-Gaussian
corrections to $c_X^{(2)}$ are at the level of $10^{-5}f_\mathrm{NL}$.
By contrast, the scale-dependent bias on very large scales
predominantly arises from the primordial non-Gaussianity.
Hence, we will neglect the subdominant corrections to the renormalized 
bias functions due to primordial non-Gaussianity for simplicity.

\begin{figure*}
\begin{center}
\includegraphics[height=18pc]{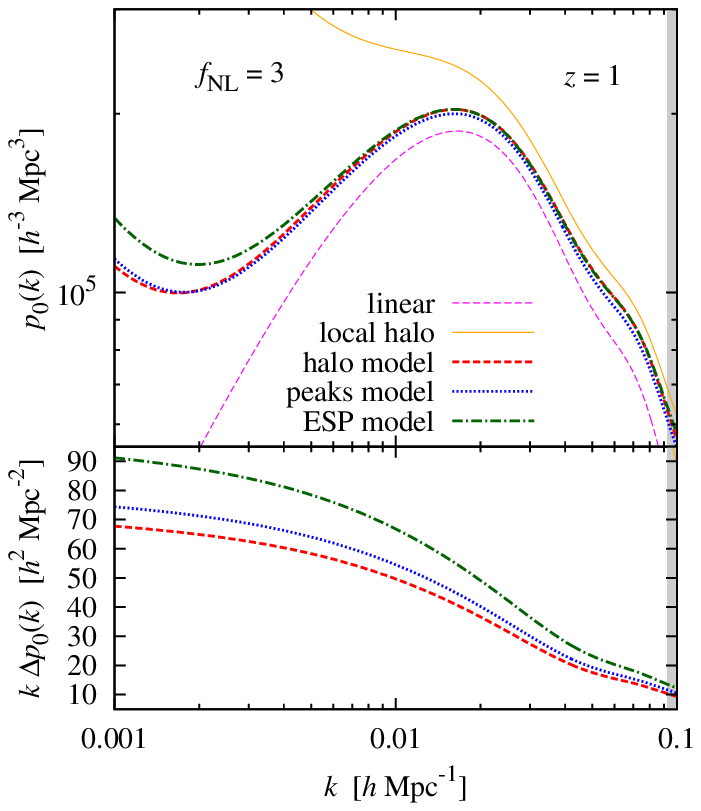}
\hspace{1pc}
\includegraphics[height=18pc]{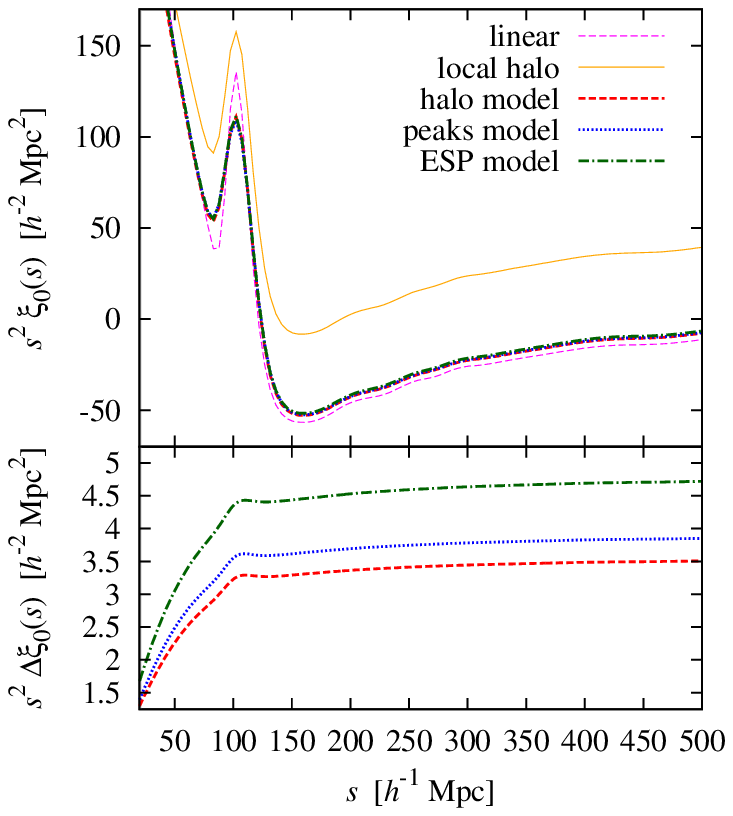}
\caption{\label{fig:nG} Effects of primordial non-Gaussianity on the
  power spectra (left) and correlation functions (right) of monopole
  components in redshift space at the redshift of $z=1$. Local-type
  non-Gaussianity with $f_\mathrm{NL} = 3$ is assumed. Upper panels
  show the monopole components of power spectra and correlation
  functions. Lower panels show the pure contributions from the
  primordial non-Gaussianity. }
\end{center}
\end{figure*}

In Fig.~\ref{fig:nG}, one-loop power spectra and correlation functions 
are shown for the different biasing schemes. We focus on the monopole
component in redshift space, as it is a quantity observed in actual
redshift surveys. The primordial non-Gaussianity is assumed to be of
local, quadratic type, so that the primordial bispectrum takes the form
\begin{equation}
  B_\mathrm{L}(k_1,k_2,k_3) = 2 f_\mathrm{NL}
  \left[
    \frac{{\cal M}(k_3)}{{\cal M}(k_1){\cal M}(k_2)}
    P_\mathrm{L}(k_1) P_\mathrm{L}(k_2)
    + \mathrm{cyc.}
  \right],
  \label{eq:4-2}
\end{equation}
where
\begin{equation}
  {\cal M}(k) = \frac{2}{3} D_+(z)
  \frac{k^2T(k)}{{H_0}^2\varOmega_\mathrm{m0}}
  \label{eq:4-3}
\end{equation}
is the transfer function between the potential deeply in matter
domination and the linear density. Here, $D_+$ is a linear growth
factor, normalized as $D_+ \rightarrow a$ in the matter-dominated
epoch, and $T(k)$ is the linear transfer function, normalized to
$T(k)\to 1$ in the limit $k\to 0$. The parameter $f_\mathrm{NL}$ is
observationally constrained to be $f_\mathrm{NL} = 0.8 \pm 5.0$ (68\%
C.L.) \cite{Planck2015nG}. For illustration purposes, we assume
$f_\mathrm{NL} = 3$ consistent with the observational bound.

In the large-scale limit, $k\rightarrow 0$, the contribution of
local-type primordial non-Gaussianity to the monopole power spectrum
is given by \cite{Mat12}
\begin{multline}
  \varDelta p_X^0(k) \approx 4 f_\mathrm{NL}
  \left(1 + c_X^{(1)}(k) + \frac{f}{3}\right)
  \frac{P_\mathrm{L}(k)}{{\cal M}(k)}
\\
  \times
  \int \frac{d^3p}{(2\pi)^3} c_X^{(2)}(\bm{p},-\bm{p}) P_\mathrm{L}(p),
  \label{eq:4-4}
\end{multline}
where $\varDelta p_X^0(k) = p_X^0(k) - p_X^\mathrm{0G}(k)$ and
$p_X^\mathrm{0G}(k)$ is the Gaussian contribution with
$f_\mathrm{NL}=0$. The simplistic local halo
[$c_X^{(2)}(\bm{p},-\bm{p}) = \mathrm{const.}$] gives a
logarithmically divergent result for the above equation if
$n_\mathrm{s} = 1$ because
$P_\mathrm{L}(k) \propto k^{n_\mathrm{s}-4}$ in the limit of
$k \rightarrow \infty$ for $\Lambda$CDM models. Since the spectral
index $n_\mathrm{s} = 0.9667$ is slightly less than unity, the above
integral in the simplistic local halo converges, although it is much
larger than other schemes in which the renormalized bias functions are
suppressed by window functions in the small-scale limit. Thus, the
effects of primordial non-Gaussianity on very large scales depend not
only on the asymptotic values of $c_X^{(n)}$ but also on their shape
at small scales. However, while the amplitude of $\varDelta p_X^0$
strongly depends on the biasing schemes, the power-law scaling of the
scale dependence in the large-scale limit,
$\varDelta p_X^0 \propto P_\mathrm{L}(k)/{\cal M}(k) \propto
k^{n_\mathrm{s}-2}$, does not depend on biasing schemes. Note that the
constant term Eq.~(\ref{eq:4-1}) also contributes to the power spectra
in the large-scale limit, $k\rightarrow 0$, in addition to the
non-Gaussian contribution with which it is partly degenerate.

The non-Gaussian bias amplitude Eq.~(\ref{eq:4-4}) is consistent with
the peak-background expectation $\partial\ln n/\partial\ln\sigma_8$
obtained by Ref.~\cite{Slo08} for the peaks and ESP implementations
considered here (Ref.~\cite{DGR13}; see, however, the discussion of
Ref.~\cite{BD15} for moving stochastic barriers). However,
substituting Eq.~(\ref{eq:3-2-8}) into Eq.~(\ref{eq:4-4}) shows that
this is generally not the case of the local and halo models, unless
the multiplicity is of the Press-Schechter form.

In the lower panel of the left figure, contributions from the
primordial non-Gaussianity $\varDelta p_X^0(k)$ are shown. Variations
among the biasing schemes can be seen. They are not significant,
except for the local halo. Still, if a nonlinear parameter
$f_\mathrm{NL}\ne 0$ were detected, the different biasing schemes
would change its estimated value by $\sim 25\%$.

In the right figure, the monopole components of the correlation
function in redshift space are shown. The primordial non-Gaussianity
slightly increases the correlation functions on large scales
$s\gtrsim 100\,h^{-1}\mathrm{Mpc}$ in a scale-dependent way,
approximately $\varDelta\xi_X^0(s) \propto s^{-2}$. The simplistic
local halo even boosts the amplitude on the BAO scales by about
$100\%$, which is much larger than what is measured in $N$-body
simulations (see, e.g., Ref.~\cite{DSI09}). The variance among other
biasing schemes in $\varDelta\xi_X^0$ is about $25\%$, in accordance
with the result of $\varDelta p_X^0$.

\section{\label{sec:Conclusions}
Conclusions
}

Using the iPT formalism, we have studied the impact of biasing schemes
on the power spectra and correlation functions of biased tracers in
the weakly nonlinear regime. In this paper, we have focused on three
representative bias schemes: the halo, peaks, and ESP models. We have
also considered a simplified version of the halo model in which the
renormalized bias functions are assumed to be scale independent. This
has allowed us to quantify the impact of the scale dependence of the
bias functions on the power spectra and correlation functions.

In the iPT, all the degrees of freedom of different biasing schemes
are contained in a series of renormalized bias functions. The biasing
schemes we considered in this paper are semilocal models, in which
the number density of biased tracers at a Lagrangian position is
determined by the smoothed linear density field and its spatial
derivatives at the same Lagrangian position. After deriving a compact
formula to evaluate the renormalized bias functions in semilocal
models of bias, these functions in individual biasing schemes are
derived up to second order. Our results agree with previous works, and
show that the coefficients of the perturbative peaks and ESP bias
expansions are associated with the iPT renormalized bias functions. In
order to efficiently evaluate the renormalized bias functions, we have
provided analytic reductions of various integrals in coefficients of
the bias functions, so that all the coefficients are given by
one-dimensional integrals with sufficiently smooth functions of
integrands.

We have compared the renormalized bias functions of different biasing
schemes. The $c_X^{(n)}$ of all the models (except for the simplistic
local halo, which is not physically motivated) converge toward zero in
the high-$k$ limit because of the window functions. While the low-$k$
limit of the first-order function, $c_X^{(1)}$, is the same for all
models by construction, differences among biasing schemes can be seen
in the low-$k$ limit of the second-order functions $c_X^{(2)}$. These
differences are, however, not very significant.

By contrast, the behaviors of the renormalized bias functions around
and below the smoothing scales, $kR \gtrsim 1$, vary noticeably among
the bias models. Notwithstanding, they all exhibit a peak around $kR
\sim 2.5$ in lower redshifts. The presence of oscillations in the
Lagrangian bias functions of low redshift halos can actually be seen
in the outcome of $N$-body simulations \cite{BDS15,CSS15}.
The amplitude of the peaks in functions $c_X^{(n)}$ strongly
depends on the biasing schemes, or, how biased tracers are identified
in simulations/observations.

However, we have found that the various schemes, including the
unphysical local halo, do not change the qualitative behavior of the
one-loop power spectra and correlation functions. While, in the power
spectra, differences are at the level of $2\%$--$4\%$, they are as
small as $1\%$ on scales $r\gtrsim 20\,h^{-1}\mathrm{Mpc}$ in the
$z=1$ correlation function, and subpercent at higher redshift. This
partly follows from the fact that the shape of the power spectra is
more affected by nonlinearities than correlation functions.
Furthermore, the simplistic local halo performs comparably well,
confirming that the scale dependence of the renormalized bias
functions is not the decisive factor governing the shape of the power
spectra and correlation functions.

These conclusions also hold in redshift space, with the caveat that
the distortions induced by peculiar velocities are accounted for by
the Kaiser formula. The quadrupole and hexadecapole components exhibit
almost the same level of differences among biasing schemes as the
monopole components. The multipoles-to-monopole ratios in the power
spectra, which are scale independent in linear theory, become scale
dependent due to nonlinear effects. In addition, the ratios are
significantly smaller than the prediction of linear theory by
$5\%$--$15\%$ even at $k\simeq 0.06\,h\,\mathrm{Mpc}^{-1}$. This
illustrates the importance of including nonlinear effects when
estimating the redshift-space distortion parameter $\beta$. Of course,
a realistic calculation should include the virial motions of galaxies
within halos.


We have also estimated the effects of local-type non-Gaussianity in the 
initial conditions for the various biasing schemes. In this case, the 
simplistic local halo biasing scheme, in which small-scale filtering 
is absent, is inappropriate. The primordial non-Gaussianity adds power 
through the mode coupling between large and small scales, such that the 
behavior of renormalized bias functions at small scales can critically 
affect the power spectrum on very large scales. The amplitude of the 
non-Gaussian bias does not differ significantly among the other bias 
schemes, with deviations no larger than $25\%$ both in the power spectra 
and correlation functions.

Before concluding, let us emphasize that, for the peaks and ESP
models, the linear velocities are biased owing to the coupling between
the velocity $\partial^{-1}\delta$ and $\partial\delta$ \cite{VD08a}.
This statistical bias affects the redshift space distortions
\cite{DS10} as well as the two-point correlation around the BAO scales
\cite{Des10}. While it is difficult to measure this effect in
numerical simulations (see, e.g., the discussion in
Ref.~\cite{ZZJ15}), several lines of evidence indicate that it is
present in the Lagrangian space \cite{ELP12,BDS15} and remains
constant throughout time \cite{BDS15}. Although we did not highlight
it explicitly, this effect is already included in the iPT. We plan to
address this important issue in more details in future work.

\begin{acknowledgments}
    T.~M.~acknowledges support from MEXT KAKENHI Grants No.~15K05074
    (2015) and No.~15H05890 (2015). V.~D.~acknowledges support from
    the Swiss National Science Foundation.
\end{acknowledgments}

\appendix


\section{\label{app:XiFunc}
The auxiliary function $\varXi(\delta_M-\delta_\mathrm{c},\sigma_M)$ in the simple halo model
}

In the simple halo model of Sec.~\ref{subsec:halobias}, we have
introduced an auxiliary function
$\varXi(\delta_M-\delta_\mathrm{c},\sigma_M)$. This function is a
phenomenological alternative to the step function $\varTheta$ designed
to produce a mass function more general than the PS one. The mass
function may not be universal. As explained in the main text, we do
not need its actual form in deriving the renormalized bias functions.
However, one may wonder whether this auxiliary function exists for an
arbitrary mass function. In this Appendix, we discuss some details of
the relation between the auxiliary function and the mass function.

The differential mass function $n(M)$ is given by
Eq.~(\ref{eq:3-2-4}). This defines the multiplicity function $f(\nu)$,
which we assume universal in what follows,
\begin{equation}
  n(M) = \frac{\bar{\rho}_0}{M} \frac{f(\nu)}{\nu}
  \frac{d\nu}{dM},
  \label{eq:a-1}
\end{equation}
where $\nu = \delta_\mathrm{c}/\sigma$, and we denote $\sigma = \sigma_M$ for 
simplicity. 
In our simple halo model, the localized differential number density of halos at 
a Lagrangian position $\bm{x}$ is given by Eq.~(\ref{eq:3-2-7}), i.e.,
\begin{equation}
  \label{eq:a-2}
  n(\bm{x},M) =
  -\frac{2\bar{\rho}_0}{M}
  \frac{\partial}{\partial M}
  \varXi\left[\delta(\bm{x})-\delta_\mathrm{c}, \sigma\right],
\end{equation}
where we denote $\delta(\bm{x}) = \delta_M(\bm{x})$ for simplicity.
Both $\delta$ and $\sigma$ depend on the mass $M$ through the
smoothing kernel, and the partial derivative $\partial/\partial M$
applies with fixed $\delta_\mathrm{c}$. The PS mass function
corresponds to the case that the function
$\varXi(\delta-\delta_\mathrm{c},\sigma)$ is given by a step function
$\varTheta(\delta - \delta_\mathrm{c})$. Substituting the step
function by the general function $\varXi$ corresponds to adopting a
fuzzy barrier for the identification of the collapsed regions.
Therefore, it is desirable to have the same asymptotes as the step
function,
\begin{equation}
  \varXi(x,\sigma) \rightarrow
  \begin{cases}
      0 & (x\rightarrow -\infty)\\
      1 & (x\rightarrow +\infty)
  \end{cases},
  \label{eq:a-2-1}
\end{equation}
while the transition between the two limits can be arbitrary.

The above model of a fuzzy barrier is closely related to the model of
square-root stochastic moving barrier \cite{PLS12,PSD13,Bia14}, where
the barrier is replaced by $B = \delta_\mathrm{c} + \beta \sigma$ and
$\beta$ is a stochastic variable with a probability distribution
function $p(\beta)$. With this model, the sharp barrier represented by
the step function $\varTheta(\delta-\delta_\mathrm{c})$ in the PS
formalism is replaced by
\begin{equation}
  \varTheta(\delta-\delta_\mathrm{c}) \rightarrow
  \int d\beta\, p(\beta)\,\varTheta(\delta-\delta_\mathrm{c}-\beta\sigma) =
  \varPhi\left(\frac{\delta-\delta_\mathrm{c}}{\sigma}\right),
  \label{eq:a-2-2}
\end{equation}
where $\varPhi(\beta) = \int_{-\infty}^\beta p(\beta')d\beta'$ is the
cumulative distribution function of $\beta$. Thus, the square-root
stochastic moving barrier corresponds to choosing the function
$\varXi(x,y) = \varPhi(x/y)$.

The mass fraction of the halos with a mass greater than $M$ is given
by
\begin{equation}
  \frac{1}{\bar{\rho}_0}
  \int_M^\infty n(M) M dM = 
  \int_\nu^\infty
  \frac{f(\nu)}{\nu}d\nu
  \equiv F(\nu),
  \label{eq:a-3}
\end{equation}
which corresponds to the filling factor of collapsed regions in
Lagrangian space. Because the ensemble average of Eq.~(\ref{eq:a-2})
should give the global mass function, $n(M) = \langle
n(M,\bm{x})\rangle$, the auxiliary function should satisfy
\begin{equation}
  \langle\varXi(\delta - \nu\sigma, \sigma)\rangle
  = \frac{1}{2} F(\nu).
  \label{eq:a-4}
\end{equation}
or
\begin{equation}
  F(\nu) =
  2 \int_{-\infty}^\infty
  \varXi(\delta-\nu\sigma, \sigma)
  \,P_\sigma(\delta)\, d\delta,
  \label{eq:a-6}
\end{equation}
where $P_\sigma(\delta)$ is the one-point probability distribution
function of $\delta$. This distribution function explicitly depends on
the mass $M$ through $\sigma$. Applying a partial differentiation
$\partial/\partial\delta_\mathrm{c}|_\sigma =
\sigma^{-1}\partial/\partial\nu|_\sigma $ to Eq.~(\ref{eq:a-6}) with
$\sigma$ fixed, and performing integration by parts, we arrive at the
relation
\begin{equation}
  \frac{f(\nu)}{\nu} =
  -2\sigma \int_{-\infty}^\infty
  \varXi(\delta-\nu\sigma, \sigma)
  \frac{\partial P_\sigma(\delta)}{\partial\delta}
  \, d\delta,
  \label{eq:a-7}
\end{equation}

The rhs of Eq.~(\ref{eq:a-6}) is a convolution integral of the
function $\varXi(x,\sigma)$ and $P_\sigma(x)$ for a fixed value of
$\sigma$. Thus, obtaining the auxiliary function $\varXi$ from the
mass function requires the deconvolution, the inverse problem of the
convolution integral. Deconvolution is an ill-posed problem, because
the solution is not unique in general: sometimes the solution does not
exist, and sometimes there are many solutions. Therefore, it is not
guaranteed that the solution of Eq.~(\ref{eq:a-6}) can be found for an
arbitrary function $F(\nu)$ [equivalently, for an arbitrary function
$f(\nu)$].

Nevertheless, numerically fitted mass functions, such as the
Sheth-Tormen (ST) mass function, are derived from a finite range of
$\nu$, i.e., $0.7 \lesssim \nu \lesssim 3.5$ \cite{ST99}. Thus, trying
to invert the convolution integral, Eq.~(\ref{eq:a-6}), from the mass
function extrapolated to all ranges of $0 < \nu < \infty$ is not what
we should do. Instead, it is sufficient to find a reasonable kernel
function $\varXi$ which can reproduce the mass function in finite
ranges of interest where a fitting formula applies. Numerically, the
deconvolution techniques are widely used in signal/image restorations,
e.g., a simple iterative method known as the Richardson-Lucy
deconvolution \cite{Ric72,Lucy74}.

For Gaussian initial conditions, the distribution function is given by
$P_\sigma(\delta) = (2\pi\sigma^2)^{-1/2}e^{-\delta^2/2\sigma^2}$.
Changing the integration variable as $\delta \rightarrow
t=\delta/\sigma$ in this case, the rhs of Eq.~(\ref{eq:a-6}) reduces
to $(2/\pi)^{1/2} \int \varXi(t\sigma-\nu\sigma,\sigma) e^{-t^2/2}
dt$. Since the lhs is a function of only $\nu$, the function
$\varXi(t\sigma-\nu\sigma,\sigma)$ in the integrand should not depend
on $\sigma$. This condition is represented by
$\partial\varXi(t\sigma,\sigma)/\partial\sigma = 0$ with $t$ fixed,
which is equivalent to a partial differential equation
$x\partial\varXi(x,y)/\partial x + y \partial\varXi(x,y)/\partial y =
0$. Its general solution is given by $\varXi(x,y) =
\hat{\varXi}(x/y)$, where $\hat{\varXi}$ is an arbitrary,
single-valued function. Therefore, we have
\begin{equation}
  \varXi(\delta-\delta_\mathrm{c},\sigma) =
  \hat{\varXi}\left(\frac{\delta-\delta_\mathrm{c}}{\sigma}\right),
  \label{eq:a-8}
\end{equation}
in order to have a universal mass function in Gaussian initial
conditions. If we use the form of Eq.~(\ref{eq:a-8}) in non-Gaussian
initial conditions, the mass function does not have the universal form
and the resulting multiplicity function has an additional dependence
of $\sigma$, which arises from the additional dependence of mass in
$P_\sigma(\delta)$ through higher-order cumulants. The model of
Eq.~(\ref{eq:a-2-2}) is consistent with the form of
Eq.~(\ref{eq:a-8}), and the function $\hat{\varXi}$ is identified as
the cumulative distribution function of the stochastic moving barrier,
$\varPhi(\beta)$. If the function
$\varXi(\delta-\delta_\mathrm{c},\sigma)$ were to not explicitly
depend on $\sigma$ and $\varXi(x,y) = \varXi(x)$ were independent of
$y$, the above differential equation would become
$x\partial\varXi(x)/\partial x = 0$. The unique solution with a
condition like Eq.~(\ref{eq:a-2-1}) is the step function
$\varXi(x) = \varTheta(x)$, which corresponds to the PS mass function.
Thus, the explicit dependence of the mass in the auxiliary function
$\varXi$ is necessary to obtain non-PS mass functions. Adopting
Eq.~(\ref{eq:a-8}) in Gaussian initial conditions, Eq.~(\ref{eq:a-6})
and (\ref{eq:a-7}) reduce to
\begin{align}
  F(\nu) &=
  \sqrt{\frac{2}{\pi}}
  \int_{-\infty}^\infty
  \hat{\varXi}(x-\nu)\, e^{-x^2/2} dx,
  \label{eq:a-9}\\
  \frac{f(\nu)}{\nu} &=
  \sqrt{\frac{2}{\pi}} \int_{-\infty}^\infty
  \hat{\varXi}(x-\nu)\,x\,e^{-x^2/2} dx.
  \label{eq:a-10}
\end{align}

\begin{figure}
\begin{center}
\includegraphics[width=20pc]{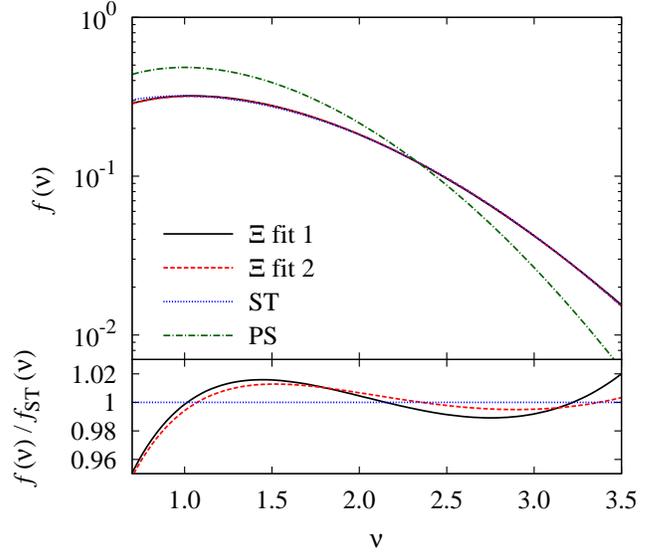}
\caption{\label{fig:fSTfit} The multiplicity functions derived by the
  model of the auxiliary function,
  $\hat{\varXi}(x) = 1/(e^{-1.802x}+1)^{1.882}$ ($\Xi$ fit 1, solid
  line) and
  $\hat{\varXi}(x) = \mathrm{erfc}[-(x-0.4778)/(0.7671\sqrt{2})]$
  ($\Xi$ fit 2, dashed line), which are fitted to give the
  Sheth-Tormen mass function (dotted line). The case of
  Press-Schechter mass function (dot-dashed line) are also shown as a
  reference. }
\end{center}
\end{figure}

Rather than deconvolving Eq.~(\ref{eq:a-7}) in some way, it is more
straightforward to find a fitting formula of $\varXi$ which can
reproduce the required mass function. As a demonstration, let us try
to find an approximate solution by assuming a simple functional form,
\begin{equation}
  \hat{\varXi}(x) = \frac{1}{(e^{-ax}+1)^b},
  \label{eq:a-18}
\end{equation}
where $a>0$ and $b>0$ are fitting parameters and Gaussian initial
conditions are assumed. This function has the desirable asymptotes of
Eq.~(\ref{eq:a-2-1}). For a given mass function with a finite range of
$\nu$, one can fit the parameters to approximately reproduce
Eq.~(\ref{eq:a-10}). We find the best fit parameters to reproduce the
ST mass function in the range $0.7 \leq \nu \leq 3.5$, which
corresponds to the fitted range of the fitting formula \cite{ST99}, to
be $a=1.802$ and $b=1.882$. The resulting mass function is shown in
Fig.~\ref{fig:fSTfit} ($\Xi$ fit 1). It is seen that the ST mass
function is precisely recovered within a few percent.

For another trial function, we consider
\begin{equation}
  \hat{\varXi}(x) = \frac{1}{2}\mathrm{erfc}
  \left[-\frac{x-\mu}{\sqrt{2}\,s}\right],
  \label{eq:a-19}
\end{equation}
where $\mu$ and $s>0$ are fitting parameters. This function is a
cumulative Gaussian distribution function with a mean $\mu$ and a
variance $s^2$ and also satisfies the property of
Eq.~(\ref{eq:a-2-1}). The best fit parameters in this case are given
by $\mu=0.4778$ and $s=0.7671$. The resulting mass function is also
shown in Fig.~\ref{fig:fSTfit} ($\Xi$ fit 2). The overall fit is
slightly better than the previous one.

If we extend the curve to the low-mass end ($\nu \lesssim 0.6$), both
fits of Eqs.~(\ref{eq:a-18}) and (\ref{eq:a-19}) somehow underestimate
the ST mass function, but in this region the ST mass function tends to
overpredict the true mass function of halos in the numerical
simulations \cite{Kly11}. It might be also possible that low-mass
halos are not described well by the simple model of Eq.~(\ref{eq:a-2})
in the first place, since the formation process of low-mass halos
could be extremely stochastic and not be described well by the local
values of the linear density field.

Finally, we comment on the difficulty in trying to analytically
deconvolve the equations by using the Fourier transformation. The
convolution integral is formally solved by the Fourier transformation,
and Eq.~(\ref{eq:a-6}) is given by
\begin{equation}
  \varXi(x, \sigma) =
  \frac{1}{2} \int_{-\infty}^\infty \frac{dk}{2\pi}
  e^{-ikx/\sigma} \frac{\tilde{F}(k)}{\tilde{P}_\sigma(k/\sigma)},
  \label{eq:a-15}
\end{equation}
where $\tilde{F}(k)$ and $\tilde{P}_\sigma(k)$ are the Fourier
transforms of $F(\nu)$ and $P_\sigma(\delta)$, respectively. For a
Gaussian distribution, we have $\tilde{P}_\sigma(k/\sigma) =
e^{-k^2/2}$, and this integral converges only if $\tilde{F}(k)$ decays
as fast as $e^{-k^2/2}$ for $k\rightarrow\infty$. Thus, the function
$F(\nu)$ should be a sufficiently smooth function in the range of
$-\infty < \nu < \infty$. Although the variable $\nu$ is a positive
number, one can apply the analytic continuation to the function
$F(\nu)$ for the negative values of $\nu$.

The Fourier transform $\tilde{F}$ can be represented directly by a
multiplicity function as
\begin{equation}
  \tilde{F}(k) = 
  \int_{-\infty}^\infty d\nu 
  \left(
    \pi \delta_\mathrm{D}(k) + \frac{i}{k}e^{-ik\nu}
  \right)
      \frac{f(\nu)}{\nu},
  \label{eq:a-16}
\end{equation}
where we assume the analytic continuation of the function $f(\nu)$
with negative argument $\nu < 0$ and use the fact that Fourier
transform of the step function is given by a formula
$\tilde{\varTheta}(k) = \pi \delta_\mathrm{D}(k) - i/k$.

For the PS mass function with a Gaussian distribution, deconvolution
with Eqs.~(\ref{eq:a-15}) and (\ref{eq:a-16}) actually works. In fact,
we have $f(\nu) = (2/\pi)^{1/2}\nu e^{-\nu^2/2}$ and
$\tilde{F}(k) = 2\pi \delta_\mathrm{D}(k) + 2ie^{-k^2/2}/k$ in this
case. Substituting the last expression and
$\tilde{P}_\sigma(k/\sigma) = e^{-k^2/2}$ into Eq.~(\ref{eq:a-15}), we
have $\varXi(x,\sigma) = \varTheta(x)$, as expected.

In the ST mass function of Eq.~(\ref{eq:3-2-5}), however, the integral
of Eq.~(\ref{eq:a-15}) does not converge. The factor $f(\nu)/\nu$ is
not regular at $\nu \rightarrow 0$ and scales as $\sim \nu^{-2p}$
near the origin. When $p>0$, the derivative of $F(\nu)$ at the origin
diverges. In the Fourier space, Eq.~(\ref{eq:a-16}) indicates that
$\tilde{F}(k) \sim |k|^{2p-2}$ for large $|k|$, and the integral of
Eq.~(\ref{eq:a-15}) does not converge for
$\tilde{P}_\sigma(k/\sigma) \sim e^{-k^2/2}$. Thus, the convolution
equation, Eq.~(\ref{eq:a-6}), does not have a regular solution when
the function $f(\nu)/\nu$ is singular at $\nu=0$, as in the case of ST
mass function. The nonexistence of the solution in this case is more
easily understood by Eq.~(\ref{eq:a-7}). According to this equation,
we have
$f(\nu)/\nu|_{\nu\rightarrow 0} = -2\sigma\int\varXi(\delta,\sigma)
[\partial P_\sigma(\delta)/\partial\delta] d\delta$. The rhs of this
equation is finite as long as the distribution function
$P_\sigma(\delta)$ is a regular function and cannot reproduce the
singularity of the lhs. This property is the reason why smooth models
of $\varXi$, such as Eq.~(\ref{eq:a-18}), tend to underestimate the ST
mass function extrapolated to the low-mass end.


\begin{widetext}

\section{On the connection between peak theory and the iPT}
\label{app:PeaksiPT}

In this Appendix, we highlight the connection that exists between the
(Lagrangian) renormalized bias function in iPT
Refs.~\cite{Mat11,Mat14} and the polynomial series expansion of
Refs.~\cite{Des13,LMD15}. Unlike in iPT, where the renormalized bias
functions are defined independently of the statistical correlators
under consideration, we shall start from the peak two-point
correlation in Lagrangian space. Therefore, our conclusions formally
apply to the two-point correlation only. However, we will argue below
that it should also hold for higher-order correlation functions.

The Lagrangian, two-point correlation $\xi_\mathrm{pk}(r)$ of the
density peaks can generically be written as
\begin{equation}
  \left[ 1 + \xi_\mathrm{pk}(r)\right]\bar{n}_\mathrm{pk}^2 =
  \int d^Ny_1 d^Ny_2\,n_\mathrm{pk}(\bm{y}_1)\,n_\mathrm{pk}(\bm{y}_2)
  P(\bm{y}_1,\bm{y}_2;r)\;,
\end{equation}
where $n_\mathrm{pk}(\bm{y})$ is the localized number density of the
biased tracers (represented here as a set of constraints applied to
the linear fluctuations fields $\bm{y}$), whereas $\bar n_\mathrm{pk}$
is the average number density.

We can write down the joint probablity distribution function (PDF)
$P(\bm{y}_1,\bm{y}_2;r)$ as the Fourier transform
\begin{equation}
  P(\bm{y}_1,\bm{y}_2;r) = 
  \frac{1}{(2\pi)^{2N}}\int d^NJ_1d^NJ_2\,
  \exp\left(-\frac{1}{2}\bm{J}^\top\Sigma\,\bm{J}\right)e^{-i\bm{J}^\top\bm{y}}\;,
\end{equation}
where $N$ is the dimension of $\bm{y}_i$ and, for shorthand
convenience, we have $\bm{J}=(\bm{J}_1,\bm{J}_2)$ and
$\bm{y}=(\bm{y}_1,\bm{y}_2)$. Moreover,
$\Sigma\equiv(\bm{M},\bm{B}^\top;\bm{B},\bm{M})$ is the covariance
matrix of $\bm{y}$. Substituting this relation into the definition of
$\xi_\mathrm{pk}(r)$, we arrive at
\begin{equation}
\label{eq:intermedxi}
  1 + \xi_\mathrm{pk}(r)
  = \prod_{a=1}^2\left(\frac{1}{(2\pi)^N\bar{n}_\mathrm{pk}}
    \int d^NJ_a\,\tilde{n}_\text{pk}(\bm{J}_a)
    e^{-(1/2)\bm{J}_a^\top\bm{M}\bm{J}_a}\right) 
  \, \exp\left(-\bm{J}_1^\top\bm{B}^\top\bm{J}_2\right) \;,
\end{equation}
where
\begin{equation}
\label{eq:CJ}
\tilde{n}_\mathrm{pk}(\bm{J}_a)\equiv 
\int d^Ny_a\,n_\mathrm{pk}(\bm{y}_a) e^{-i\bm{J}_a^\top\bm{y}_a}
\end{equation}
is the Fourier transform of the localized number density. 

We will now expand $\exp(-\bm{J}_1^\top\bm{B}^\top\bm{J}_2)$ in series
and exploit the fact that the covariance matrix $\bm{M}$ can be block
diagonalized, i.e.
$\bm{M}={\rm diag}(\bm{M}_1,\dots,\bm{M}_i,\dots,\bm{M}_p)$. Let
$\bm{J}_a=(\bm{J}_{a,1},\dots,\bm{J}_{a,i},\dots,\bm{J}_{a,p})$ be the
corresponding decomposition of $\bm{J}_a$ in the frame in which
$\bm{M}$ is diagonal (not necessarily unique block diagonal
decomposition, but there is certainly a unique frame in which the
number of blocks is maximal). Substituting the expression of
$n_\mathrm{pk}(\bm{J})$, Eq.(\ref{eq:CJ}), into
Eq.(\ref{eq:intermedxi}), we obtain
\begin{align}
  \xi_\mathrm{pk}(r) &=
  \frac{1}{\bigl[(2\pi)^N\bar{n}_\mathrm{pk}\bigr]^2}
  \int\!\!d^NJ_1\left\{\int\!\!d^Ny_1\,n_\mathrm{pk}(\bm{y}_1)
    e^{-i\bm{J}_1^\top\bm{y}_1}\right\} 
  e^{-\frac{1}{2}\bm{J}_1^\top\bm{M}\bm{J}_1}
\nonumber \\
  &\hspace{5pc} \times
  \int\!\!d^NJ_2\left\{\int\!\!d^Ny_2\, n_\mathrm{pk}(\bm{y}_2)
    e^{-i\bm{J}_2^\top\bm{y}_2}\right\} 
  e^{-\frac{1}{2}\bm{J}_2^\top\bm{M}\bm{J}_2} 
  \left(\sum_{n=1}^\infty\frac{(-1)^n}{n!}
  \left(\bm{J}_1\bm{B}^\top\bm{J}_2\right)^n \right)
\nonumber \\
  &= \sum_{n=1}^\infty \frac{(-1)^n}{n!}\sum_{I_1,L_1=1}^p\cdots
  \sum_{I_n,L_n=1}^p
  \frac{1}{\bar{n}_\mathrm{pk}}
  \int\!\!d^Ny_1\, n_\mathrm{pk}(\bm{y}_1)\left\{\frac{1}{(2\pi)^N}\int\!\!d^NJ_1\,
  \bm{J}_{1,I_1}^\top\times\dots\times \bm{J}_{1,I_n}^\top
  e^{-\frac{1}{2}\bm{J}_1^\top\bm{M}\bm{J}_1}
  e^{-i\bm{J}_1^\top\bm{y}_1}\right\}
  \bm{B}_{I_1L_1}^\top \times \dots \times \bm{B}_{I_n L_n}^\top
 \nonumber \\
  &\hspace{5pc} \times
  \frac{1}{\bar{n}_\mathrm{pk}}
  \int\!\!d^Ny_2\,
  n_\mathrm{pk}(\bm{y}_2)\left\{\frac{1}{(2\pi)^N}\int\!\!d^NJ_2\, 
  \bm{J}_{2,L_1}\times\dots\times \bm{J}_{2,L_n}
  e^{-\frac{1}{2}\bm{J}_2^\top\bm{M}\bm{J}_2} 
  e^{-i\bm{J}_2^\top\bm{y}_2}\right\}  
\nonumber \\
  &= \sum_{n=1}^\infty\frac{(-1)^n}{n!} \sum_{I_1,L_1=1}^p\cdots
  \sum_{I_n,L_n=1}^p\bigg\{\frac{1}{\bar{n}_\mathrm{pk}} \int\!\!d^Ny_1\, 
  n_\mathrm{pk}(\bm{y}_1)\,
  i^n\frac{\partial^\top}{\partial\bm{y}_{1,I_1}}
  \cdots\frac{\partial^\top}{\partial\bm{y}_{1,I_n}}
  P(\bm{y}_1)\bigg\} 
   \bm{B}_{I_1L_1}^\top \times \dots \times
  \bm{B}_{I_nL_n}^\top
\nonumber \\
&\hspace{5pc} \times
  \bigg\{\frac{1}{\bar{n}_\mathrm{pk}} \int\!\!d^Ny_2\,
  n_\mathrm{pk}(\bm{y}_2)\, 
  i^n\frac{\partial}{\partial\bm{y}_{2,L_1}}
  \cdots\frac{\partial}{\partial\bm{y}_{2,L_n}} 
  P(\bm{y}_2) \bigg\}\;. 
\end{align}
Here, $I_\alpha$ (respectively, $L_\alpha$) designates the subsets of
variables $\bm{y}_{1,I_\alpha}\in\bm{y}_1$ (respectively,
$\bm{y}_{2,L_\alpha}\in\bm{y}_2$) that correlate {\it at a given
  spatial location}. The block diagonalization implies that we have
$p\leq N$ such subsets. Through a suitable change of variables, we can
also write $\bm{y}_{1,I_\alpha}$ in the form
$\bm{y}_{1,I_\alpha}=(\bm{w}_{I_\alpha},\bm{\Omega}_{I_\alpha})$,
where $\bm{\Omega}_{I_\alpha}$ are angles which we want to integrate
out. For illustration, in the case of the peak constraint for which
$\bm{y}_1=\{\nu,\eta_i,\zeta_{ij}\}$, we can split $\bm{y}_1$ into
three subsets,
$\bm{y}_1=(\bm{y}_{1,I=1},\bm{y}_{1,I=2},\bm{y}_{1,I=3})$ such that
\begin{align}
\bm{y}_{1,I=1} &= \{\nu , J_1\} \\
\bm{y}_{1,I=2} &= \{\eta_1,\eta_2,\eta_3\} = \{\eta^2, \mbox{2 angles}\} \\
\bm{y}_{1,I=3} &= \{\tilde{\zeta}_{ij}\} = \{J_2, J_3, \mbox{3 angles}\} \;,
\end{align}
where $\tilde\zeta_{ij}$ are the five independent components of the
Hessian; the two angles in $\bm{y}_{1,2}$ and the three angles in
$\bm{y}_{1,3}$ describe the orientation of the vector $\bm{\eta}$ and
the principal axis frame of the tensor $\tilde\zeta_{ij}$,
respectively; and the invariants $J_i$ are defined in
Ref.~\cite{LMD15}.

Furthermore, the cross-covariance matrix $\bm{B}_{IL}^\top$ is of the
form
\begin{equation}
\bm{B}_{IL}^\top = \int\!\!\frac{d^3k}{(2\pi)^3}\,
\bm{\mathcal{U}}_I(-\bm{k})\, \bm{\mathcal{U}}_L^\top\!(\bm{k})\, P_\mathrm{L}(k)\,e^{i\bm{k}\cdot\bm{r}} \;,
\end{equation}
where $\delta(\bm{k})$ are the Fourier mode of the {\em unsmoothed}
linear density field, $P_0(k)$ is its power spectrum and
$\bm{\mathcal{U}}_I(\bm{k})$ are functions of the wave number
analogous to those introduced in Eq.(\ref{eq:3-5}). For instance,
\begin{equation}
  \bm{\mathcal{U}}_{I=1}(\bm{k})
  = \left(\frac{1}{\sigma_0},\frac{k^2}{\sigma_2}\right) W_s(kR_s)
\end{equation}
for the peak height $\nu$ and curvature $J_1$, and
\begin{equation}
\bm{\mathcal{U}}_{I=2}(\bm{k})
=\frac{i}{\sigma_1}\left(k_1,k_2,k_3\right)W_s(kR_s)
\end{equation}
for $\bm{y}_{1,2}$ corresponding to the vector components $\eta_i$, whereas 
\begin{equation}
\bm{\mathcal{U}}_{I=3}(\bm{k})=\frac{1}{\sigma_2}
\left(-k_1^2+\frac{k^2}{3},-k_2^2+\frac{k^2}{3},-k_3^2+\frac{k^2}{3},-k_1k_2,-k_1k_3,-k_2k_3\right)  
W_s(kR_s)
\end{equation}
for the components $\tilde{\zeta}_{ij}$ of the traceless matrix. Here,
$k^2=k_1^2+k_2^2+k_3^2$ and $W_s(kR_s)$ is the Fourier transform of
the filtering kernel. We use the same notation as Ref.~\cite{Mat11} to
emphasize that we are talking about the same quantity.

Substituting this relation into the expression of
$\xi_\mathrm{pk}(r)$, we obtain
\begin{align}
\xi_\mathrm{pk}(r) &= \sum_{n=1}^\infty\frac{1}{n!}
\sum_{I_1,L_1=1}^p\cdots\sum_{I_n,L_n=1}^p 
\int\!\!\frac{d^3k_1}{(2\pi)^3}\cdots\int\!\!\frac{d^3k_n}{(2\pi)^3}\\ 
&\qquad \times
\bigg\{\frac{1}{\bar{n}_\mathrm{pk}} \int\!\!d^Ny_1\, 
n_\mathrm{pk}(\bm{y}_1)\,
\frac{\partial^\top}{\partial\bm{y}_{1,I_1}}
\bm{\mathcal{U}}_{I_1}\!(-\bm{k}_1)\cdots
\frac{\partial^\top}{\partial\bm{y}_{1,I_n}}
\bm{\mathcal{U}}_{I_n}\!(-\bm{k}_n)P(\bm{y}_1)
\bigg\}
\nonumber \\
&\qquad \times
\bigg\{\frac{1}{\bar{n}_\mathrm{pk}} \int\!\!d^Ny_2\, n_\mathrm{pk}(\bm{y}_2)\,
\bm{\mathcal{U}}_{L_1}^\top\!(\bm{k}_1)\frac{\partial}{\partial\bm{y}_{2,L_1}}\cdots
\bm{\mathcal{U}}_{L_n}^\top\!(\bm{k}_n)\frac{\partial}{\partial\bm{y}_{2,L_n}}
P(\bm{y}_2) \bigg\} 
\nonumber \\
&\qquad \times
P_\mathrm{L}(k_1) \dots P_\mathrm{L}(k_n)\, 
e^{i(\bm{k}_1+\dots+\bm{k}_n)\cdot\bm{r}}\nonumber \;.
\end{align}
It is not difficult to see that the partial derivatives with respect
to the variables $\bm{y}_{1,I}$ and $\bm{y}_{2,L}$ correspond to the
renormalized bias functions of iPT. Namely, we have
\begin{align}
c_n^L(\bm{k}_1,\dots,\bm{k}_n) &\equiv \sum_{I_1,\dots,I_n=1}^p
\bigg\{\frac{1}{\bar{n}_\mathrm{pk}} \int\!\!d^Ny\, n_\mathrm{pk}(\bm{y})\,
\bm{\mathcal{U}}_{I_1}^\top\!(\bm{k}_1)\frac{\partial}{\partial\bm{y}_{I_1}}\cdots
\bm{\mathcal{U}}_{I_n}^\top\!(\bm{k}_n)\frac{\partial}{\partial\bm{y}_{I_n}}
P(\bm{y}) \bigg\}
\end{align}
For example, considering only the variables relevant to a peak
constraint and on writing $P(\bm{y})=\prod_I
P(\bm{y}_I)=P(\bm{w})P(\bm{\Omega}_\eta,\bm{\Omega}_{\tilde\zeta})$,
where $\bm{\Omega}_\eta$ and $\bm{\Omega}_{\tilde\zeta}$ are the
angles associated with $\bm{\eta}$ and $\tilde\zeta_{ij}$ and
$\bm{w}=(\nu,J_1,3\eta^2,5J_2,J_3)$, we find that the linear
renormalized bias function is
\begin{align}
\sum_{I=1}^p
\frac{1}{\bar{n}_\mathrm{pk}} &\int\!\!d^Ny\, n_\mathrm{pk}(\bm{y})\,
\bm{\mathcal{U}}_I^\top\!(\bm{k})\frac{\partial}{\partial\bm{y}_I}P(\bm{y}) 
\nonumber \\
&=\frac{1}{\bar{n}_\mathrm{pk}}\int\!\!d^Ny\, n_\mathrm{pk}(\bm{y})\,P(\bm{y})
\sum_{I=1}^p\left( P(\bm{y}_I)^{-1}\bm{\mathcal{U}}_I^\top\!(\bm{k})\frac{\partial}{\partial\bm{y}_I}P(\bm{y}_I)\right) 
\nonumber \\
&=\frac{1}{\bar{n}_\mathrm{pk}}\int\!\!d\bm{w}\, n_\mathrm{pk}(\bm{w})\,P(\bm{w})
\bigg\{\mathcal{N}(\nu,J_1)^{-1}
\frac{1}{\sigma_0}\left(\frac{\partial}{\partial\nu}
+\frac{k^2}{\sigma_2}\frac{\partial}{\partial J_1}\right)\mathcal{N}(\nu,J_1) 
\nonumber \\
&\qquad
+ \int\!\!\bm{\Omega}_\eta \, e^{3\eta^2/2} \frac{i}{\sigma_1}\sum_i k_i\frac{\partial}{\partial\eta_i} e^{-3\eta^2/2} 
+ \int\!\!\bm{\Omega}_{\tilde\zeta}e^{5J_2/2} \frac{1}{\sigma_2} \sum_{i\leq j}
\left(-k_i k_j+\frac{1}{3}\delta_{ij}k^2\right)
\frac{\partial}{\partial\tilde\zeta_{ij}} e^{-5J_2/2} \bigg\}
W_s(kR_s) \nonumber \\ 
&\equiv \left(b_{10}+ b_{01} k^2\right) W_s(kR_s) \;,
\nonumber 
\end{align}
which coincides indeed with the linear bias of peaks. We have
exploited the fact that the localized peak number density depends only
on the variables $\bm{w}$ to average the derivative operators over the
angular variables $(\bm{\Omega}_\eta,\bm{\Omega}_{\tilde\zeta})$. This
way we follow the same logic as Ref.~\cite{LMD15} and our discussion
in Sec.~\ref{sec:seminonlocal}. We have also checked that the
agreement also holds at second order, though the calculation is
already much more involved.

Therefore, this clearly suggests that the peak two-point correlation
$\xi_\mathrm{pk}(r)$ can also be written as
\begin{align}
\xi_\mathrm{pk}(r) &= \sum_{n=1}^\infty\frac{1}{n!} 
\int\!\!\frac{d^3k_1}{(2\pi)^3}\dots\int\!\!\frac{d^3k_n}{(2\pi)^3}\,
\left[c_X^{(n)}(\bm{k}_1,\dots,\bm{k}_n)\right]^2 P_\mathrm{L}(k_1)
\cdots P_\mathrm{L}(k_n)\, 
e^{i(\bm{k}_1+\dots+\bm{k}_n)\cdot\bm{r}},
\end{align}
which agrees with the iPT result in the absence of gravitationally
induced motions.
\\

\end{widetext}

\renewcommand{\apj}{Astrophys.~J. }
\newcommand{\aap}{Astron.~Astrophys. }
\newcommand{\aj}{Astron.~J. }
\newcommand{\apjl}{Astrophys.~J.~Lett. }
\newcommand{\apjs}{Astrophys.~J.~Suppl.~Ser. }
\newcommand{\apss}{Astrophys.~Space Sci. }
\newcommand{\jcap}{J.~Cosmol.~Astropart.~Phys. }
\newcommand{\mnras}{Mon.~Not.~R.~Astron.~Soc. }
\newcommand{\mpla}{Mod.~Phys.~Lett.~A }
\newcommand{\pasj}{Publ.~Astron.~Soc.~Japan }
\newcommand{\physrep}{Phys.~Rep. }
\newcommand{\ptp}{Progr.~Theor.~Phys. }
\newcommand{\ptep}{Prog.~Theor.~Exp.~Phys. }
\newcommand{\jetp}{JETP }


\begin{thebibliography}{10}

\bibitem{Dav85} M.~Davis, G.~Efstathiou, C.~S.~Frenk and
  S.~D.~M.~White, \apj \textbf{292}, 371 (1985).
\bibitem{Kai84} N.~Kaiser, \apjl \textbf{284}, L9 (1984).
\bibitem{Kai87} N.~Kaiser, \mnras \textbf{227}, 1 (1987).
\bibitem{Ham92} A.~J.~S.~Hamilton, \apjl \textbf{385}, L5 (1992).
\bibitem{Ber02} F.~Bernardeau, S.~Colombi, E.~Gazta{\~n}aga, and
    R.~Scoccimarro, \physrep {\bf 367}, 1 (2002).
\bibitem{Mat11} T.~Matsubara, \prd {\bf 83}, 083518 (2011).
\bibitem{Mat14} T.~Matsubara, \prd \textbf{90}, 043537 (2014).

\bibitem{Dal08} N.~Dalal, O.~Dor{\'e}, D.~Huterer and A.~Shirokov,
    \prd \textbf{77}, 123514 (2008).

\bibitem{buc89} T.~Buchert, \aap \textbf{223}, 9 (1989).
\bibitem{mou91} F.~Moutarde, J.-M.~Alimi, F.~R.~Bouchet, R.~Pellat,
  and A.~Ramani, \apj {\bf 382}, 377 (1991).
\bibitem{buc92} T.~Buchert, \mnras \textbf{254}, 729 (1992).
\bibitem{cat95} P.~Catelan, \mnras \textbf{276}, 115 (1995).
    R.~Juszkiewicz, \aap \textbf{298}, 643 (1995).
\bibitem{bha96} S.~Bharadwaj, \apj \textbf{472}, 1 (1996).
\bibitem{cat98} P.~Catelan, F.~Lucchin, S.~Matarrese and C.~Porciani,
  \mnras \textbf{297}, 692 (1998).
\bibitem{por98} C.~Porciani, S.~Matarrese, F.~Lucchin and P.~Catelan,
  \mnras \textbf{298}, 1097 (1998).
\bibitem{RB12} C.~Rampf and T.~Buchert, \jcap \textbf{6}, 21 (2012).
%
\bibitem{BBKS} J.~M.~Bardeen, J.~R.~Bond, N.~Kaiser \& A.~S.~Szalay,
  \apj \textbf{304}, 15 (1986).
\bibitem{LMD15} T.~Lazeyras, M.~Musso \& V.~Desjacques, \prd \textbf{93},
  063007 (2016).

\bibitem{BM96} J.R.~Bond and S.T.~Myers, \apjs \textbf{103}, 1 (1996).
\bibitem{OKT04} Y.~Ohta, I.~Kayo and A.~Taruya, \apj \textbf{608},
  647 (2004).
\bibitem{SCS13} R.K.~Sheth, K.C.~Chan and R.~Scoccimarro, \prd
    \textbf{87}, 083002 (2013).

\bibitem{MW96} H.~J.~Mo and S.~D.~M.~White, \mnras {\bf 282}, 347
    (1996).
\bibitem{MJW97} H.~J.~Mo, Y.~P.~Jing, and S.~D.~M.~White, \mnras {\bf
      284}, 189 (1997).
%
\bibitem{AP90} L.~Appel and B.~J.~T.~Jones, \mnras \textbf{245}, 522 (1990).
\bibitem{PS12} A.~Paranjape and R.~K.~Sheth, \mnras \textbf{426}, 2789
    (2012).

\bibitem{VD08a} V.~Desjacques, \prd \textbf{78}, 103503 (2008).
\bibitem{Des10} V.~Desjacques, M.~Crocce, R.~Scoccimarro and
  R.K.~Sheth, \prd \textbf{82}, 103529 (2010).

\bibitem{Mat08a} T.~Matsubara, \prd \textbf{77}, 063530 (2008).

\bibitem{Mat15} T.~Matsubara, \prd \textbf{92}, 023534 (2015).

\bibitem{Mat12} T.~Matsubara, \prd \textbf{86}, 063518 (2012).

\bibitem{CSS15} K.~C.~Chan, R.~K.~Sheth, and R.~Scoccimarro,
    arXiv:1511.01909.

\bibitem{PGP09} D.~Pogosyan, C.~Gay, and C.~Pichon, \prd \textbf{80}, 081301
  (2009); \prd \textbf{81}, 129901(E) (2010). 

\bibitem{GPP12} C.~Gay, C.~Pichon, and D.~Pogosyan, \prd \textbf{85},
  023011 (2012).

\bibitem{DGR13} V.~Desjacques, J.-O.~Gong, and A.~Riotto, \jcap
    \textbf{9}, 006 (2013).

\bibitem{PS74} W.H.~Press and P.~Schechter, \apj \textbf{187}, 425
    (1974).

\bibitem{ST99} R.K.~Sheth and G.~Tormen, \mnras \textbf{308}, 119
    (1999).

\bibitem{Dor70} A.~G.~Doroshkevich, Astrofiz. \textbf{6}, 581 (1970).

\bibitem{Des13} V.~Desjacques, \prd \textbf{87}, 043505 (2013).

\bibitem{BCDP14} M.~Biagetti, K.~C.~Chan, V.~Desjacques, and
  A.~Paranjape, \mnras \textbf{441}, 1457 (2014).

\bibitem{PSD13} A.~Paranjape, R.~K.~Sheth, and V.~Desjacques, \mnras
    \textbf{431}, 1503 (2013).

\bibitem{Dizgah+15} A.~M.~Dizgah, K.~C.~Chan, J.~Nore{\~n}a,
  M.~Biagetti, and V.~Desjacques, arXiv:1512.06084.

\bibitem{Planck2015} Planck Collaboration, arXiv:1502.01589.

\bibitem{BDS15} T.~Baldauf, V.~Desjacques, and U.~Seljak, \prd
  \textbf{92}, 123507 (2015). 

\bibitem{DEFW85} M.~Davis G.~Efstathiou C.~S.~Frenk, and
    S.~D.~M.~White, \apj \textbf{292}, 371 (1985).

\bibitem{LC94} C.~Lacey and S.~Cole \mnras \textbf{271}, 676 (1994).

\bibitem{EH99} D.~J.~Eisenstein and W.~Hu, \apj {\bf 511}, 5 (1999).

\bibitem{CRW13} J.~Carlson, B.~Reid, and M.~White, \mnras
  \textbf{429}, 1674 (2013).

\bibitem{SM11} M.~Sato and T.~Matsubara, \prd {\bf 84}, 043501 (2011)

\bibitem{SM13} M.~Sato and T.~Matsubara, \prd {\bf 87}, 123523 (2013)

\bibitem{Mat08b} T.~Matsubara, \prd \textbf{78}, 083519 (2008).

\bibitem{Guz08} L.~Guzzo \textit{et al.}, Nature (London)
    \textbf{451}, 541 (2008).
\bibitem{Oku15} T.~Okumura \textit{et al.}, \pasj \textbf{68}, 47 (2016).

\bibitem{MV08} S.~Matarrese and L.~Verde, \apjl \textbf{677}, L77 (2008).
\bibitem{Slo08} A.~Slosar, C.~Hirata, U.~Seljak, S.~Ho, and N.~Padmanabhan,
  \jcap \textbf{08}, 31 (2008).

\bibitem{Planck2015nG} Planck Collaboration, arXiv:1502.01592.

\bibitem{BD15} M.~Biagetti and V.~Desjacques, \mnras \textbf{451},
    3643 (2015). 

\bibitem{DSI09} V.~Desjacques, U.~Seljak and I.T.~Iliev, \mnras
  \textbf{396}, 85 (2009).

\bibitem{DS10} V.~Desjacques and R.K.~Sheth, \prd \textbf{81}, 023526 (2010).

\bibitem{ZZJ15}, Y.~Zheng, P.~Zhang, and Y.~Jing, \prd \textbf{91},
    123512 (2015).

\bibitem{ELP12} A.~Elia, A.D.~Ludlow, and C.~Porciani, \mnras
    \textbf{421}, 3472 (2012).

\bibitem{PLS12} A.~Paranjape, T.~Y.~Lam, and R.~K.~Sheth, \mnras
    \textbf{420}, 1429 (2012).

\bibitem{Bia14} M.~Biagetti, K.~C.~Chan, V.~Desjacques, and
    A.~Paranjape, \mnras \textbf{441}, 1457 (2014).

\bibitem{Ric72} W.~H.~Richardson, J.~Opt.~Soc.~Am. \textbf{62}, 55
    (1972).

\bibitem{Lucy74} L.~B.~Lucy, \aj \textbf{79}, 745 (1974).

\bibitem{Kly11} A.~A.~Klypin, S.~Trujillo-Gomez, and J.~Primack \apj
    \textbf{740}, 102 (2011).


\end{thebibliography}

\end{document}